\documentclass{emulateapj}

\usepackage{epsfig}
\usepackage{longtable}

\begin{document}

\title{The Evolution of Dusty Star Formation and Stellar Mass Assembly in
  Clusters:  Results from the IRAC 3.6$\micron$, 4.5$\micron$, 5.8$\micron$
  and 8.0$\micron$ Cluster Luminosity Functions}

\author{Adam Muzzin\altaffilmark{1,2,3}, Gillian
  Wilson\altaffilmark{4,5}, Mark Lacy\altaffilmark{4},
  H.K.C. Yee\altaffilmark{6}, \& S. A. Stanford\altaffilmark{7,8}}

\altaffiltext{1}{Department. of Astronomy \& Astrophysics, University
  of Toronto, 50 St. George St., Toronto, Ontario, Canada, M5S 3H4} 
\altaffiltext{2}{Visitor, Spitzer Science Center, California Institute
  of Technology, 220-6, Pasadena, CA, 91125} 
\altaffiltext{3}{Current Address: Department of Astronomy, Yale
  University, New Haven, CT, 06520-8101; adam.muzzin@yale.edu} 
\altaffiltext{4}{Spitzer Science Center, California Institute of
  Technology, 220-6, Pasadena, CA, 91125} 
\altaffiltext{5}{Department of Physics and Astronomy,
University of California, Riverside, CA 92521}
\altaffiltext{6}{Department. of Astronomy \& Astrophysics, University
  of Toronto, 50 St. George St., Toronto, Ontario, Canada, M5S 3H4} 
\altaffiltext{7}{University of California, Davis, CA, 95616} 
\altaffiltext{8}{Institute of Geophysics and Planetary Physics,
  Lawrence Livermore National Laboratory, Livermore, CA, 94551} 




\begin{abstract}
We present a catalogue of 99 candidate clusters and groups of galaxies in the
redshift range 0.1 $<
z_{phot} <$ 1.3 discovered in the $Spitzer$ First
Look Survey (FLS).  The clusters are selected by their R$_{c}$ - 3.6$\micron$
galaxy color-magnitude relation using the cluster red sequence algorithm.
Spectroscopic redshifts from numerous FLS followup projects confirm
the photometric redshifts of 29 clusters and demonstrate that the R$_{c}$ -
3.6$\micron$ red sequence color
provides photometric redshifts with an accuracy of $\Delta$z = 0.04 in
the redshift range 0.1 $< z <$ 1.0.  Using this cluster sample we
compute the 3.6$\micron$, 4.5$\micron$, 5.8$\micron$, \&
8.0$\micron$ cluster luminosity
functions (LFs).  Similar to previous studies, we find that for the bands
that trace stellar mass at these redshifts (3.6$\micron$, 4.5$\micron$) the evolution in
M$^*$ is consistent with a passively evolving population of galaxies
with a high formation redshift ($z_{f}$ $>$ 1.5).  
Using the 3.6$\micron$ LF as a proxy for stellar luminosity we
remove this component from the MIR (5.8$\micron$ \& 8.0$\micron$) cluster LFs and measure the
LF of dusty star formation/AGN in clusters. We find that at $z <$ 0.4 the
bright end of the cluster 8.0$\micron$ LF is well-described by a composite population of quiescent galaxies and
regular star forming galaxies with a mix consistent with typical cluster
blue fractions; however, at $z >$ 0.4, regular star forming galaxies
are insufficient to
account for the excess of 8.0$\micron$ galaxies,
and an additional population of dusty starburst galaxies is required to properly model the 8.0$\micron$ LFs.  
Comparison to field studies at similar redshifts shows a strong
differential evolution in the field and cluster 8.0$\micron$ LFs
with redshift. At $z \sim$ 0.65 
8.0$\micron$-detected galaxies are more abundant in clusters compared to the
field, but thereafter the number of 8.0$\micron$ sources
in clusters declines with decreasing redshift and by  $z\sim$ 0.15,
clusters are underdense relative to the
field by a factor of $\sim$ 5.  The rapid differential evolution
between the cluster and field LFs is qualitatively consistent with recent
field galaxy studies that show the star formation rates of galaxies in high density environments are larger than those in low density
environments at higher redshift.
\end{abstract}


\section{Introduction}

Since the compilation of the first large samples of galaxy clusters
almost 50 years ago (Zwicky 1961; Abell 1958), clusters have been used as fundamental 
probes of the effect of environment on the evolution of galaxies.
Over this time, our understanding of this phenomenon has grown significantly, and a basic picture 
of the formation and evolution of cluster galaxies between 0 $< z <$ 1 has
emerged.  Studies of the stellar populations of cluster galaxies via
the fundamental plane (e.g., van Dokkum et al. 1998; van Dokkum \& Stanford 2003; Holden et
al. 2005) and the evolution of the
cluster color-magnitude relation (e.g., Ellis et al. 1997; Stanford et al. 1998; Gladders et
al. 1998; Blakeslee et al. 2003; Holden et al. 2004; Mei et al. 2006;
Homeier et al. 2006; Tran et al. 2007) have shown
that the majority of stars in cluster galaxies are formed at high-redshift
($z >$ 2) and that most of the evolution thereafter is the passive
aging of these stellar populations.  Studies of the evolution of the
near-infrared (NIR)
luminosity functions (LFs) of clusters have shown that not only are the stellar
populations old, but that the bulk of the stellar mass is already
assembled into massive galaxies at high-redshift (e.g., De Propris et
al. 1999; Toft et al. 2003; Strazullo et al. 2006; Lin et al. 2006; Muzzin et
al. 2007a).  Furthermore, it appears that the cluster
scaling relations seen locally ($z <$ 0.1, e.g., Lin et al. 2004;
Rines et al. 2004; Lin et al. 2003), such as the Halo Occupation
Distribution, Mass-to-Light ratio, and the galaxy number/luminosity density profile are already in place by at least
$z \sim$ 0.5 (e.g., Muzzin et al. 2007b; Lin et al. 2006).
\newline\indent
These studies suggest a picture where the formation of the stars
in cluster galaxies, as well as the
assembly of the galaxies themselves occurs at a higher redshift
than has yet been studied in detail; and that, other than the passive
aging of the stellar populations, clusters and cluster
galaxies have changed relatively little since $z \sim$ 1.  This 
picture appears to be a reasonable
zeroth-order description of the evolution of cluster galaxies;
however, there are still properties of
the cluster population which cannot be explained within this context. 
In particular, there are significant changes in the morphology
(Dressler et al. 1997; Postman et al. 2005, Smith et al., 2005),
color (e.g., Butcher \& Oemler 1984; Rakos \&
Schombert 1995; Smail et al. 1998; Ellingson et al. 2001; Margoniner
et al. 2001; Loh et al. 2008) and star-formation properties (e.g., Balogh et
al. 1999; Dressler et al. 1999;
Poggianti et al. 1999; Dressler et al. 2004; Tran et al. 2005a; Poggianti et
al. 2006, although see Kodama et al. 2004) of cluster 
galaxies since $z \sim$ 1.  The fraction of
blue, star forming galaxies increases from almost zero at $z =$ 0 to
as much as 50\% at
$z \sim 0.5$ (the so-called Butcher-Oemler Effect), and correspondingly,
the fraction of S0
galaxies in clusters drops by a factor of 2-3, with similar increase
in the number of spiral/irregular galaxies over the same redshift
range (Dressler et al. 1997).  Naively, these results suggest that gas-rich, star-forming
galaxies at high-redshift have their star-formation truncated by the
cluster environment at moderate redshift and become the
dominate S0 population seen locally.  How such a transformation occurs,
and how it avoids leaving a notable imprint on the stellar
populations, is still not well-understood.  
\newline\indent
Citing an abundance of post-starburst
(k+a) galaxies in clusters at $z \sim$ 0.4, Poggianti et al. (1999) and Dressler et
al. (2004) suggested that
there may be an abundance of dusty starburst galaxies in clusters at
moderate redshift,
and that the dusty starburst and k+a galaxies may represent the intermediate stages between regular
star forming late-type galaxies and S0 galaxies (e.g., Shioya et al. 2004; Bekki \& Couch 2003).  In
particular, they suggested that the cluster e(a)\footnote{An e(a)
  galaxy is defined as a galaxy with EW([OII]) $<$ -5\AA$ $ and
  EW(H$\delta$) $>$ 4\AA$ $ by Dressler et al. (1999). These are
  emission line galaxies with a
  strong A star component to their spectrum suggesting a recent,
  possibly obscured, burst
  of star formation.} galaxies
would be the best candidates for dusty starburst galaxies because
their inferred star formation rates appear larger from H$\alpha$
emission than from [OII] emission.  If the cluster environment
excites a dusty starburst from harassment, tidal interaction, or
ram-pressure stripping, then this may quickly deplete a star forming
galaxy of its gas, transforming it first into a k+a galaxy, and then leaving it an S0.  More detailed work on two $z \sim$
0.5 clusters by Moran et al. (2005) also showed an abundance of
starbursting galaxies conspicuously near the cluster virial radius,
suggesting a environmental origin to their ``rejuvenation''. $ISO$
observations of relatively nearby clusters have detected significant amounts of
dust-obscured star formation (e.g., Fadda et al. 2000; Duc et
al. 2002; Biviano et al. 2004; Coia et al. 2005), and this has recently been confirmed at
even higher redshift ($z =$ 0.2 - 0.8) by $Spitzer$ observations
(Geach et al. 2006; Marcillac et al. 2007; Bai et al. 2007; Fadda et
al. 2007; Saintonge et al. 2008; Dressler et al. 2008).  Despite this, it is currently unclear whether
there is a  population of dusty starbursts which is sufficiently abundant to be the progenitors of the
large number of cluster k+a galaxies.  
\newline\indent
Alternatively, there is evidence from other cluster samples that the
S0 population may simply be the result of the truncation of star formation in
infalling late-type galaxies via gas strangulation (e.g., Abraham et
al. 1996; Balogh et al. 1999; Treu et al. 2003; Moran et al. 2006) and that no
accompanying starburst occurs.  Most likely, the star formation and
morphology of galaxies are transformed
both ``actively'' (as in a starburst triggered from merging/harassment/tidal
forces) and ``passively'' (from gas strangulation or ram-pressure
stripping), and that the magnitude of each effect varies significantly from cluster-to-cluster
and possibly by epoch, which may explain why studies of small numbers of
clusters have found discrepant results.  Interestingly, both
processes can be active within massive clusters as was
demonstrated by Cortese et al. (2007), who found two interesting
galaxies in Abell 1689 and Abell 2667, one of which seems to be undergoing gas
strangulation and ram-pressure stripping, while the other is experiencing
an induced starburst. There is evidence that galaxies in clusters
that are less dynamically relaxed have larger star formation rates
(e.g.,  Owen et al. 1999; Metevier et al. 2000; Moss \& Whittle 2000; Owen et al. 2005; Moran et al. 2005; Coia et
al. 2005, and numerous others) and that
the accretion of large substructures induces starbursts from harassment
and tidal forces.
\newline\indent
The most obvious way to understand whether dusty starbursts
are important in the evolution of cluster galaxies is to observe their
abundances directly in the mid-infrared (MIR).  In particular, differences in the
MIR LFs of the cluster and field environments can be used to
determine if dusty starbursts are more common in the cluster
environment.  If so, it would suggest that environmental processes may
be responsible for triggering these events.
\newline\indent
The InfraRed Array Camera (IRAC) onboard $Spitzer$ provides a unique tool for studying this
problem.  IRAC
images in 4 bands simultaneously (3.6$\micron$, 4.5$\micron$,
5.8$\micron$, 8.0$\micron$) and this is particularly advantageous because 3.6$\micron$ and 4.5$\micron$
observations are a good proxy for the stellar mass of cluster galaxies
between $0 < z < 1$, and 5.8$\micron$ and 8.0$\micron$ are sensitive to
emission from warm dust (i.e., from dusty star forming regions) over
the same redshift range.  In particular, the Polycyclic Aromatic
Hydrocarbons (PAHs) emit strong line emission at rest frame 3.3$\micron$,
6.2$\micron$, 7.7$\micron$, 8.6$\micron$, and 11.3$\micron$
(e.g., Gillett et al. 1973; Willner et al. 1977).  These features,
in addition to the warm dust continuum, are sensitive indicators of dusty
star formation, and several studies have already shown a good
correlation between 8.0$\micron$ flux and star formation
rate (SFR\footnote{Although there is a direct correlation between
  8$\micron$ flux and SFR, the scatter in the correlation is
  approximately a factor of 2 for metal rich galaxies in the local universe, and metal
  poor galaxies can deviate by as much as a factor of 50 (e.g., Calzetti et
  al. 2007).  Because of the large scatter and metallicity dependence,
  throught this paper we do not use
  the 8$\micron$ data to quantitatively measure SFRs. Instead, we use
  the presence of enhanced
  8$\micron$ flux as a qualitative indicator of increased dusty star formation.};
e.g., Calzetti et al. 2005; Wu et al. 2005; Calzetti et al. 2007).  Therefore,
examining the suite of IRAC cluster LFs at redshifts 0 $< z <$ 1 shows
both the evolution of the majority of stellar mass in cluster
galaxies, as well as the evolution of dusty star formation in the same
galaxies.
\newline\indent
The obvious approach to measuring the presence of dusty star formation in
clusters is to observe
a handful of ``canonical'' galaxy clusters with IRAC.  However, given
that determining the LF from a single cluster suffers
significantly from Poisson noise, and perhaps most importantly,  is not necessarily representative of the average
cluster population at a given mass/epoch, a better approach would be
to stack large numbers of clusters in
order to improve the statistical errors, and avoid peculiarities
associated with individual clusters.  This approach requires targeted
observations of numerous clusters, which is time-consuming compared to other
alternatives.  For example, large-area $Spitzer$ surveys such as the
50 deg$^2$ $Spitzer$ Wide-area
Infrared Extragalactic Survey (SWIRE\footnote{SWIRE data are publically available at
  http://swire.ipac.caltech.edu/swire/}, Lonsdale et al. 2003), the
8.5 deg$^2$ IRAC Shallow
Survey (Eisenhardt et al. 2004), and the 3.8 deg$^2$ $Spitzer$ First Look
Survey (FLS\footnote{The FLS data are publically available at
  http://ssc.spitzer.caltech.edu/fls/}, Lacy et al. 2005) are
now, or soon-to-be, publically available and these 
fields already contain significant amounts of optical photometry.  These wide optical-IRAC
datasets can be employed to find clusters in the survey area itself 
using optical cluster detection methods such as the cluster red sequence (CRS) technique
(Gladders \& Yee 2000, hereafter GY00), or photometric redshifts
(e.g., Eisenhardt et al. 2008; Brodwin et al. 2006).  Subsequently, the IRAC survey data can be used to
study the LFs of clusters at a much larger range of masses and
redshifts than could be reasonably followed up by $Spitzer$.
Furthermore, these surveys also provide panoramic
imaging of clusters out to many virial radii, something that has thus far
rarely been attempted because it is time-consuming.
\newline\indent
Finding clusters with the CRS algorithm is relatively straightforward
with the ancillary data available from these surveys.  The technique exploits the fact that
the cluster population is dominated by early type galaxies, and that
these galaxies form a tight red sequence in color-magnitude
space.  If two filters which span the 4000\AA$ $ break are used to construct color-magnitude diagrams, 
early types are always the brightest, reddest galaxies at any redshift
(e.g., GY00) and therefore provide significant contrast
from the field.  The CRS technique is well-tested and provides
photometric redshifts accurate to $\sim$ 5\% (Gilbank et al. 2007a; Blindert et
al. 2004) as well as a low false-positive rate ($<$5\%, e.g., Gilbank et al. 2007a; Blindert et al. 2004;
Gladders \& Yee 2005).  The method has been used for the 100
deg$^2$ red sequence Cluster Survey (RCS-1, Gladders \& Yee 2005) and is also being used for
the next generation, 1000 deg$^2$ RCS-2 survey (Yee et al. 2007).  Variations of
the red sequence method have also been used to detect clusters in the
Sloan Digital Sky Survey (the ``BCGmax'' algorithm, Koester et
al. 2007; Bahcall et
al. 2003) as well as in the fields of X-ray surveys (e.g. Gilbank et
al. 2004; Barkhouse et al. 2006).  
\newline\indent
In this paper we combine the $Spitzer$ FLS R$_{c}$-band and 3.6$\micron$
photometry and use it to detect clusters with the CRS algorithm.
Given the depth of the data, and that the R$_{c}$ - 3.6$\micron$ filter combination
spans the rest-frame 4000\AA$ $ break to $z >$ 1, we are capable of
detecting a richness-limited sample of clusters out to $z \sim$ 1.
Using the sample of clusters
discovered in the FLS we compute the 3.6$\micron$, 4.5$\micron$, 5.8$\micron$, and 8.0$\micron$
LFs of clusters 0.1 $< z <$ 1.0 and study the role of dusty
star formation in cluster galaxy evolution.  A second paper on the
abundance of dusty starburst galaxies detected at 24$\micron$ in the same clusters using
the FLS MIPS data is
currently in preparation by Muzzin et al. (2008).
\newline\indent
The structure of this paper is as follows.  In \S 2 we give a brief
overview of the optical, IRAC, and spectroscopic data used in the
paper.  Section 3 describes the cluster-finding algorithm used to
detect clusters and \S 4 contains the FLS cluster catalogue, and a
basic description of its properties.  In \S 5 we present the IRAC
cluster LFs and \S 6 contains a discussion of these results as well as
a comparison of the cluster and field LFs.  We
conclude with a summary in \S 7.  Throughout this paper we assume an 
$\Omega_{m}$ = 0.3, $\Omega_{\Lambda}$ = 0.7, H$_{0}$ = 70 km s$^{-1}$
  Mpc $^{-1}$ cosmology.  All magnitudes are on the Vega
  system.

\section{Data Set}
\subsection{{\it Spitzer} IRAC Data and Photometry}
The IRAC imaging data for this project  was observed as part
of the publically available, {\it Spitzer} First Look Survey (FLS; see Lacy et
al. 2005 for details of the data acquisition and reduction).  The FLS
was the first science
survey program undertaken after the telescope's in-orbit-checkout was
completed.  It covers 3.8 square degrees and has imaging in the four IRAC
bandpasses (3.6$\micron$, 4.5$\micron$, 5.8$\micron$,
8.0$\micron$).  The FLS is a shallow survey with a total
integration time of only 60 seconds per pixel.  Because IRAC images all four channels
simultaneously,  the total integration time is identical in each channel.
The resulting 5$\sigma$ limiting flux densities are
20, 25, 100, and 100 $\mu$Jy in the 3.6$\micron$, 4.5$\micron$, 5.8$\micron$, 8.0$\micron$ bandpasses, respectively.
These flux densities correspond to Vega magnitudes of 18.0, 17.2,
15.2, and 14.6 mag, respectively.  The 50\% completeness limits for
the 4 channels are 18.5, 18.0, 16.0, 15.4 mag and hereafter we use
these limits for the cluster finding algorithm (\S 3) and computing
the cluster LFs (\S 5).  The data was corrected for completeness using
a third-order polynomial fit to
the survey completeness as a function of magnitude determined by
Lacy et al. (2005).  Lacy et al. compared their completeness
estimates, made
using artifical galaxies, to  completeness estimates determined
by comparing the recovery of sources in the FLS to a deeper
``verification strip''.  The completeness was similar using both
methods; however, in some cases the latter suggested it might be higher
by $\sim$ 10-15\%.  When counting galaxies we have
multiplied the formal uncertainties by an additional $\pm$
20\% of the completeness correction to account for this additional uncertainty.  
\newline\indent
Photometry for the IRAC data was performed using the SExtractor (Bertin \& Arnouts 1996)
package.  For each channel, four aperture magnitudes plus an isophotal
magnitude are computed.  The four apertures used are 3, 5, 10, and 20
IRAC pixels in diameter (3.66$''$, 6.10$''$, 12.20$''$, 24.40$''$).  The
aperture magnitudes are corrected for the flux lost outside the
aperture due to the large diffraction-limit of the telescope and 
the significant wings of the IRAC point spread function (PSF).  The aperture
corrections are computed from bright stars within the FLS field and are
listed and discussed further in Lacy et al. (2005).  
The majority of galaxies with 3.6$\micron$ $>$ 15.0 mag are unresolved, or only slightly
resolved at the resolution of the 3.6$\micron$ bandpass and therefore the 3 pixel aperture corrected
magnitude provides the best total magnitude.  For
galaxies which are extended and resolved, this small aperture is an
underestimate of their total magnitude.  
For these galaxies, a ``best'' total magnitude is measured by
estimating an optimum photometric aperture
using the isophotal magnitudes.  The
geometric mean radius of the isophote ({\it r$_{m}$} =
(A/$\pi$)$^{0.5}$, where A is the isophotal area) is compared to the radius of
each of the 4 apertures used for the aperture magnitudes ({\it r$_{1}$},{\it r$_{2}$},{\it r$_{3}$},{\it
  r$_{4}$}).  If {\it r$_{m}$} $<$ 1.1 {\it r$_{ap}$}, then that
aperture magnitude is chosen as the best total magnitude.  For objects with {\it
  r$_{m}$} $>$ 1.1 {\it r$_{4}$} the isophotal magnitude is used as
the best total magnitude.  When measuring the R$_{c}$ - 3.6$\micron$ colors, we always use
the 3 pixel aperture-corrected magnitude, even for resolved galaxies
(see discussion in \S 2.3).    
\newline\indent
Object detection was performed separately in all 4 channels and these
catalogues were later merged using a 1.8$''$ search radius.  Tests of
this matching (Lacy et al. 2005) show that this radius provides
the most reliably matched catalogues. 
\subsection{Optical Data}
\indent
The ground-based Cousins R$_{c}$-band (hereafter ``R-band'') imaging
used in this study  was obtained as part of the FLS campaign and is also publically
available.  R-band imaging covering the entire FLS IRAC and MIPS
fields was observed on the Kitt Peak 4m Mayall telescope using the
MOSAIC-1 camera.  MOSAIC-1 consists of eight 4096 $\times$
2048 CCDs, and has a field-of-view of 36 $\times$ 36 arcmin with a
pixel scale of 0.258 arcseconds per pixel.  Data reduction was performed
using the NOAO IRAF $mscred$ package and procedures, and galaxy photometry was performed using the
SExtractor (Bertin \& Arnouts 1996) package.  Typical seeing for the
images was $\sim$ 1.1 arcseconds and the 5$\sigma$ limiting magnitude
in an aperture of 3 arcseconds is 24.7 mag. The 50\% completeness
limit in the same aperture is $\sim$ 24.5 mag.  A complete discussion
of the data reduction, object finding, and photometry can 
be found in Fadda et al. (2004).  For this study we
performed additional photometry to that publically available in
order to measure fluxes in a slightly larger 3.66$''$ aperture which matches
with the smallest aperture of the IRAC data (D. Fadda, private communication).
\newline\indent
The mean absolute positional error in the astrometry for the R-band
data is 0.35$''$ (Fadda et al. 2004) and the mean positional error in
the astrometry
of bright (faint) sources in the IRAC catalogue is 0.25$''$ (1.0$''$) (Lacy et al. 2005).   Given
these uncertainties, as well as the large IRAC pixel scale,  the R-band catalogue was matched to the IRAC
catalogue by looking for the closest object within 1.5 IRAC pixels (1.8$''$) of
each IRAC detection.  Tests of  matching radii ranging between 0.3 and
3.0 IRAC pixels (0.37$''$ - 3.66$''$) showed that
the number of matches increased rapidly using progressively larger
radii up to $\sim$ 1.5 IRAC pixels and thereafter the gain in the number of
matches with increasing radius was relatively modest, suggesting that
the majority of additional matches were likely to be chance associations.
Given that the IRAC
astrometry is calibrated using bright stars from the Two Micron All
Sky Survey (2MASS, Strutskie et al. 2006), whereas the R-band data was
astrometrically calibrated using the USNO-A2.0 catalogue (Monet et
al. 1998) an additional concern was the possibility of a systematic
linear offset between the two
astrometric systems.  We attempted to iteratively correct for any systematic offset by
shifting the IRAC astrometry by the median offset of all
matched sources and then rematching the catalogues; however, 
multiple iterations could not converge to a solution significantly better than
the initial 0.2$''$ offset seen between the two systems.  Given that
this offset is less than the quoted positional errors in the two systems, 
it suggests that any systematic offset between the 2MASS and USNO-A2.0 system in the FLS
field is less than the random positional error in the R-band and IRAC
data themselves.  The iterative refinements increased/decreased the
total number of matches by $\pm$ 0.05 - 0.3\% depending on which
iteration.  Given these small variations, and the lack of
further evidence for a systematic offset between the coordinate
systems, the final matched catalogue uses the original IRAC and R-band astrometry.
\newline\indent
In approximately 4\% of cases more 
than one R-band object was located within the search radius.  In these
cases, the object closest to the IRAC centroid was taken as the match.
The space density of R-band sources is approximately 5 times higher
than the number of IRAC sources at these respective depths.  This
suggests that at most, 20\% of R-band sources have an IRAC
counterpart at the respective depths.  
\newline\indent
When there are multiple R-band matches for an IRAC detection, the
majority of cases will be where only one of the R-band detections
is the counterpart of the IRAC detection, and our approach will
provide correct colors.  Nevertheless, 
a certain percentage of the multiple matches will be when two R-band objects,
both of which have IRAC counterparts, have these counterparts blended
together into a single IRAC detection due to the large IRAC PSF.  Because the IRAC
source is a blend of two objects, but we use only one R-band
counterpart, these objects 
will be cataloged as brighter and redder than they truly are.
However, because only 4\% of IRAC sources have multiple R-band matches, and the
probability that both of those R-band sources have an IRAC counterpart
is roughly, 20\%$^2$ = 4\%, this suggests that only 4\% x 4\% = 0.16\%
of all IRAC sources are blended sources where only one R-band galaxy
has been identified as the counterpart.  
\newline\indent
Although this estimated contamination is  small, clusters have
greater surface densities of galaxies than the field, and therefore it might be
expected that cluster
galaxies are blended more frequently than field galaxies.  We measured the frequency of multiple matches for galaxies in the fields
of the clusters (\S 4) and found that 
6.5\% of IRAC sources had multiple R-band counterparts making blending
about 1.5 times more common in cluster fields.  Even
though the rate of blends is higher, it should not have a significant effect on the LFs.  Even in
the worst case that all 6.5\% of IRAC-detected galaxies with multiple R-band
matches are blended (not just coincidentally aligned with a faint
R-band galaxy in the foreground), and those blends are with a galaxy of
comparable luminosity, the values of M$^{*}$ measured from the LFs
would be only $\sim$ 0.05 mag brighter.  Given the Schechter function shape of the LF, it
is more probable that most galaxies are blended with a   fainter
galaxy and therefore 0.05 mag is likely to be the upper limit of how significantly
blending affects the LFs.  This effect is smaller than the statistical
errors in the measurement of M$^{*}$ for the LFs (\S 5) and therefore we make no attempt
to correct for it, but note that our M$^{*}$ values could be
systematically high by as much as 0.05 mag.
\newline\indent
The large IRAC PSF means that star-galaxy separation
using these data is difficult and therefore the classification of each matched
object is determined from the R-band data using the CLASS\_STAR parameter
from SExtractor.  This is done using the criteria suggested in Fadda et
al. (2004).  All objects with R $<$ 23.5 with CLASS\_STAR $<$ 0.9 are
considered galaxies.  For fainter objects with R $>$ 23.5,  
those with CLASS\_STAR $<$ 0.85 are considered galaxies.  Most stars
have R - 3.6$\micron$ colors of $\sim$ 0 in the Vega system.  The
R-band data is $\sim$ 5 mag deeper than the IRAC data and therefore most
stars detected by IRAC should be robustly removed using this classification.
\newline\indent

\subsection{Galaxy Colors}
The most important ingredient in the cluster red sequence algorithm is
the measurement of accurate colors.  Excess noise in the colors causes
scatter in the cluster red sequence and reduces the probability that
a cluster will be detected.  For images with large differences in seeing,
PSF shape, and pixel size such as the R-band and 3.6$\micron$,
measuring accurate colors can be problematic.  To this end,
significant effort was invested in finding the most appropriate way to
measure colors with this filter combination.  
\newline\indent
Studies of the
cluster red sequence using telescopes/filters with equivalently large
angular resolution differences (e.g., HST + ground based, Holden et
al., 2004; Optical and low-resolution IR, Stanford et al., 1998) have
typically measured colors by degrading the highest
resolution images using the PSF of the lowest resolution images.  This
is the most accurate way to measure colors, and is feasible for a
survey of several
clusters; however, it is time consuming for a survey the size of
the FLS that has more than a million sources detected in the R-band.
More importantly, because there are so many more
galaxies detected in R-band 
than in 3.6$\micron$, degrading those images
causes numerous unnecessary blends of R-band galaxies resulting in
an increased number of catastrophic color errors.  Degrading the resolution also
inhibits the potential for detecting distant clusters
because the signal-to-noise ratio of the faintest R-band objects becomes
much worse when they are smeared with a large PSF.
\newline\indent
The compromise is to use a fixed
aperture that provides accurate colors, yet is as large as possible
for the IRAC data (to reduce the need for aperture corrections), and
yet is as small is possible to reduce the excess sky noise in
the R-band measurement.  It is important to use the same diameter apertures for
both 3.6$\micron$ and R-band 
so that the colors of bright resolved galaxies are measured
properly.  Galaxies which are small and
mostly unresolved require an aperture of only 2-3 times the seeing
disk to measure a correct color.  In principle, colors for such
galaxies can be measured correctly using a different sized apertures for both
3.6$\micron$ and R-band (i.e., optimized apertures).  
However, because measuring the color correctly for large galaxies that
are resolved in both filters requires that the
aperture must be the same size in both filters, 
we use the same aperture for all
galaxies.  
\newline\indent
After experimenting with apertures ranging in diameter between
one to ten IRAC pixels (1.22$''$ to 12.2$''$) we determined
that the three IRAC pixel diameter aperture (3.66$''$) was the optimum
aperture because it requires a relatively small aperture correction at 3.6$\micron$
($\sim$ 10\%) yet is only slightly larger than three R-band seeing
disks, resulting in only a marginal excess sky noise being added to the R-band
aperture magnitudes.  Using this large fixed aperture
means that the photometry of faintest R-band galaxies
is not optimized because much of the aperture contains sky. As a
result, some potential in discovering the
most distant clusters is sacrificed because the faintest red galaxies
(i.e., distant red sequence galaxies) may
have excessively large photometric errors.  However, most importantly, accurate
colors are determined for all galaxies, and overall the approach
provides much better photometry than degrading the
entire survey data.
\newline\indent
As an illustration of the quality of colors achievable with this approach we show the color-magnitude
diagram of FLS J171648+5838.6, the richest cluster in the survey, in
Figure 1.  The typical intrinsic scatter of early type galaxies on the red sequence at the
redshift of this cluster ($z_{spec} =$ 0.573) is $\sim$ 0.075 (
Stanford et al. 1998; Holden et al. 2004).  As a comparison we
measure the intrinsic scatter for FLS J171648+5838.6 by subtracting the mean
photometric error from the total scatter in quadrature.  This is
slightly less rigorous than the Monte-Carlo methods used by other
authors, but provides a reasonable estimate of the scatter. For galaxies with 3.6$\micron$ $<$ 17 mag (3.6$\micron$ $>$
17 mag) the observed scatter
of the red sequence is 0.149(0.225) mag, and the mean photometric color error is 0.118(0.167)
mag, resulting in an intrinsic scatter of 0.091(0.151) mag.  We note that without knowing the
morphologies of the galaxies we are
unable to properly separate early type galaxies from bluer disk
galaxies, and therefore this measurement of the scatter is almost
certainly inflated by Sa or Sb galaxies bluer than the red sequence.  In
particular, these galaxies are generally more
prevalent at fainter magnitudes (e.g., Ellingson et al. 2001).
However, even
without morphological separation, the scatter in the color of
red sequence galaxies are in
fair agreement with scatter in the colors of typical red sequences  and
demonstrates that the 3.66$''$ aperture   works
well for measuring colors.  

\begin{figure}
\epsscale{1.1}
\plotone{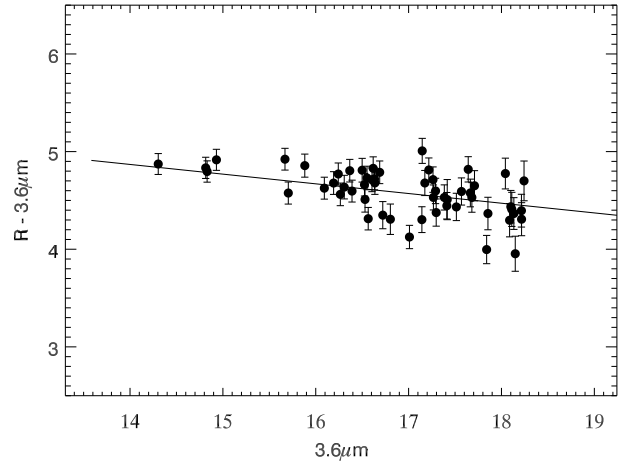}
\caption{\footnotesize Color-Magnitude diagram within a 1 Mpc (2.5
   arcmin) diameter of FLS J171648+5838.6
  (cluster 44, $z_{spec}$ = 0.573), the richest cluster in the FLS.
   Several field
   galaxies with R - 3.6$\micron$ colors $>$ 5.5 have been removed for clarity.
   The solid line is the best red sequence model for the cluster (\S
  3.1).  The intrinsic scatter in the red sequence for this cluster 
  is 0.091 mag for galaxies with 3.6$\micron$ $<$ 17 mag, and 0.151
  mag for galaxies with 3.6$\micron$ $>$ 17 mag, and is comparable
  to the scatter in other clusters at this redshift. }
\end{figure}

\subsection{Keck, WIYN, \& SDSS Spectroscopic Data}
A large number of spectroscopic redshifts are available for
galaxies in the FLS field from
several spectroscopic campaigns.    A sample of 642 redshifts were obtained using the
HYDRA spectrograph on the Wisconsin-Illinois-Yale-NOAO (WIYN) 3.6m
telescope as part of a program to followup radio sources in the FLS
(Marleau et al. 2007).  A set of 1373 redshifts
in the FLS field were obtained for galaxies selected by their red R -
K$_{s}$ colors using the DEIMOS spectrograph on the
10m KECK II telescope by Choi et al. (2006).   Lastly,
1296 redshifts were obtained using the Hectospec Fiber Spectrograph on
the 6.5m MMT by Papovich et al. (2006).  The
primary target of that survey were galaxies that are
detected in the FLS MIPS 24$\micron$ imaging and have R $<$ 21.5 mag. 
In addition to redshifts from these projects, 1192 redshifts in the
FLS field are also available from the Sloan Digital Sky
Survey (SDSS) DR5 database (Adelman-McCarthy et al. 2007).  In total there are 4503 redshifts at
various positions available in the FLS.  Of these, 26 are
likely to be cluster red sequence galaxies (see \S 3.7). 
\subsection{Palomar Spectroscopy}
In addition to the spectroscopic catalogues available, we also
obtained our own longslit spectroscopy for bright red sequence galaxies in three clusters with
0.4 $< z_{phot} <$ 0.6 in the FLS using the Double-Spectrograph (Doublespec)
on the 200-inch Hale telescope at Palomar
Mountain (P200).   We also obtained multi-object spectroscopy using the
COSMIC Spectrograph on the P200 for an additional three clusters with
$z_{phot} <$ 0.3.  These six clusters were chosen for followup because they were amongst the richest clusters in
our preliminary cluster catalogues.
\subsubsection{Double-Spectrograph Data}
Spectroscopy of bright red sequence galaxies in clusters FLS J171241+5855.9,
FLS J172122+5922.7, and FLS J171648+5838.6 (clusters 16, 38, and 44
listed in Table 1) was performed on 17, 18, and 19 August 2004 with Doublespec
on the P200.  The
observations were made with the ``Red'' camera using the 316 l/mm
grating blazed at 7150$ $\AA $ $ and a 0.5$''$ wide slit,
giving a spectral resolution of 2.6$ $\AA$ $ ($\sim$ 150 km s$^{-1}$).  The Doublespec longslit is $\sim$ 1.5
arcmin long and the angle of the slit on the sky can be rotated.  In
all 3 clusters we centered the slit on the brightest cluster galaxy
(BCG) and then chose a rotation angle so that we could get at least 2
other bright objects (preferentially red sequence galaxies) on the
slit.  
\newline\indent
For FLS J172122+5922.7 and FLS J171648+5838.6 we obtained 
spectra of 3 objects in the field, and in FLS J171241+5855.9 we managed 4.  We
obtained three 20 minute exposures for  FLS J172122+5922.7 and FLS
J171648+5838.6, which have
photometric redshifts of 0.57 and 0.55, respectively, and one 20 minute
exposure for FLS J171241+5855.9, which has a photometric redshift of 0.39.  We
also observed a spectroscopic standard, calibration lamps, dome flats
and twilight flats at the beginning of each night.  Data reduction and
wavelength calibration were performed using the standard IRAF
techniques.  After 1-d spectra were extracted, 7 of the 10 objects had a
signal-to-noise ratio suitable for cross-correlation.  One of the spectra in
FLS J171241+5855.9 has a strong emission line at 7056\AA$ $ and no possible
identification that puts it near the photo-{\it z} of the cluster.  This
object was therefore considered a field interloper.  The remaining 6
spectra (two per cluster) showed significant absorption features typical of early type
galaxies and redshifts were obtained by cross correlating them with an
elliptical galaxy spectrum.  The redshifts of the galaxies within each cluster were
similar ($\Delta$$z$ $<$ 0.01) and are in excellent agreement with
the cluster photometric redshift. These spectroscopic redshifts are
listed in the cluster catalogue (Table 1). 
\subsubsection{COSMIC Data}
Multi-object spectroscopy of both red sequence galaxies and MIPS
24$\micron$-detected galaxies in the fields of clusters FLS J171059+5934.2,
FLS J171639+5915.2, FLS J171505+5859.6, and FLS J172449+5921.3
(clusters 1, 2, 8, and 10
listed in Table 1) were performed on 26, 27, 28, 29 May 2006, and 15, 16,
17 June 2007 using the
COSMIC Spectrograph on the 200$''$ Hale Telescope at Palomar Mountain.  These observations were made with the 300 l/mm grating blazed
at 5500$ $\AA$ $ with 1$''$ wide
slits giving a spectral resolution of 8$ $\AA$ $ ($\sim$ 450 km
s$^{-1}$).  These data are part of a
larger campaign to study cluster 24$\micron$ sources and full details of
the data reduction, calibration and cross-correlation will be
presented in a future paper (Muzzin et al. 2008, in preparation).  We
obtained 17, 16, 12 and 20 good-quality spectra in the fields of FLS J171059+5934.2,
FLS J171505+5859.6, and FLS J172449+5921.3 
respectively, and redshifts were determined using cross-correlation.
Including the data from the other spectroscopic campaigns, the field of FLS J171059+5934.2 has 10 galaxies with
$\overline z$ = 0.126, the field of FLS J171639+5915.2 has 7 galaxies with  $\overline z$ =  0.129,
the field of FLS J171505+5859.6 has 9 galaxies with  $\overline z$ =  0.252, and the field
of FLS J172449+5921.3 has 12 galaxies with $\overline z$ = 0.253.  These redshifts are
included in the cluster catalogue (Table 1).
\subsection{Keck/DEIMOS Spectroscopy of FLS J172126+5856.6}
Spectroscopy was obtained of the candidate cluster FLS J172126+5856.6
(cluster 93 in Table 1) with the Deep Imaging
Multi-Object Spectrograph (DEIMOS, Faber et al. 2003) on the 10~m Keck~II telescope.
On the night of 1 September 2005, we obtained three 1800s exposures on the same mask in
non-photometric conditions with $\sim$1.3$''$ seeing.   The 600ZD grating ($\lambda_{\rm blaze} = 7500$ \AA;
$\Delta \lambda_{\rm FWHM} = 3.7$ \AA) and a GG455 order-blocking filter
were used.  The DEIMOS data were processed using a slightly modified version
of the pipeline developed by the DEEP2 team at UC-Berkeley\footnote{\tt
http://astro.berkeley.edu/~cooper/deep/spec2d/}.  Relative flux calibration was achieved from
observations of standard stars from Massey \& Gronwall (1990).
\newline\indent
Slits were preferentially placed on candidate red sequence galaxies,
allowing a total of 10 slits on likely cluster members.
Of the 10 candidate red sequence galaxies, five had sufficient S/N for determining
redshifts, and four had redshifts $\Delta z$ $<$ 0.01 from each other,
with the $\overline z$ =  1.045.  These redshifts are included in the
cluster catalogue (Table 1).
\section{Cluster Finding Algorithm}
The cluster finding algorithm employed in this study is essentially
the CRS algorithm of Gladders \& Yee (2000,
2005) with some minor modifications.  Here we outline only the major
steps, and refer to those papers
for a more detailed explanation of the procedures.  
\newline\indent
The CRS algorithm is motivated by the observation that early type
galaxies
dominate the bright end of the cluster LF and that 
these galaxies always follow a tight red sequence in the
color-magnitude plane.  At increasing redshift the observed
red sequence color becomes redder\footnote{The observed-frame color of the
  red sequence becomes redder with increasing redshift 
  because of band shifting.  The rest-frame color change due to passive
  evolution actually makes galaxies bluer at higher redshift,  but is
  a small effect for a single-burst population formed at
  high redshift.  Because the  change in observed-frame color
  is dominated by the k-correction from an old stellar population, it
  increases monotonically with redshift and provides a good estimate
  of the cluster redshift.}
and because this change in color follows closely the predictions from a
passively evolving stellar population, the color can be used as a robust
photometric redshift estimate for a cluster.  In order to apply the
CRS algorithm, slices are made in the color-magnitude plane of a
survey.  Galaxies are then assigned weights based on the probability
that they belong to a particular slice.  This probability is determined
by the color and the photometric error in the color.  Once color
weights for each galaxy in each slice have been assigned, each galaxy
is also assigned a magnitude weight.  Magnitude weighting is
done because bright red sequence galaxies are more likely to be
members of clusters than faint ones.  
\newline\indent
Once each galaxy is assigned a color and magnitude weight for each
slice, the positions of each galaxy are plotted for each slice with
their respective weights.  The resulting ``probability map'' for each
slice is then smoothed and peaks in these maps represent likely 
cluster candidates.  In the following subsections we discuss in more
detail the steps in our version of the algorithm.
\subsection{Red Sequence Models}
The first step in finding clusters with the CRS is to model the color,
slope, and intercept of the
cluster red sequence as a function of redshift.  This was done by making
simulated single-burst galaxies using all available metallicities from the Bruzual
\& Charlot (2003) spectral synthesis code.   The models are constructed
with 50\% of the stars forming in a single-burst
at t = 0, and the remainder forming with an exponentially
declining star formation rate of $\tau$ = 0.l Gyr.  Using a range of
metallicities causes the color of 
each galaxy to be slightly different at $z$ = 0, with the most metal-rich
galaxies being the reddest.  The absolute magnitude of each
galaxy with a different metallicity is normalized using the U-V, V-I,
and J-K red sequences of Coma (Bower et al. 1992) assuming that
a metallicity gradient with magnitude is the primary source of the slope of
the red sequence.
Normalizing the absolute magnitude of each galaxy this way allows us
to reproduce models with the correct red sequence color and
slope with redshift.  
\newline\indent
There is increasing evidence that the slope of the red sequence is not only
caused by a metallicity sequence, but is also the product of an
age sequence, with the less
luminous galaxies being both more metal poor and younger (e.g.,
Nelan et al. 2005; Gallazzi et al. 2006).  Examination of
spectroscopically confirmed clusters in
the FLS shows that the pure metallicity sequence used in our models
reproduces the red sequence slope and color quite well,
and because we are only interested in a fiducial red sequence model
for detecting clusters and determining photometric redshifts,
no further tuning of the ages of galaxies along the
sequence is done.  
\newline\indent
Once the absolute magnitude of each model galaxy is normalized using
the Coma red sequences,
linear fits to the R - 3.6$\micron$ vs. 3.6$\micron$ color-magnitude
relations of the model galaxies between 0.1 $< z <$ 1.6 are made.   A
high density of redshift models are fit so that there is
significant overlap in color space (185 slices between 0.1 $< z <$ 1.6).  This assures that no clusters are missed
because they have colors between the finite number of models
and it also allows for increased precision in the photometric redshifts.
\newline\indent
We computed two sets of single-burst models, one with a formation
redshift ({\it z}$_{f}$) = 2.8, and
another with {\it z}$_{f}$ = 5.0.  These two sets of models produce nearly
identical observed red sequences at {\it z} $<$ 1.1, but begin to
diverge at higher redshifts.  There is evidence from
previous studies of the fundamental plane (e.g., van Dokkum et
al. 1998; van Dokkum \& Stanford 2003, and many others), evolution of
the color-magnitude diagram (Stanford et al. 1998; Holden et al. 2004), and K-band
luminosity function (De Propris et al. 1999; Lin et al. 2006; Muzzin et al. 2007a) that a
{\it z}$_{f}$ $\sim$ 3 model is appropriate for cluster early types;
however, the uncertainties in these studies are fairly large.  There is also
evidence that many of the most massive field early types formed the
majority of their stars at {\it z} $>$ 5
(McCarthy et al. 2004; Glazebrook et al. 2004), so the possibility remains that
a {\it z}$_{f}$ = 5.0 is more appropriate.  Regardless, the majority
of the systems we have discovered are at {\it z} $<$ 1.1, and therefore
the $z_{f}$ uncertainty does
not affect the photometric redshifts of these systems.  For
systems at {\it z} $>$ 1.1, the redshift can be considered an upper
limit.  For example, the photo-{\it z} for a cluster at {\it z} = 1.3
in the {\it z}$_{f}$ = 2.8 model would be {\it z} $\sim$ 1.2 in the
{\it z}$_{f}$ = 5.0 model.  
\newline\indent
To illustrate the depth of the survey, and the location of the
red sequence models, Figure 2 shows the R - 3.6$\micron$ vs. 3.6$\micron$
color-magnitude diagram for all galaxies in the FLS with some of the
{\it z}$_{f}$ = 2.8 red sequence models overlaid.   The density of
galaxies with M
$\sim$ M$^{*}$ begins to drop off significantly for the $z >$ 1.2
red sequence models
because of the depth of the R-band data (M$^{*}$(3.6$\micron$)
$\sim$ 17.0 mag at $z =$ 1.2) and therefore we consider $z
\sim$ 1.2 the upper limit at which we can reliably detect clusters.
Remarkably, the red sequence models are even well separated in
color space at $z <$ 0.5 where the R - 3.6$\micron$ filters do not
span the 4000\AA$ $ break.  This is caused by the large wavelength
separation between the bands, and the wide 3.6$\micron$ filter
which has a strongly redshift-dependent negative
k-correction.  Although the k-correction in R-band evolves slowly with
redshift out to $z \sim$ 0.5, the k-correction for 3.6$\micron$ is
significant and therefore the R - 3.6$\micron$ color is still a sensitive
redshift indicator.  
\begin{figure*}
\plotone{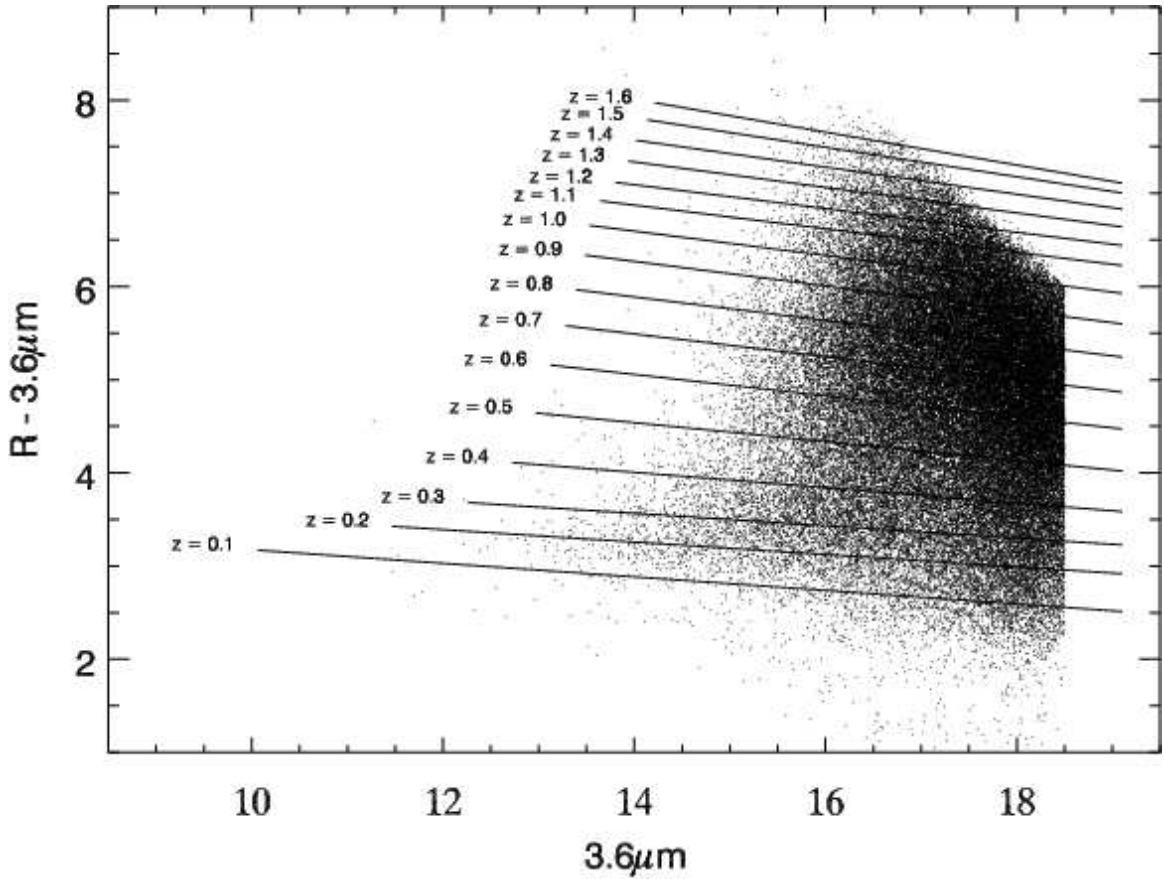}
\caption{\footnotesize Observed color-magnitude diagram for all galaxies in the
FLS.  The solid lines are fiducial red sequence models at different
redshifts generated using
the Bruzual \& Charot code.  The redshift of each model is labelled in
the figure.  The bulk of the shift in color  with redshift of the
models is due to
bandpass shifting or ``k-correction'', not because of evolution in the
rest-frame colors of the galaxies.}
\end{figure*}
\subsection{Color Weights}
Once red sequence models have been made, weights based on the
probability that a galaxy belongs within a color slice are computed.
The typical 1$\sigma$ scatter in the local cluster color-magnitude
relation is $\sim$ 0.075 mag (e.g., Lopez-Cruz et al. 2004; Bower et
al. 1992).  The scatter has been measured in clusters to z
$\sim$ 1 where it remains remarkable consistent (e.g., Stanford et
al. 1998; Gladders et al. 1998; Blakeslee et al. 2003).  Thereafter, it may become somewhat more
scattered (Holden et al. 2004).  Assuming that this relation holds to
$z \sim$ 1.3, color weights (with values ranging from 0 to 1) are assigned by computing the 
overlapping area of a galaxy's color with the red sequence assuming a
red sequence intrinsic dispersion of 0.075 mag and assuming the galaxy's color is
represented by a Gaussian centered on the measured color with a 1$\sigma$ dispersion equal to
the color error (see e.g. GY00, Figure 3 for an
example).  Using this method, the weight of a bright
galaxy lying directly on the red sequence with a color error significantly narrower than the width of the
red sequence is 1.0.  The same galaxy,
with a color error equal to the dispersion in the red sequence, has a 
weight of 0.67.  Color weights are computed for all galaxies in all 185
color slices. 
\subsection{Magnitude Weights}
In addition to the color weights, galaxies are also weighted based on
their magnitude relative to a fiducial M$^{*}$ value.  Cluster
early types are usually the brightest, reddest galaxies at a given
redshift and therefore, the brightest galaxies within a color slice
are more likely to be cluster galaxies and should be given
extra weight.  The distribution of magnitude weights was defined as
P(M) by GY00 (see their \S 4.3).  We compute the P(M)
using the data themselves, as suggested by those authors, and when doing
so we consider objects within the one-percentile highest density regime as
``cluster'' galaxies.  This is a slightly more strict cut than the ten-percentile cut
used by GY00; however, the fact that 
IR-selected galaxies are more strongly clustered than
optically-selected galaxies justifies using a
more stringent cut.  
\subsection{Probability Maps} 
Once the magnitude and color weights for all galaxies in each of the individual color slices
have been computed, a probablility map of each slice is created.  The map
is a spatial galaxy density map of the survey within each redshift slice.  The map is made
using pixels which are 125 kpc in physical size at the redshift of each slice.  The
probablility flux from each pixel is determined by placing each galaxy on the pixel
that corresponds to its location in the survey, weighted by the
product of its color
and magnitude weights.  Once each slice is constructed this way, it is
smoothed with the exponential kernel suggested in GY00 (their equation
3).  
\subsection{Noise Maps}
The noise properties of the probability maps of different color/redshifts slices
are usually different.  In particular, the maps of the highest redshift slices
tend to have large noise peaks because the survey is
only as deep as $\sim$ M$^{*}$ in those slices.  
The lower redshift probability maps have a smoother background because there are
numerous M $>$ M$^{*}$ galaxies which are more evenly distributed
spatially and
have a low probability of belonging to any slice because they have a
large color error.  The higher redshift maps are shallower, thereby
lacking the M $>$ M$^{*}$
galaxies which provide this smooth background. 
If peak finding is run on all probability 
maps using similar detection parameters, it produces significantly
different numbers of detections in different slices.  In particular, almost any noise in
the highest redshift maps results in the detection of a ``cluster''.    
\newline\indent
To circumvent this problem the parameters of the peak finding for each map can
be tuned
individually in order to produce a reasonable number of detections in
each slice;
however, the resulting cluster catalogue is clearly biased by
what is considered a ``real'' detection in a given map.
It is preferrable to have a cluster catalogue which is as
homogenously-selected as possible and based on a quantitative selection.  Therefore ``noise''
maps are constructed and are added to each probability map to homogenize their noise
properties.    
\newline\indent
The noise maps are constructed by adding fake red sequence galaxies to
each pixel of the
probability maps.  Adding a constant background of fake
galaxies does not change the
noise properties of a map because it is the variance in the number of background
galaxies that determines the noise.  We experimented with a variety of variances to add,
but settled that the variance from the photometric color errors of six M$^{*}$ red sequence galaxies per
pixel provided the best results.  This level of noise removes the spurious detections
in the highest redshift slices, but does not add so much noise as to
wash-out the majority of the poorer clusters in the lower redshift slices.  
\newline\indent
The noise in each pixel is calculated by first
determining the average color error of an M$^{*}$ red sequence
galaxy using the survey data.  Once the average color error per slice
is tabulated, the weights of six M$^{*}$ red sequence galaxies are 
Monte Carlo simulated for each pixel of a noise map assuming that the colors are
normally distributed around the red sequence with a dispersion equal
to the mean color error.  These simulated weights are
then assigned to each pixel of the noise map and each noise
map is added to the appropriate cluster probability map.  This approach thereby
implictly defines a ``cluster'' as an overdensity detectable above the
Poisson noise from six M$^{*}$ background
red sequence galaxies at any redshift.  The noise+clusters maps
have  similar noise properties for every slice and peak finding can be run using
identical parameters for all maps. 
\newline\indent
We note that in our simple empirical method for homogenizing the
noise in the probability maps the added noise is Poissionian, not clustered like the underlying background galaxy
distribution.  Despite this, the noise maps technique works extremely well, effectively smoothing out
spurious noise spikes in the highest redshift probability maps.  In principle, a more sophisticated method which includes
the clustering properties of background galaxies could be
implemented; however, for our purposes such an approach is unecessary.  Only the detection probability of poorest clusters near the
significance limit are affected by different choices in noise maps.
Galaxies from the poorest clusters do not contribute significantly to
the LFs, which are dominated by counts from more massive systems, and therefore we do not consider this issue further.
\subsection{Cluster Detection}
Once the combined noise-probability maps have been made, peaks are detected in
each map using SExtractor.  The peak-finding is done differently
from GY00 in that the individual 2d slices are searched instead
of merging the slices into a 3d datacube and searching for
3d peaks.  It is unclear how
these two methods compare; however, they are likely to be similar and 
searching the slices individually permits easy visual inspection of
the sources on each map which allows us to check any problems that
have occured
with peak finding or in the generation of the map.  Pixels 5$\sigma$ above the
background are flagged  and 25 connected pixels are
required to make a detection.  
\newline\indent
The slices are close in color space and therefore clusters
(particularly rich ones) are detected in more than one color slice.
The same cluster is identified in multiple color slices
by merging the slice catalogues using a matching radius of 8
pixels (1 Mpc).  Clusters found across as many as 20 color slices
are connected as being the same object.  The color slices are not linear in
redshift, but 20 slices correspond to $\Delta$$z \sim \pm$0.06.
These combined spatial and
color limits for connecting clusters imply that 
clusters with separations $>$ 1 Mpc in transverse distance and
$>$ 0.06 in redshift space can be resolved into distinct
systems\footnote{The color slices are closer together at $z >$ 1 and
   only systems with $\Delta z$ $>$ 0.12 can be resolved at this
   redshift.  We note that although the overall level of projections
   is likely to be low, because of the bunching up of the color
   slices, the highest redshift clusters will be the most suseptible to
   projection effects.}.  This level of sensitivity is similar to that found by
Gladders \& Yee (2005) using R - z$^{\prime}$ colors to select
clusters.  They also demonstrated that subclumps at redshift spacings much
less than this are likely to be associated subclumps or infalling
structures related to the main body of the cluster.
\subsection{Photometric Redshifts}
Each cluster is assigned the photometric redshift of the
color slice in which it is most strongly detected.  The strength of the
detection is determined by using
SExtractor to perform aperture photometry of each cluster on each
probability map.  This provides a ``probability flux'', and the cluster is
assigned to the slice in which it has the largest
probability flux.  
\newline\indent
The large number of spectroscopic redshifts available for the
FLS can be used to verify the accuracy of the red sequence photometric
redshifts.
Examining the spectroscopic catalogue for galaxies within a 1 Mpc circle around each cluster
shows that there are numerous galaxies with spectroscopic redshifts in the
field of many of the clusters.  The spectroscopic targets were chosen
with a variety of selection criteria (none of which preferentially
select early type galaxies) and therefore the majority of galaxies with redshifts are
foreground or background galaxies.  We use only 
the spectroscopic redshifts for 
galaxies which have a combined magnitude and color weight of $>$ 0.2
in order to preferentially select likely cluster members.
This cut in weight is used because it corresponds to the typical
combined magnitude and color weight of
M $<$ M$^{*}$ red sequence galaxies.  Once the cut is made there are 23 clusters which have at least one spectroscopic
redshift for a likely cluster red sequence galaxy.  Remarkably, there
are 26 galaxies which meet this criteria and 24 of these have a
spectroscopic redshift $<$ 0.1 from the photometric redshift of the
cluster.  This illustrates the effectiveness of the red sequence color at
estimating photometric redshifts provided that the galaxy has a
high-probability of being  a cluster early type.  
\newline\indent
In Figure 3 we plot spectroscopic vs. photometric redshift for these
23 clusters plus the additional 6 for which we obtained our own
spectroscopic redshifts ($\S$ 2.5).  The straight line marks a one-to-one
correlation.  Large points represent clusters with more than one
galaxy with a redshift consistent with the being in the cluster.
Small points represent clusters with a single spectroscopic redshift.
Excluding the large single outlier with $z_{spec} \sim$ 0.9 (which is likely to be a
bluer galaxy at high-redshift based on its spectrum and IRAC colors, see $\S$6.3) the rms scatter in the cluster spectroscopic
vs. photometric redshift is $\Delta$$z$ = 0.04, demonstrating that the
photometric redshifts from the red sequence algorithm work extremely
well.  
\newline\indent
The accuracy of the photometric redshifts from the FLS sample is
comparable to the accuracy of the RCS surveys (Yee et al. 2007; Gladders
\& Yee 2005) which use R - z$^{\prime}$
color selection, even though the R - 3.6$\micron$ colors
have larger photometric errors than the R - z$^{\prime}$ colors.  It
is likely this is because 
the model red sequence colors change much
more rapidly with redshift in R
- 3.6$\micron$ than in R - z$^{\prime}$ (R - 3.6$\micron$ spans 2
magnitudes between $z$ = 0.5 and $z$ = 1.0, whereas it spans only 1
magnitude in R - z$^{\prime}$).  The larger change in the R -
3.6$\micron$  colors with
redshift means that photometric measurement errors should correspond to smaller errors in photometric
redshift.  
\begin{figure}
\plotone{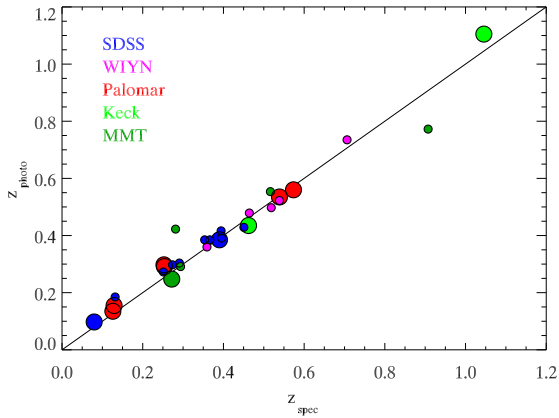}
\caption{\footnotesize Photometric vs. spectroscopic
  redshift for clusters in the FLS field.  The color of the circle
  corresponds to the telescope/project where the spectroscopic redshifts were obtained (see
  $\S$2).  Large circles denote clusters with more than one
  spectroscopic redshift, small circles denote clusters with only one
  spectroscopic redshift.  Excluding the one large outlier at $z_{spec} \sim$
  0.9, the rms scatter is $\Delta$z = 0.04.}
\end{figure}
\subsection{B$_{gc}$ Richness Parameter}
The final step in the selection of the cluster sample  is to cut
low-richness detections from the catalogue.  The false postive rate is
higher for low richness systems (i.e., galaxy groups) and
we prefer restrict our analysis of the cluster LFs to a high confidence sample
of massive clusters.  The cluster richnesses are measured quantitatively using
the B$_{gc}$ richness parameter (Longair \& Seldner 1979; for a
detailed look at the application of B$_{gc}$ to measuring cluster
richnesses see Yee \& L\'{o}pez-Cruz
1999).  B$_{gc}$ is the
amplitude of the 3-dimensional correlation function between cluster
galaxies and the cluster center.  B$_{gc}$ is measured within a fixed
aperture (typically 500 kpc radius) and is well-correlated with cluster physical parameters such
as velocity disperion ($\sigma$), X-ray temperature (T$_{x}$), and
the radius as which the mean density of the cluster exceeds the
critical density by a factor of 200 (R$_{200})$ (e.g., Yee \&
Lopez-Cruz 1999; Yee \&
Ellingson 2003; Gilbank et al. 2004; Muzzin et al 2007b).  
\newline\indent
Gladders \& Yee (2005) introduced a new form of the B$_{gc}$
parameter, counting the overdensity of red sequence galaxies within a
fixed aperture, rather
than all galaxies and defined this new parameter as B$_{gc,R}$.  This form of richness suffers less from cosmic
variance in the background because red sequence galaxies provide
better contrast with the field, and therefore it is a more
robust estimate of the cluster richness.  We use B$_{gc,R}$ rather
than B$_{gc}$ for determing the richnesses of the FLS
clusters.  The net
number  of 3.6$\micron$ red sequence galaxies  with M $<$ M$^{*}$ + 1.0 (where
M$^{*}$ is determined from the data itself, see $\S$5.1) are counted within a fixed
aperture of 500 kpc radius.  The
model red sequences from $\S$3.1 are used and galaxies within $\pm$ 0.3
in color are considered to belong to the red sequence.  
\newline\indent
Systems with B$_{gc,R}$ $<$ 200 are removed from the cluster catalogue.  The B$_{gc}$-M$_{200}$ 
relation of Yee \& Ellingson (2003) implies that this corresponds to
removing groups with
M$_{200}$ $<$ 6.6 x 10$^{13}$ M$_{\odot}$, where M$_{200}$ is defined
as the mass contained with R$_{200}$.  Groups with masses below
this typically have only $\sim$ 5-10 bright galaxies (e.g., Balogh et
al. 2007), making them difficult to  select robustly with the CRS
algorithm.  
 The systems that are removed by the richness cut are typically tight compact groups
of 3-4 extremely bright galaxies that are the same color.  Although they
are not rich, they have a
strong probability of being detected by the CRS algorithm because of
their luminosity and compactness.  It is likely that the majority of these systems are bona fide
low-richness galaxy groups; however, we have no way of verifying the
false-positive rate for these systems.
\newline\indent
Before these low-richness systems are cut from the catalogue there are 134
cluster candidates between 0.1 $< z <$ 1.4 in the FLS field.  Removing
systems with B$_{gc,R}$ $<$ 200 leaves a total of 99 candidate clusters in the
sample.  
\subsection{Cluster Centroids}
\indent
Defining a centroid for clusters can be a challenging task, yet is extremely
important because properties determined within some aperture around
the cluster (such as
richness, or LF) can vary
strongly with the choice of centroid.  In many
cluster studies the location of the BCG is used as the center of the cluster.  This is
a reasonble definition as frequently the BCG lies at the center of the
dark matter halo and X-ray emission; however, there are also many
examples where it does not.  Furthermore, not all clusters have an
obvious BCG, particularly at higher redshift.
\newline\indent
Given these issues, two centroids are
computed for the FLS
clusters, one based on the location of the peak of the red sequence probability flux
in the probability maps, and the other based on the location of the BCG within 500 kpc of this
centroid.  In order to avoid bright foreground galaxies the brightest galaxy in the field
with a red sequence weight $>$ 0.4 is designated as the BCG.  Eye examination of the
clusters shows that this criteria is effective at choosing what
appears visually to be the
correct galaxy; however, because it chooses only a single galaxy this technique is still potentially
suseptible to red low-redshift field interlopers.
\newline\indent
When computing the cluster LFs, only one of the centroids can be used.
We define an ``optimum'' centroid for each cluster
using the B$_{gc,R}$ parameter.  B$_{gc,R}$ is computed at both
centroids and the optimum  centroid is the centroid which
produces the maximum value of B$_{gc,R}$.  This approach is
simplistic, but because B$_{gc}$ is the correlation amplitude between
the cluster center and galaxies, the centroid which produces the
largest value should be the best centroid of the galaxy population.
\section{Properties of the Cluster Catalogue}
\indent
The final cluster catalogue of 99 clusters and groups is presented in
Table 1.  Where spectroscopic redshifts are available for
high-probability cluster members they are listed
in column 3, with the number of redshifts in parenthesis.  For each
cluster we also compute an estimate of R$_{200}$ and M$_{200}$.  The M$_{200}$ values are estimated using the
correlation between B$_{gc}$ and M$_{200}$ measured by Muzzin et
al. (2007b) for 15 X-ray selected clusters at $z \sim$ 0.3 in the
K-band.  The K-band and 3.6$\micron$ bandpasses sample similar parts
of a galaxy's spectrum at 
0.1 $< z <$ 1.5 and therefore it is reasonable to assume that
B$_{gc}$ values measured in both these bands will be comparable.  The
best-fit relation between M$_{200}$ and B$_{gc}$ is
\begin{equation}
Log(M_{200}) = (1.62 \pm 0.24)Log(B_{gc}) + (9.86 \pm 0.77).
\end{equation}
Muzzin et al. (2007b) did not measure the correlation between B$_{gc}$
and R$_{200}$ in the K-band, although Yee \& Ellingson (2003) showed a
tight correlation for the same clusters using $r$-band selected
B$_{gc}$.  Using the Muzzin et al. (2007b) K-band data we fit
the correlation between these parameters for those clusters and find
that the best fit relation is
\begin{equation}
Log(R_{200}) = (0.53 \pm 0.09)Log(B_{gc}) - (1.42 \pm 0.29).
\end{equation}
The rms scatter in the M$_{200}$ - B$_{gc}$ relation is 35\% and for
the R$_{200}$ - B$_{gc}$ relation it is 12\%.  These scatters are
similar to that measured between M$_{200}$ and K-band selected richness
(parameterized by N$_{200}$) at $z \sim$ 0 by Lin et al. (2004).  The
values of M$_{200}$ and R$_{200}$ derived from these equations are
listed in columns 9 and 10 of Table 1, respectively.
\newline\indent
We caution that these equations have only been calibrated using rich
clusters, and that extrapolating to  lower
richness clusters such as those in the FLS may not be appropriate.  The lowest richness
cluster in the Muzzin et al. (2007b) 
sample has a richness of Log(B$_{gc}$) = 2.8, yet the majority of
clusters in the FLS (70/99) have lower richnesses than this.
There is evidence from both observations 
(e.g., Lin et al. 2004) and numerical simulations (e.g., Kravtsov et
al. 2004) that the same power-law correlation between 
cluster galaxy counts (which are closely related to B$_{gc}$) and M$_{200}$
extend to richnesses well lower than our B$_{gc,R}$ $>$ 200 cut, and
therefore it probably not too unreasonable to extrapolate equations (1) and (2)
to lower richnesses.
\newline\indent
Using an indirect method to estimate M$_{200}$ and R$_{200}$ means that
reliable errors in R$_{200}$ and M$_{200}$ can not be computed for
individual clusters; however,  
the rms scatters in the correlations are at least indicative of the
average uncertainty in the measurement of the parameters for the sample.
Therefore, we suggest that the average error in the M$_{200}$ and R$_{200}$ values listed in Table 1
are $\pm$ 35\% and 12\% respectively, but that the error in a
particular cluster can be several times larger or smaller.  
\newline\indent
In Figure 4 we plot a histogram of the number of clusters as a
function of redshift in the FLS.  The solid histogram shows the
distribution of all clusters and the dot-dashed histogram shows the
distribution of clusters with M$_{200}$ $>$ 3 x 10$^{14}$ M$_{\odot}$
(B$_{gc,R} >$ 700).  Similar to the predictions of
numerical simulations (e.g., Haiman et al. 2003) the number of
clusters peaks at $z \sim$ 0.6.  Qualitatively, the distribution of clusters is also
similar to that found  by Gladders \& Yee (2005) in comparable size
patches; however, the cosmic variance in the number of clusters in 
$\sim$ 4 deg$^2$ patches is  too large to make a meaningful
comparison between the selection of clusters in the R - $z^{\prime}$
bandpasses versus the R - 3.6$\micron$ bandpasses.
\newline\indent
We plot the locations of the clusters superposed on the 3.6$\micron$
image of the FLS field in Figure 5 as
open circles.  Large and small circles represent 
clusters with M$_{200}$ $>$ 3 x 10$^{14}$ M$_{\odot}$ and M$_{200}$ $<$ 3 x 10$^{14}$ M$_{\odot}$, respectively,
and clusters with photometric redshifts 0.1 $< z <$ 0.4, 0.4 $< z <$ 0.8,
z $>$ 0.8 are plotted as blue, green, and red circles,
respectively.  The clusters themselves are clearly clustered;
demonstrating the need for wide-field surveys when searching for
representative samples of galaxy clusters.  
\newline\indent
We show a few examples of some of the richest cluster candidates in
Figures 6 - 11.  The top left panel for each Figure is the R-band
image; the top right is the 3.6$\micron$ image, and the bottom left
panel is the 8.0$\micron$ image.  All images are 1 Mpc across at
the cluster redshift.  The bottom right panel of each figure shows a
histogram of the color distribution of galaxies with M $<$ M$^{*}$
within  a 1 Mpc diameter aperture.
The color of the red sequence for the photometric redshift is marked
with an arrow.  The dashed histogram is the mean color background in
that aperture measured from the entire survey.  The error bars on the
dashed histogram are computed as the 1$\sigma$ variance in each bin from
200 randomly selected 1 Mpc apertures within the survey.  Galaxies are
clustered, and therefore assuming the variance is Gaussian-distributed
is probably an overestimate of the true variance (because there will
be large wings in the distribution due to clustering); however, it provides a first-order demonstration
of the overdensity of the cluster relative to the field.
\newline\indent 
Overall, the cluster catalogue is qualitatively similar in both redshift, and
richness distributions to catalogues selected with the same technique
in different bandpasses (e.g. Gladders \& Yee 2005, Gilbank et
al. 2004), demonstrating that clusters can be reliably selected with the
CRS method on IRAC data despite the limited spatial resolution of the instrument. 
\begin{figure}
\plotone{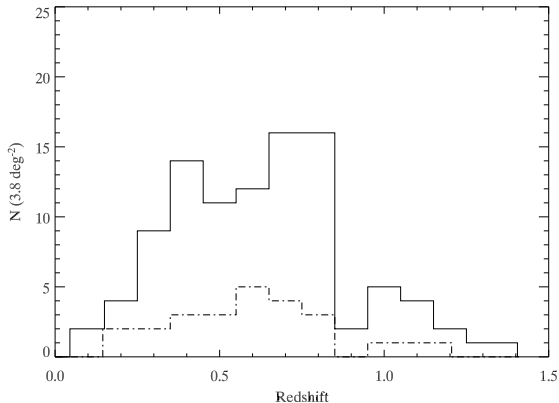}
\caption{\footnotesize Redshift distribution of clusters in the FLS.
  The solid histogram is for all clusters and the dot-dashed histogram
  is for clusters with M$_{200}$ $>$ 3 x 10$^{14}$
  M$_{\odot}$.}
\end{figure}
\begin{figure*}
\plotone{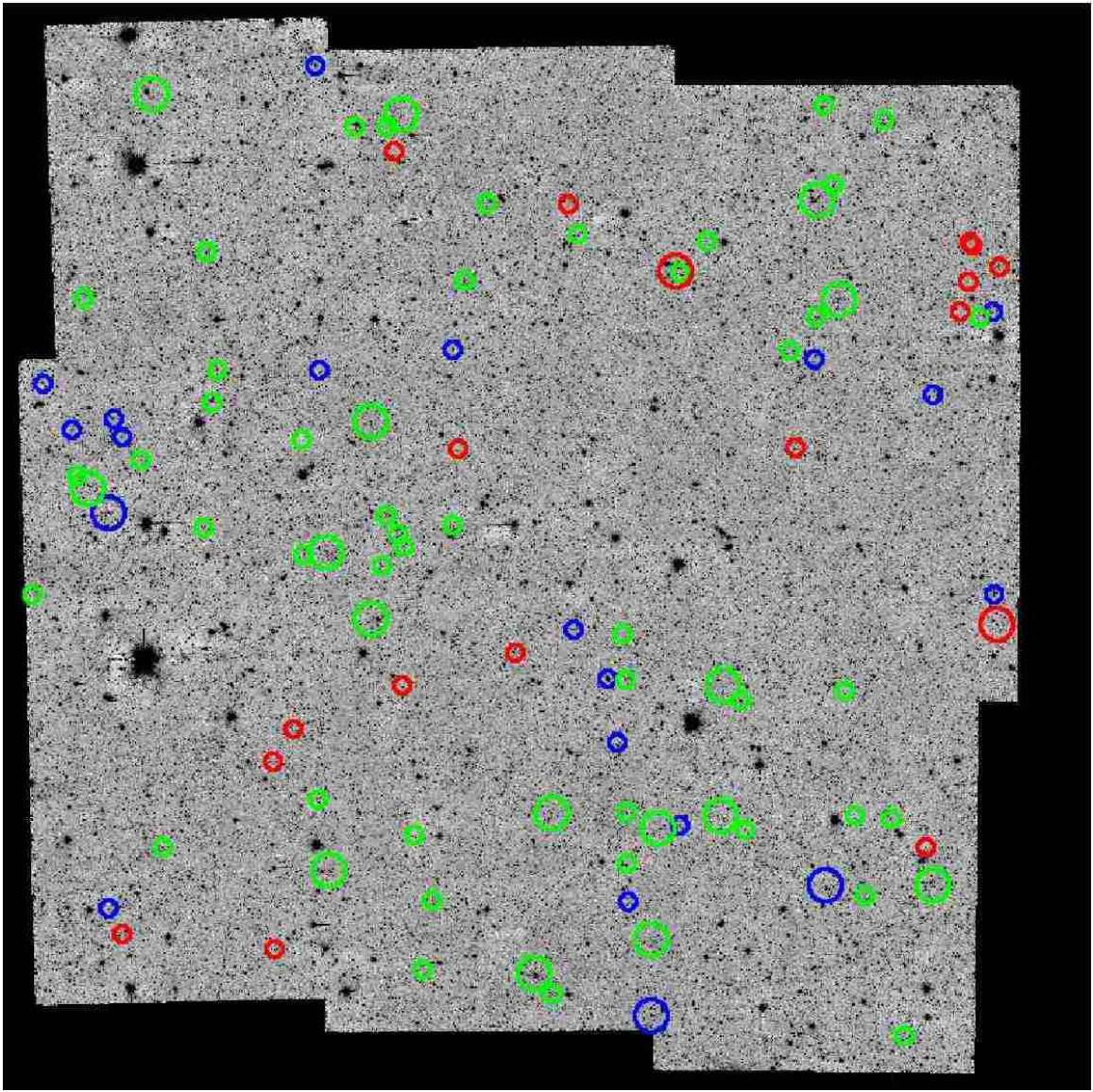}
\caption{\footnotesize The 3.6$\micron$ image of the FLS with the
  positions of clusters superposed.  The blue, green, and red circles denote clusters with
  0.1 $< z <$ 0.4, 0.4 $< z <$ 0.8, and $z >$ 0.8 respectively.  Large circles represent clusters
  with M$_{200}$ $>$ 3 x 10$^{14}$ M$_{\odot}$ and small circles
  represent clusters with M$_{200}$ $>$ 3 x 10$^{14}$
  M$_{\odot}$.  The size of the circles is arbitrarily chosen for
  clarity and is not related to the projected size of R$_{200}$ for the clusters.  }
\end{figure*}
\begin{figure}
\plotone{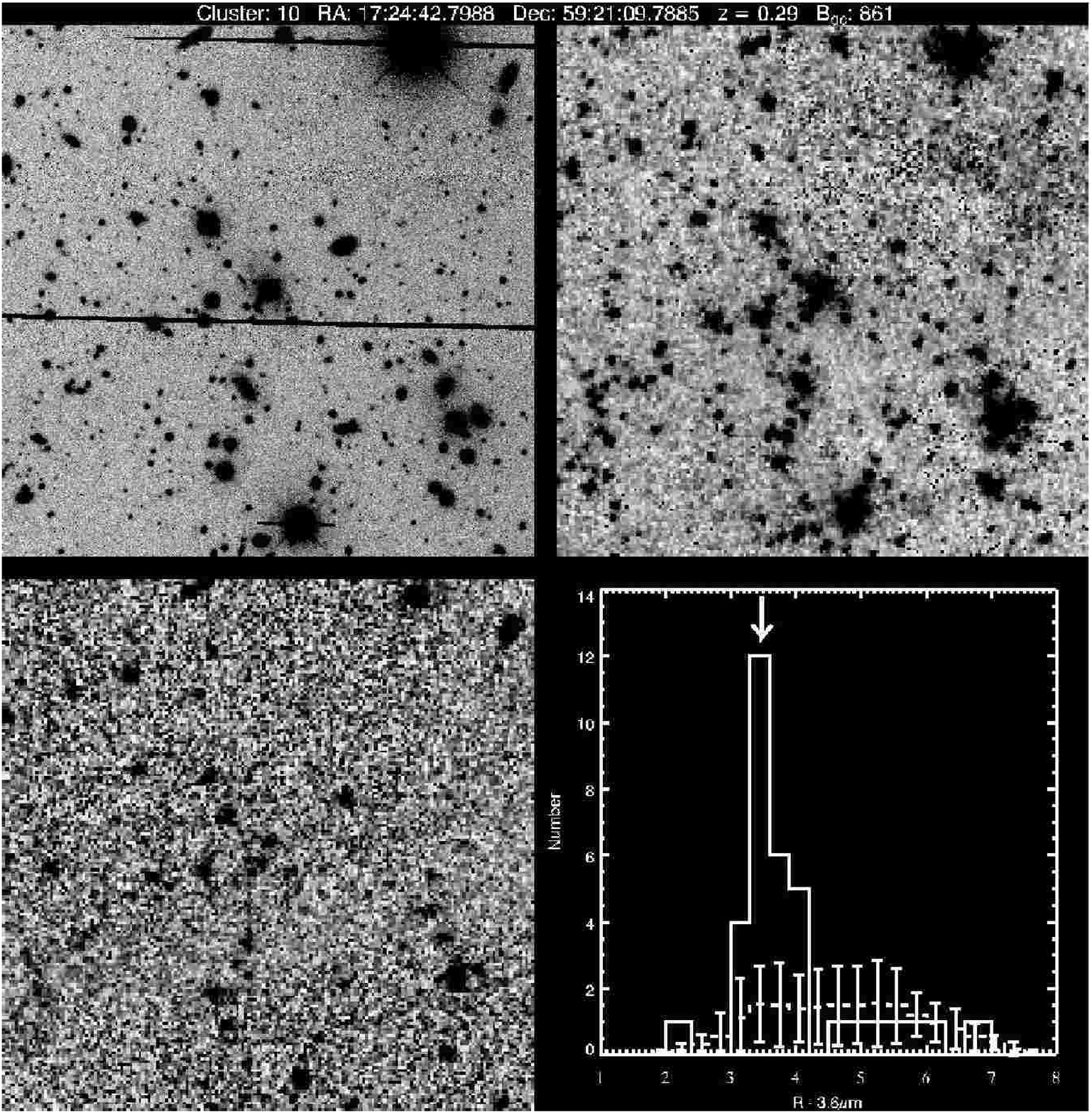}
\caption{\footnotesize Multi-wavelength images of FLS J172449+5921.3
  at $z_{spec}$ = 0.252 (Cluster \#10 from Table 1).  The top
  left, top right, and bottom left panels are the R-band, IRAC
  3.6$\micron$, and IRAC 8.0$\micron$ respectively.  In each image
  the field-of-view is 1 Mpc across at the redshift of the cluster.  The
  solid histogram in the bottom
  right panel shows the color distribution of galaxies with M $<$
  M$^{*}$ in the same field.  The dashed histogram is the background distribution in
  the same aperture and the error bars show the average variance in
  the background.  The arrow marks the color of the red sequence from
  the color-redshift models.  The cluster red sequence is clearly
  detected at many sigma above the background.}
\end{figure}
\begin{figure}
\plotone{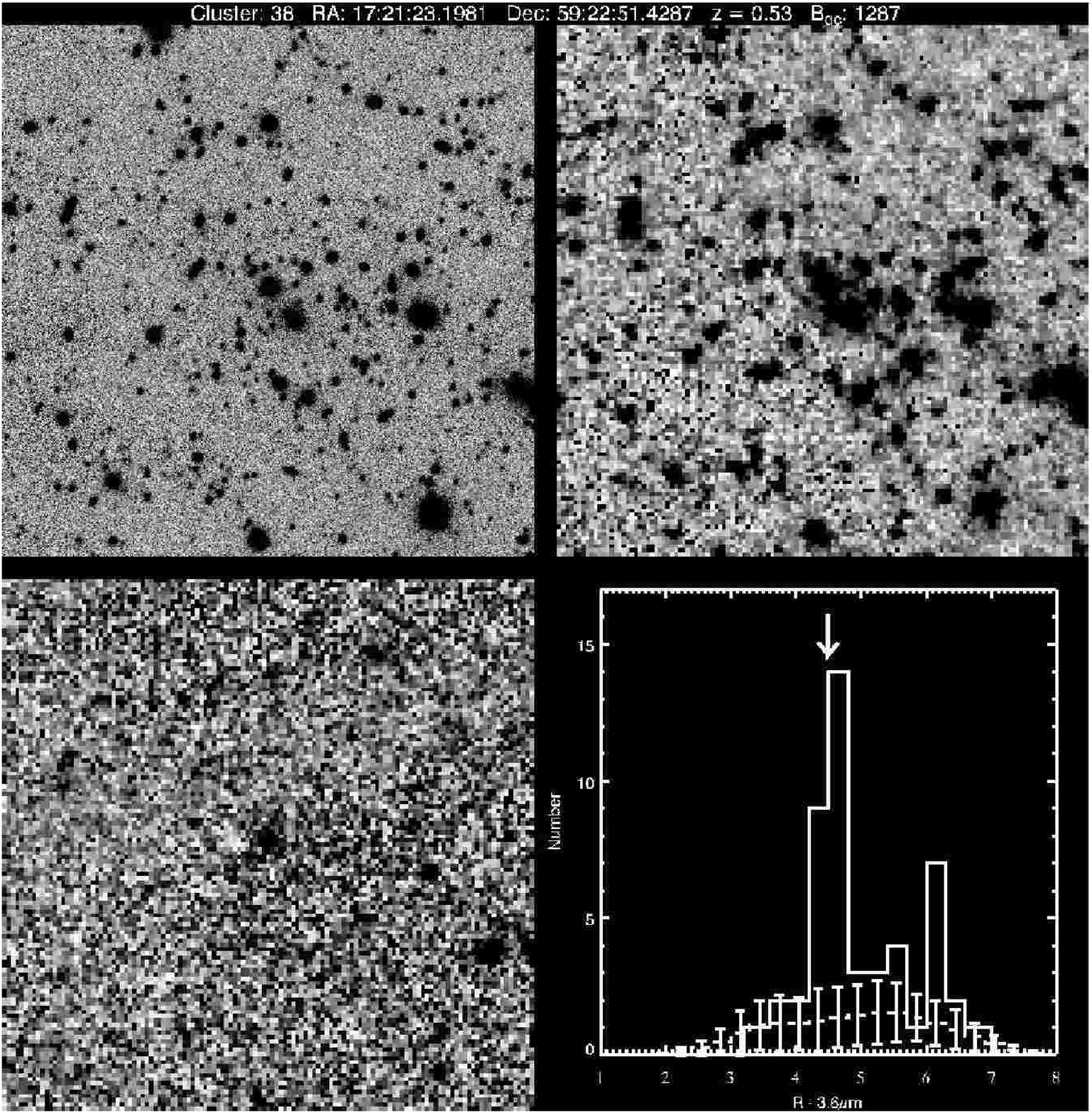}
\caption{\footnotesize Same as for Figure 6, but for FLS
  J172122+5922.7 at $z_{phot}$ = 0.53 (Cluster \#38 from Table 1).}
\end{figure}
\begin{figure}
\plotone{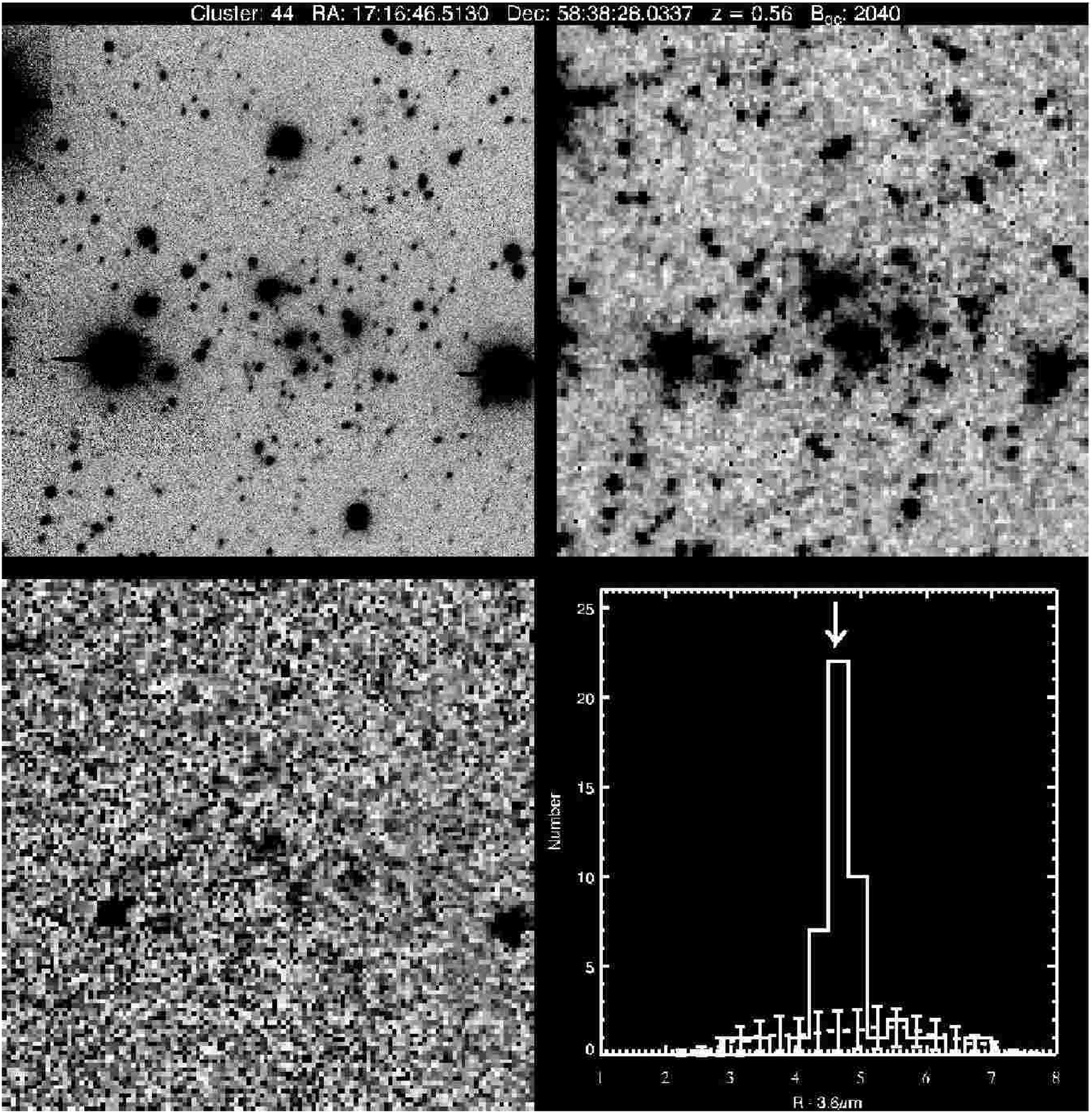}
\caption{\footnotesize Same as for Figure 6, but for FLS J171648+5838.6 at $z_{phot}$ = 0.56 (Cluster \#44 from Table 1).}
\end{figure}
\begin{figure}
\plotone{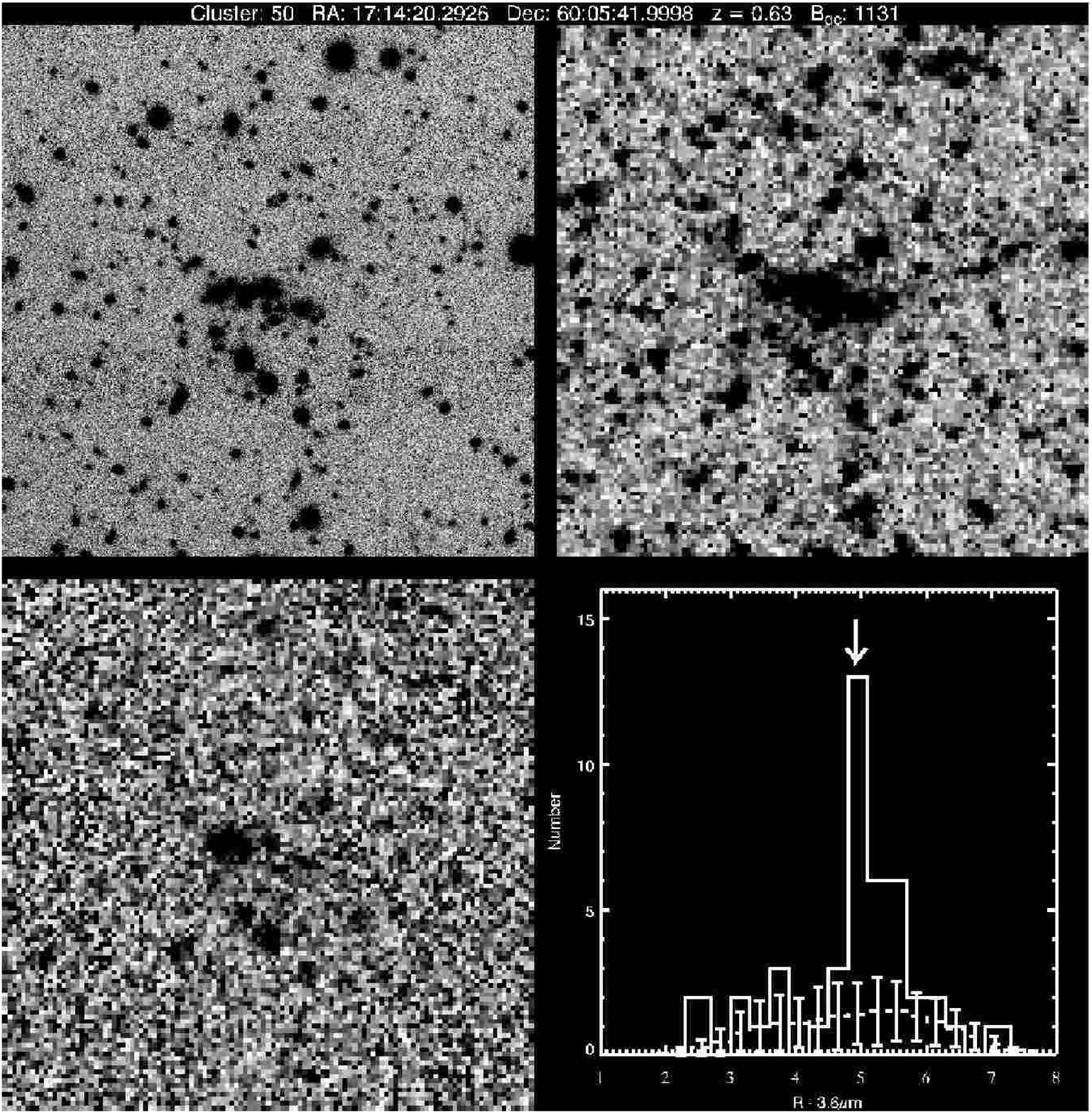}
\caption{\footnotesize Same as for Figure 6, but for FLS J171420+6005.5 at $z_{phot}$ = 0.63 (Cluster \#50 from Table 1).}
\end{figure}
\begin{figure}
\plotone{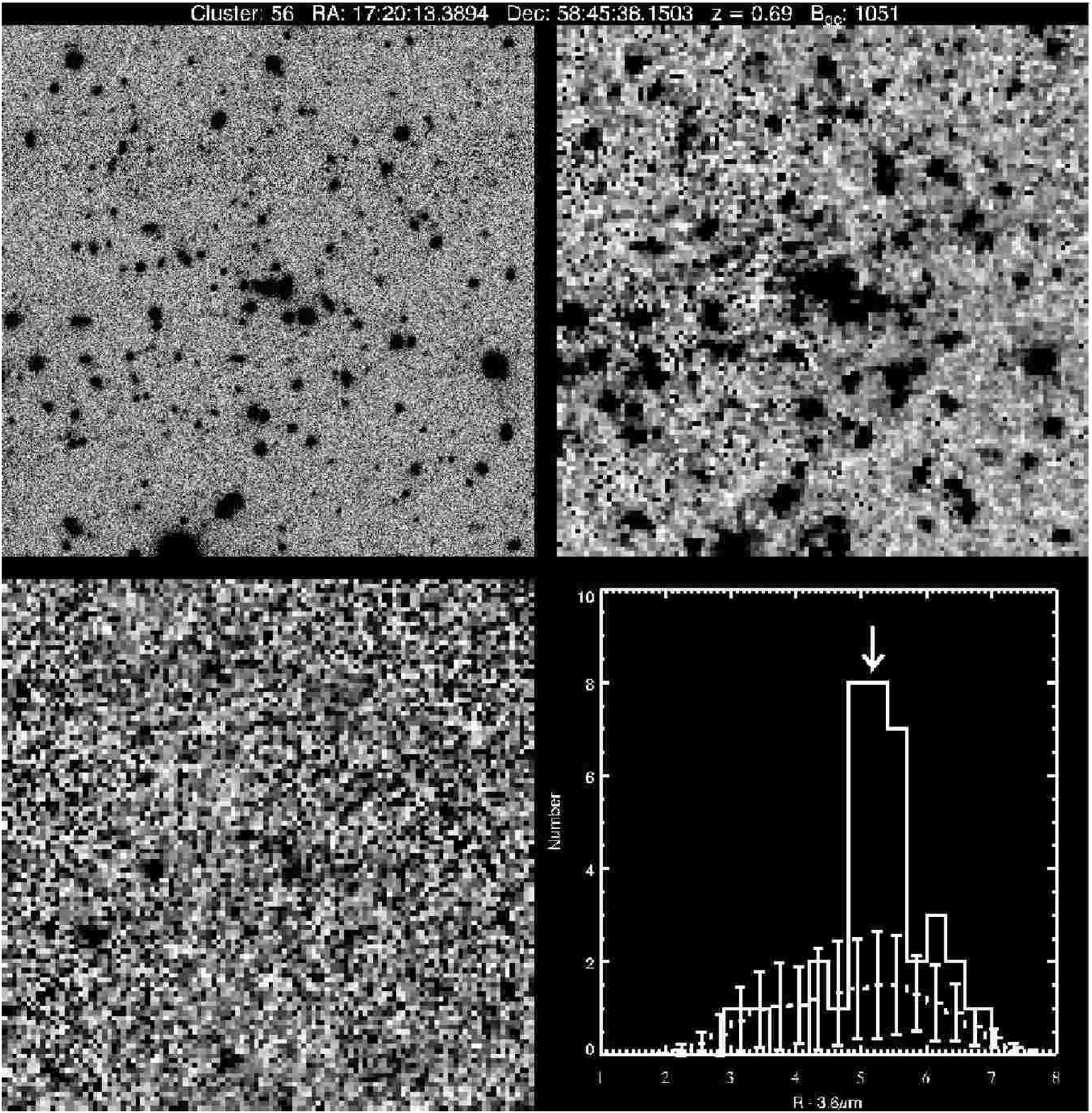}
\caption{\footnotesize Same as for Figure 6, but for FLS J172013+5845.4 at $z_{phot}$ = 0.69 (Cluster \#56 from Table 1).}
\end{figure}
\begin{figure}
\plotone{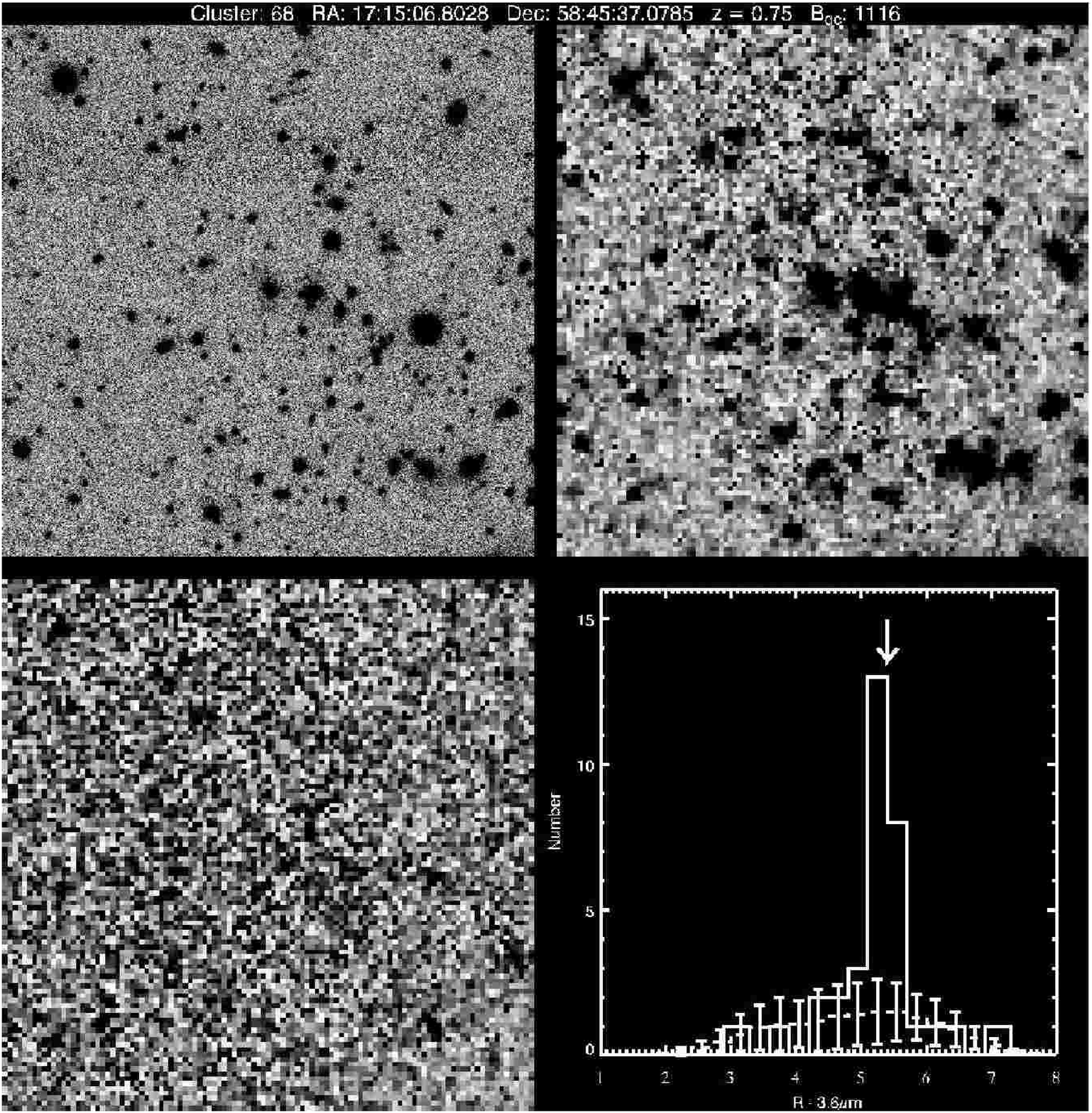}
\caption{\footnotesize Same as for Figure 6, but for FLS J171508+5845.4 at $z_{phot}$ = 0.75 (Cluster \#68 from Table 1).}
\end{figure}
\section{Cluster Luminosity Functions}
\indent
In this section we measure the IRAC luminosity functions of the FLS
cluster sample and use these to study the
evolution of stellar mass assembly and dusty star formation/AGN activity in clusters.
\subsection{The 3.6$\micron$ and 4.5$\micron$ Luminosity Functions}
The luminosity of galaxies at 3.6$\micron$ and 4.5$\micron$ over
the redshift range 0.1 $< z <$ 1.5 is dominated by emission from low mass stars and
is fairly insensitive to ongoing star formation or dust.   Consequently, the
3.6$\micron$ and 4.5$\micron$ cluster LFs provide an  estimate of the
stellar mass function of cluster galaxies, and their redshift evolution can constrain
the mass assembly history of cluster galaxies.  One concern with using
these LFs as a proxy for the stellar mass function
is that at $z <$ 0.5 the 3.3$\micron$ PAH feature found in strongly star forming galaxies
can contaminate the stellar emission observed at 3.6$\micron$ and 4.5$\micron$; however,
it is likely that such contamination will be small for cluster galaxies in this
redshift range.  In a study of Luminous Infrared Galaxies
(LIRGs, L$_{IR}$ $>$ 10$^{11}$ L$_{\odot}$) with estimated star formation rates of $\sim$ 100
M$_{\odot}$ yr$^{-1}$, Magnelli et al. (2008) found that the excess emission in the IRAC
bands due to the 3.3$\micron$ PAH feature was only $\sim$ 30\%.  Given
that such luminous LIRGs are fairly rare at $z <$ 0.5 (e.g.,
P\'{e}rez-Gonz\'{a}lez et al. 2005), and the increase in flux is
small, even for strongly star forming galaxies, contamination of the 3.6$\micron$
and 4.5$\micron$ bandpasses by 3.3$\micron$ PAH emission should be negligible.  
\newline\indent
Another concern is that in the worse cases there can be variations in the stellar mass-to-light ratio
(M$_{*}$/L) of galaxies in similar bandpasses
as large as a factor of 5-7  (such as in the K-band, e.g., Brinchmann
1999; Bell \& de Jong, 2001; Bell et al. 2003).  These variations are
smaller for evolved populations such as those found in clusters
and in general the luminosity of a
galaxy at 3.6$\micron$ and 4.5$\micron$ is still a 
reasonable proxy for its stellar mass. 
\newline\indent
Exhaustive studies of both the K-band (e.g., De Propris et al. 1999, Lin et
al. 2006, Muzzin et al. 2007a) and 3.6$\micron$ \& 4.5$\micron$ 
(Andreon 2006, De Propris et al. 2007) LFs of cluster galaxies have shown that the evolution of
M$^{*}$ in these bands is consistent with a passively evolving stellar
population formed at high-redshift ($z_{f} >$ 1.5), suggesting that
the majority of the stellar mass in bright cluster galaxies is already
assembled
into massive galaxies by at least $z \sim$ 1.  Here we compute the
LFs in the 3.6$\micron$ and 4.5$\micron$ bands for the FLS clusters  to confirm that the FLS cluster sample provides
similar results, and to demonstrate that these LFs can be used to estimate the
stellar contribution to the MIR cluster LFs ($\S$5.2).
\newline\indent
The LFs are measured by stacking clusters in redshift bins of
$\Delta$z = 0.1 starting from $z = 0.1$.  For each cluster, the number of galaxies within
R$_{200}$ in 0.25 mag bins is tabulated and the expected number of background galaxies within
R$_{200}$
is subtracted from these counts.  The background counts are
determined from the entire 3.8 deg$^2$ survey area and are well
constrained.   Each background-subtracted cluster LF is then ``redshifted'' to the mean redshift of the
bin using a differential distance modulus and a differential
k-correction determined from the single-burst model
($\S$3.1).  At 3.6$\micron$ and 4.5$\micron$ the k-corrections
for galaxies are almost independent of
spectral-type (e.g., Huang et al. 2007a) and therefore using only the
single-burst k-correction rather than a k-correction based on
spectral-type  does not affect the 
LFs.  Furthermore, the differential k-corrections and
distance moduli are small (typically $<$ 0.1 mag) and do not affect the
LFs in a significant way.  
\newline\indent
The final stacked LFs are constructed by summing the individual LFs
within each bin.  The errors for each magnitude bin of the final LF
are computed by adding the Poisson error of the total
cluster counts to the Poisson error of the total background counts in
quadrature.
\newline\indent
In Figure 12 and Figure 13 we plot the 3.6$\micron$ and
4.5$\micron$ cluster LFs respectively.  The 3.6$\micron$ LFs are
fit to a Schechter (1976) function of the form
\begin{equation}
\phi(\mbox{M}) = (0.4 \mbox{ln} 10)\phi_{*}(10^{0.4(\mbox{\scriptsize
    M$^{*}$-M})})^{1+\alpha}\mbox{exp}(-10^{0.4(\mbox{\scriptsize $M^{*}$-M})}),
\end{equation}
where $\alpha$ is the faint-end slope; $\phi_{*}$, the normalization;
and M$^{*}$ is the ``characteristic'' magnitude, which indicates the
transition between
the power-law behavior of the faint-end and the
exponential behavior of the bright end.
The functions are fit using the Levenberg-Marquardt algorithm for
least-squares (Press et al. 1992) and errors are estimated from the fitting
covariance matrix.
The data are not deep enough to provide good constraints on $\alpha$,
$\phi^{*}$ and M$^{*}$ simultaneously, and therefore the faint-end slopes
of the LFs are assumed to be fixed at $\alpha$ = -0.8.  This
value is similar to the $\alpha$ = -0.84 $\pm$ 0.08 measured in the K-band for clusters at $z
\sim$ 0.3 by Muzzin et al. (2007a) as
well as the value measured in the K-band for local clusters ($\alpha$ = -0.84 $\pm$
0.02) by Lin et al. (2004).  Although assuming a fixed value of $\alpha$ precludes
measuring any evolution of the faint-end slope of the LFs with
redshift, it
removes the strong correlation between M$^{*}$ and $\alpha$ in the
fitting and, provided the evolution in $\alpha$ is
modest, it is 
the best way to measure the luminosity evolution of the
cluster galaxies via the evolution of M$^{*}$.
The fitted values of M$^{*}$ and the 1$\sigma$ errors are listed in the
upper left of the panels in Figure 12.  
\newline\indent
We plot the evolution of M$^{*}$ at 3.6$\micron$ as a function of
redshift in Figure 14 as filled circles.  Figure 14 also shows the predicted evolution
of M$^{*}$ for single-burst models with $z_{f}$ = 1.0, 1.5,
2.0, 2.8, and 5.0.  These models are normalized to M$^{*}$ = -24.02
at $z =$ 0 in the K-band, the result obtained by Lin et al. (2004) for 93 local
clusters.  This corresponds to a normalization of M$^{*}$ = -24.32 at
3.6$\micron$, assuming a K-3.6$\micron$ color from the z$_{f}$ =
2.8 passive
evolution model.  The FLS values of M$^{*}$ are consistent with most of
these models, except the $z_{f}$ = 1.0 model, for which they are
clearly too faint.  Therefore, similar to the majority of previous
studies we can conclude that the bulk of the stellar mass in bright cluster
galaxies is consistent with having been both formed and assembled at $z >$ 1.5 and has passively
evolved since then.  As a comparison, the values measured at 
3.6$\micron$ by De Propris et al. (2007) and Andreon (2006) are
overplotted as open squares and open diamonds respectively.  These
values are from spectroscopically confirmed samples of $\sim$ 40 clusters
(the majority of which are X-ray detected clusters) and both
agree well with the FLS values demonstrating that passive
evolution appears to be the ubiquitous conclusion regardless of
cluster sample.
\newline\indent
Similar to the 3.6$\micron$ LFs, the 4.5$\micron$ LFs can be
fit using a Schechter function; however, we do not perform fitting of the
4.5$\micron$ LFs.
Instead, as a demonstration of the technique presented in $\S$5.2.3
and $\S$5.2.4, we use the measured 3.6$\micron$ LFs to predict the
4.5$\micron$ LFs.  Unlike colors from the
redder IRAC channels, the
3.6$\micron$-4.5$\micron$ colors of galaxies are nearly identical for
most spectral-types over the redshift range 0.1 $< z <$ 1.5.
As a consequence, the 4.5$\micron$  LFs can be predicted from the
3.6$\micron$ LFs
using the 3.6$\micron$-4.5$\micron$
colors from almost any stellar population model.  
For simplicity, we use the passive
evolution model to predict the 4.5$\micron$ LFs.  
The inferred 4.5$\micron$ LFs are overplotted as solid lines in Figure
13.  The predicted 4.5$\micron$ LFs are consistent with the
measured ones and this demonstrates that the 3.6$\micron$ LFs
combined with
simple models for the color evolution of galaxies can
predict the LFs in other bandpasses.  Furthermore, the self-consistency
between the 3.6$\micron$ and 4.5$\micron$ LFs at $z$ = 0.15, where the 3.3$\micron$ PAH would
contaminate the 3.6$\micron$ band, and at $z$ = 0.33 where it would
contaminate the 4.5$\micron$ band suggests that the primary source of
the emission in these bandpasses at $z <$ 0.5 is stellar.
\begin{figure*}
\plotone{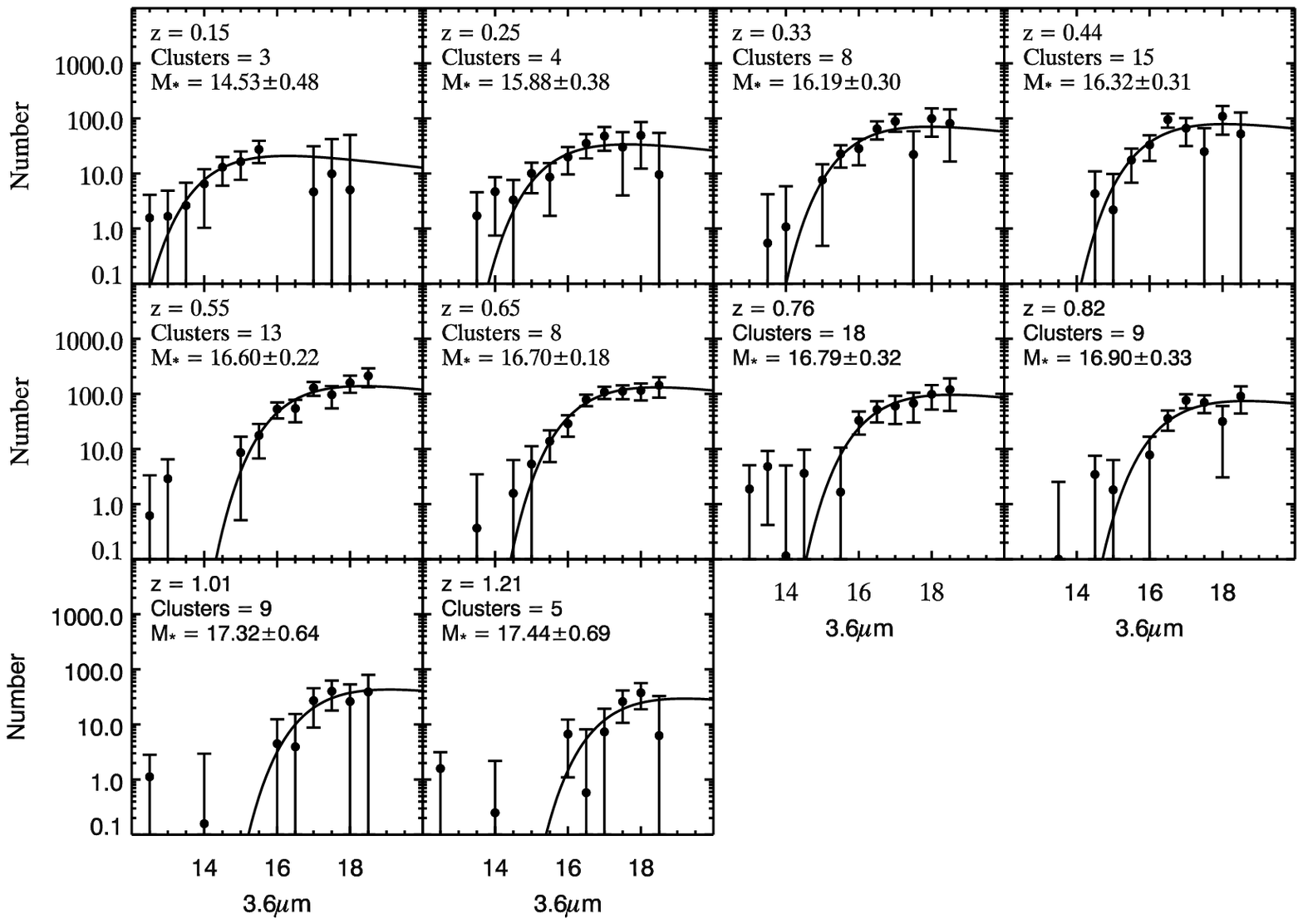}
\caption{\footnotesize The 3.6$\micron$ LFs of clusters in the FLS.
The solid line shows the best-fit Schechter function assuming $\alpha$
= -0.8.  The redshift, the value of
M$^{*}$, and the number of clusters combined to make the LF are listed in
the upper left corner of each panel.}
\end{figure*}
\begin{figure*}
\plotone{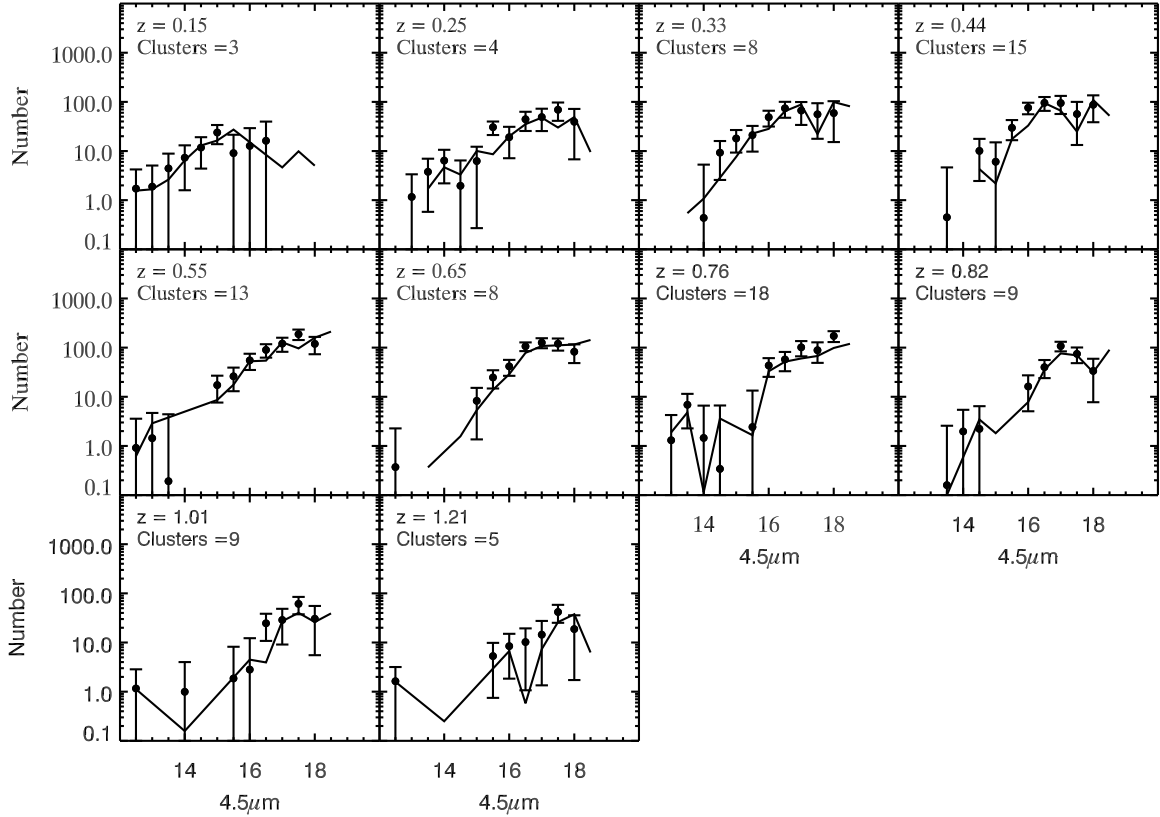}
\caption{\footnotesize The 4.5$\micron$ LFs of clusters in the FLS
  in the same redshift bins as Figure 12.
The solid line is the 4.5$\micron$ LF that is predicted from the
3.6$\micron$ LF assuming that galaxies have the
3.6$\micron$ - 4.5$\micron$ colors of a passively evolving population
formed at high redshift.}
\end{figure*}
\begin{figure}
\plotone{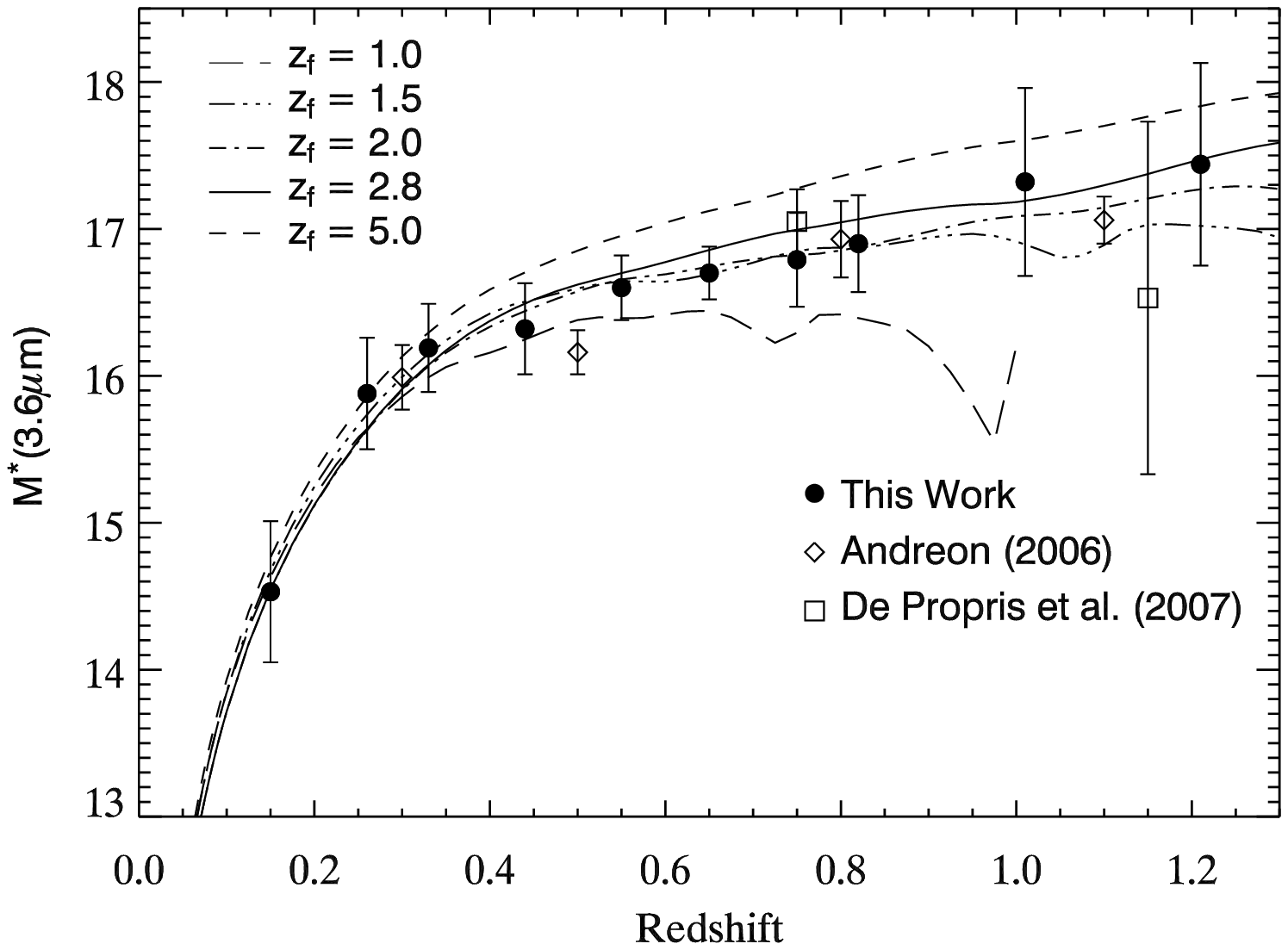}
\caption{\footnotesize Evolution in M$^{*}$ from the 3.6$\micron$
  LFs as a function of redshift.  The long dashed, dash-dotted,
  dash-dot, solid, and dashed lines show
  models where the stars form in a single burst at $z =$ 1.0, 1.5, 2.0, 2.8,
  and 5.0 respectively.  The filled circles are the FLS clusters and
  the open diamonds and open squares are the M$^{*}$ values from the Andreon (2006) and De
  Propris et al. (2007) cluster samples.}
\end{figure}
\subsection{The 5.8$\micron$ and 8.0$\micron$ Luminosity
  Functions}
Unlike the 3.6$\micron$ and 4.5$\micron$ bandpasses where the
 luminosity of galaxies is dominated by emission from low mass stars, the
luminosity of galaxies at 5.8$\micron$ and 8.0$\micron$ comes from
several sources.   It can have contributions
from warm dust continuum, PAH emission, and low mass stars.  In particular,
if warm dust (heated by intense star formation or an AGN) or
PAH emission is
present, it typically dominates the luminosity at these wavelengths.  Therefore, the 5.8$\micron$ and 8.0$\micron$ LFs 
can be useful probes of the amount of dusty star formation and AGN
activity in clusters if the contribution from stellar emission is
properly accounted for.
\newline\indent
The main challenge in modeling the LFs at these wavelengths is that a massive, dust-free
early type galaxy produces relatively the same flux at
5.8$\micron$ and 8.0$\micron$ from pure stellar emission as a much lower mass starburst
galaxy or AGN produces from PAH emission or warm dust continuum.
Determining the relative abundance of each of these populations
in a LF is more challenging for a statistically defined sample such 
as this cluster sample where individual galaxies are not identified as
field/cluster or star forming/non-star forming.  Despite this challenge,
we showed in $\S$5.1 that the 
3.6$\micron$ LFs can be used as a diagnostic of the average
stellar emission from the cluster galaxies and that with a
 model for galaxy colors they can predict the
4.5$\micron$ LFs extremely well.  The 3.6$\micron$-5.8$\micron$
and 3.6$\micron$-8.0$\micron$ colors of different
spectral-types vary significantly more than the
3.6$\micron$-4.5$\micron$ colors; however, if  
these colors are modeled correctly  the same technique can
be used to model the LFs in the 5.8$\micron$ and
8.0$\micron$ bandpasses and provide constraints on the number and
type of star forming galaxies in clusters.
\newline\indent
Put another way, the 3.6$\micron$ LF provides
effectively a ``stellar mass budget'' for predicting the 5.8$\micron$ and
8.0$\micron$ LFs.
Subtracting this stellar mass budget at 5.8$\micron$ and
8.0$\micron$ leaves an excess which can be modeled with different
populations of star forming galaxies or AGN.  Unfortunately,
such models are unlikely to be completely unique in the sense that
there will be a degeneracy
between the {\it fraction} of star forming galaxies or AGN, and the
{\it intensity} of the star formation or AGN activity within those galaxies;
however, we will show that using only rough empircal
constraints on the fraction of star forming/non-star forming
galaxies in clusters places interesting
constraints on the intensity of star formation in cluster galaxies, and
the relative percentages of ``regular'' star forming galaxies and
dusty starbursts.
\subsubsection{Measuring the 5.8$\micron$ and 8.0$\micron$ LFs}
Before models of the cluster population are made we measure the 5.8$\micron$ and 8.0$\micron$ LFs
using the same
stacking and background subtraction methods as for the 3.6$\micron$
and 4.5$\micron$ LFs.  The LFs 
are plotted in Figures 15 and 16 in the same redshift bins as the 3.6$\micron$
and 4.5$\micron$ LFs.  IRAC is
significantly less sensitive at 5.8$\micron$ and 8.0$\micron$
than at 3.6$\micron$ and 4.5$\micron$ and therefore these LFs
are much shallower.  Only the bright end of the LF (roughly M $<$
M$^{*}$, assuming a dust-free, pure stellar emission early type model) can be measured with
these data; however, this shallow depth is still sufficient to be a good
diagnostic of the presence of luminous dusty starbursts.  For example,
at 0.1 $< z <$ 0.4, an M82-type starburst is
roughly 3 magnitudes brighter at 8.0$\micron$ than an early type
model (e.g., Huang et al. 2007a; Wilson et al. 2007, see also $\S$6.3) and
therefore, even a galaxy with M $\sim$ M$^{*}$ + 3 from the
3.6$\micron$ LF would be detected at 8.0$\micron$ if undergoing
an M82-like dusty starburst.
\begin{figure*}
\plotone{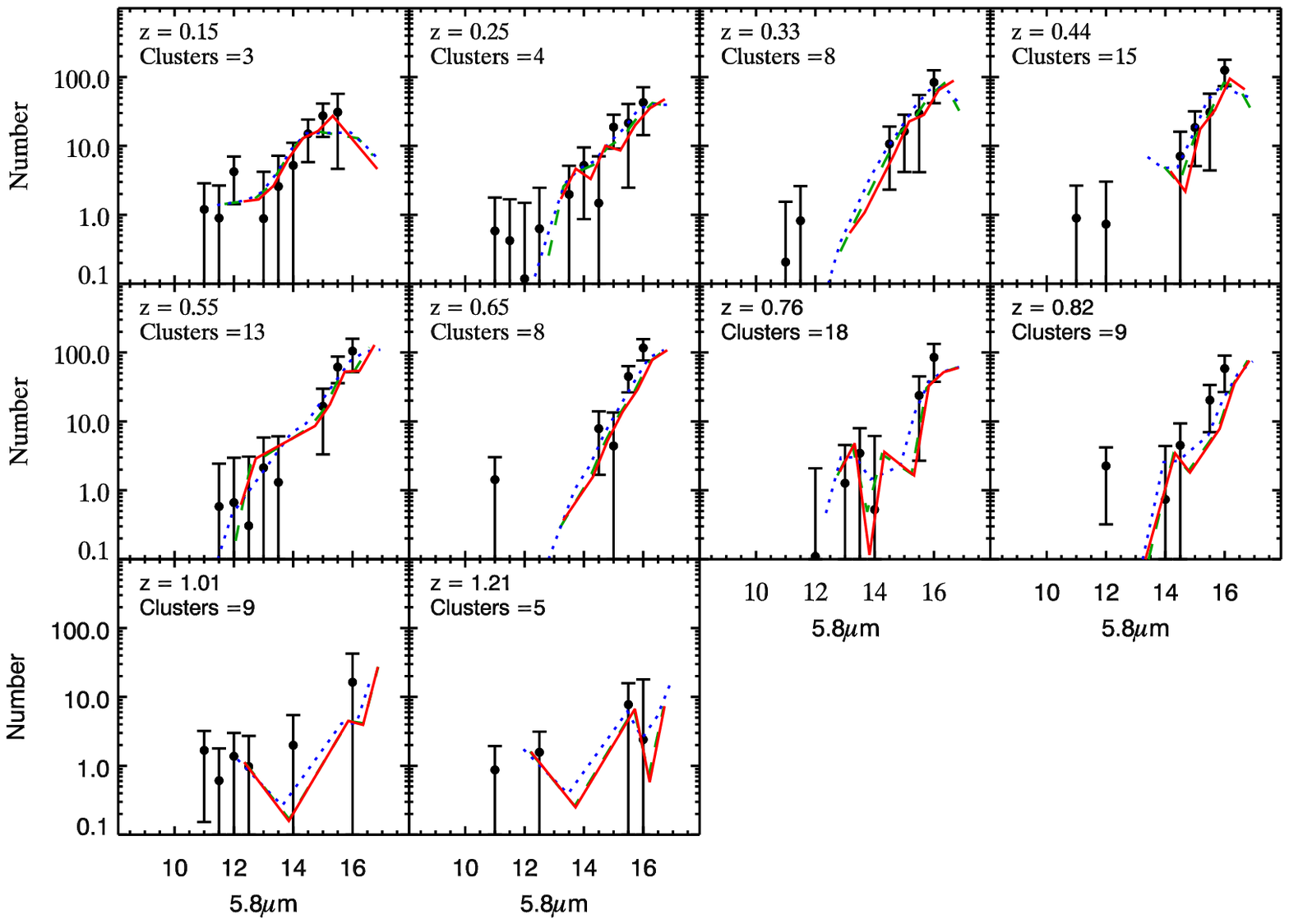}
\caption{\footnotesize The 5.8$\micron$ LFs of clusters in the FLS.
The solid red line shows the 5.8$\micron$ predicted
from the 3.6$\micron$ LF assuming all galaxies have the colors of
the passive evolution model.  The dashed green lines and dotted blue
lines are the regular+quiescent and starburst+regular+quiescent models
described in $\S$5.2.4 respectively; however, 5.8$\micron$ is not
sensitive to PAH emission or warm dust at $z >$ 0.3 and therefore these models are not
notably different from the passive evolution model.}
\end{figure*}
\begin{figure*}
\plotone{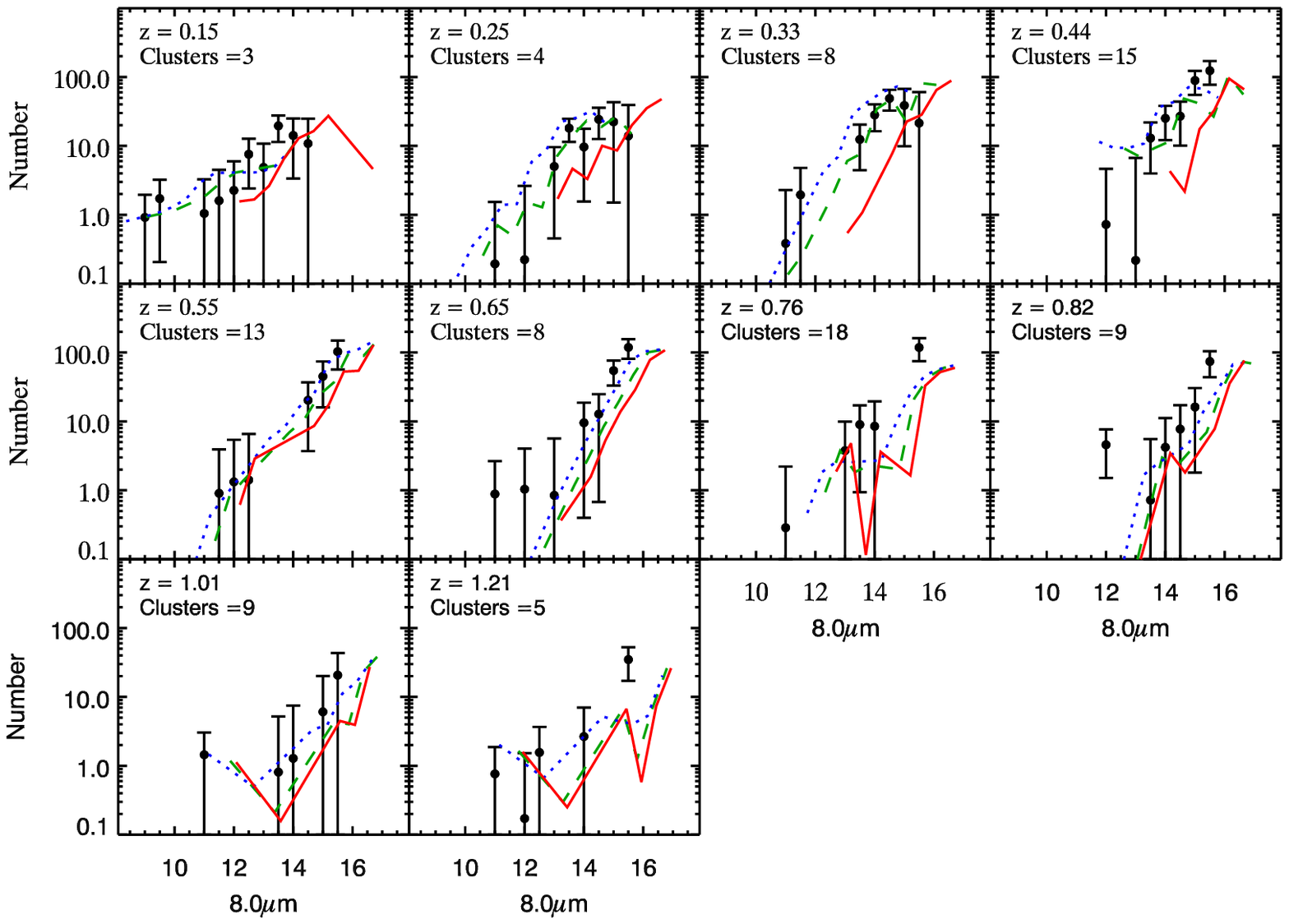}
\caption{\footnotesize The 8.0$\micron$ LFs of clusters in the FLS.
The solid red line, dashed green line, and dotted blue line are the
8.0$\micron$ LFs predicted using the 3.6$\micron$ LF and the quiescent,
regular+quiescent, and starburst+regular+quiescent models described in
$\S$5.2.4.  At lower redshift ($z <$ 0.4) the LFs are most similar to
the predictions from the regular+quiescent model whereas at higher
redshift ($z >$ 0.4) the LFs are better described by the
starburst+regular+quiescent model.}
\end{figure*}
\subsubsection{Contamination from AGN}
\indent
In order to draw conclusions from models of the MIR cluster LFs, it is
important to have some constraints on the  fraction of cluster MIR sources
that are AGN and the fraction that are star forming galaxies.    The
fraction of galaxies in clusters at $z <$ 0.6 identified as AGN 
based on their optical spectra in clusters is low ($<$ 2\%,
e.g., Dressler et al. 1985, Dressler et al. 1999); whereas the fraction of 
star forming galaxies can be quite large  (5 - 80\%, e.g., Butcher \&
Oemler 1984; Dressler et al. 1999; Ellingson et al. 2001; Poggianti et
al. 2006).  Therefore, it might be
expected that star forming galaxies will dominate the overall
number of cluster MIR sources.  It is possible
that the AGN fraction in clusters may have been underestimated because some cluster AGN 
are missed by  optical selection.  X-ray observations of moderate redshift clusters
have found an additional population of cluster X-ray AGN that do not have
AGN-like optical spectra (e.g., Martini et al. 2006; 2007, Eastman et
al. 2007).  Martini et al. (2007) showed that this population is roughly as large the optical AGN
population, making the overall AGN fraction  $\sim$ 5\%  for cluster
galaxies at $z \sim$ 0.2 with moderate luminosity AGNs (broad-band X-ray luminosities L$_{X}$ $>$ 10$^{41}$ erg
s$^{-1}$), but only $\sim$ 1\% for those with bright AGN (L$_{X}$ $>$ 10$^{42}$ erg
s$^{-1}$).  If the analysis is
restricted to galaxies with hard X-ray luminosities $>$ 10$^{42}$ erg
s$^{-1}$, then the fraction is about an order of magnitude lower
(0.1\%, Eastman et al. 2007).
\newline\indent
Although these studies suggest the  AGN fraction in clusters is low,
particularly for bright AGN, 
it is unclear how many of the optical and X-ray selected cluster AGN will
have detectable MIR emission, and what fraction of the cluster MIR
population they comprise.  Previous MIR studies of clusters have
detected only a few AGN in spectroscopic samples of $\sim$ 30-80
cluster MIR sources (e.g., Duc et al. 2002; Coia et
al. 2005; Marcillac et al. 2007; Bai et al. 2007) suggesting that $>$90\%
of cluster galaxies detected in the MIR are star forming galaxies.  
One way to estimate the fraction of cluster MIR-bright AGN is to use the 
IRAC and MIPS color-color diagrams suggested by Lacy et al. (2004) and
Stern et al. (2005).  Although these simple color cuts fail to
identify complete samples of AGN  because
they only identify those that have red power-law slopes in the MIR (e.g.,
Cardamone et al. 2008); these are precisely the type of AGN
that will be included in the 5.8$\micron$
and 8.0$\micron$ LFs and therefore the color cuts should provide a
reasonable estimate of the contamination of those LFs from AGN.
\newline\indent
In the left panels of Figure 17 we plot the IRAC colors of all
galaxies brighter than the 50\% completeness limits using
the color spaces suggested by Stern et al. (2005) (top) and Lacy et
al. (2004) (bottom).  The dashed lines in each panel represent the
portion of color space used to select AGN in the MIR by these
authors.  FLS galaxies which satisfy the color criteria are plotted as
grey circles. The right panels of Figure 17 shows the same plot for all galaxies
with R $<$ R$_{200}$ for clusters at $z <$ 0.7 in the FLS (59
clusters). 
\newline\indent
The entire FLS (left panels) can be used to estimate the surface density of MIR
selected AGN in these color spaces.  Subtracting this background from the cluster
fields we find and excess of 26 $\pm$ 22 galaxies using the Stern et
al. (2005) color cut and and excess of 30 $\pm$ 30 galaxies using the Lacy et
al. (2004) color cut.  Summing the background subtracted 3.6$\micron$ LFs to the same limit
implies that there are 2466 total cluster galaxies in these 59 clusters, and that the overall
fraction of cluster galaxies that candidate MIR-bright AGN (to our 3.6$\micron$
detection limit) is 1$^{+1}_{-1}$\%, where all error bars have been calculated using 
Poisson statistics.  Integrating the 5.8$\micron$ and 8.0$\micron$ LFs
shows there are 869 and 959 cluster galaxies detected in these bands
and that the fraction of cluster sources detected in the MIR that are candidate AGN is
$\sim$ 3$^{+3}_{-3}$\%.
\newline\indent
Although this crude estimate is almost certainly an incomplete census of the total fraction of
AGN in clusters, it is remarkably similar to the AGN fractions
measured with optical spectroscopy or by X-ray selection and is consistent with
the fraction of spectroscopically confirmed MIR-bright AGN seen in previous cluster MIR studies.  Based on
the low estimated AGN fraction, and for the sake of simplicity in interpretation, we do not model
an AGN component in the 5.8$\micron$ and 8.0$\micron$ LFs in this analysis.
We do note that the X-ray, spectroscopic and MIR selection do show
clearly that the fraction of MIR cluster sources that are AGN is {\it not} zero, and
therefore some of the sources in the 5.8$\micron$ and
8.0$\micron$ LFs will certainly be AGN.  
\begin{figure*}
\plotone{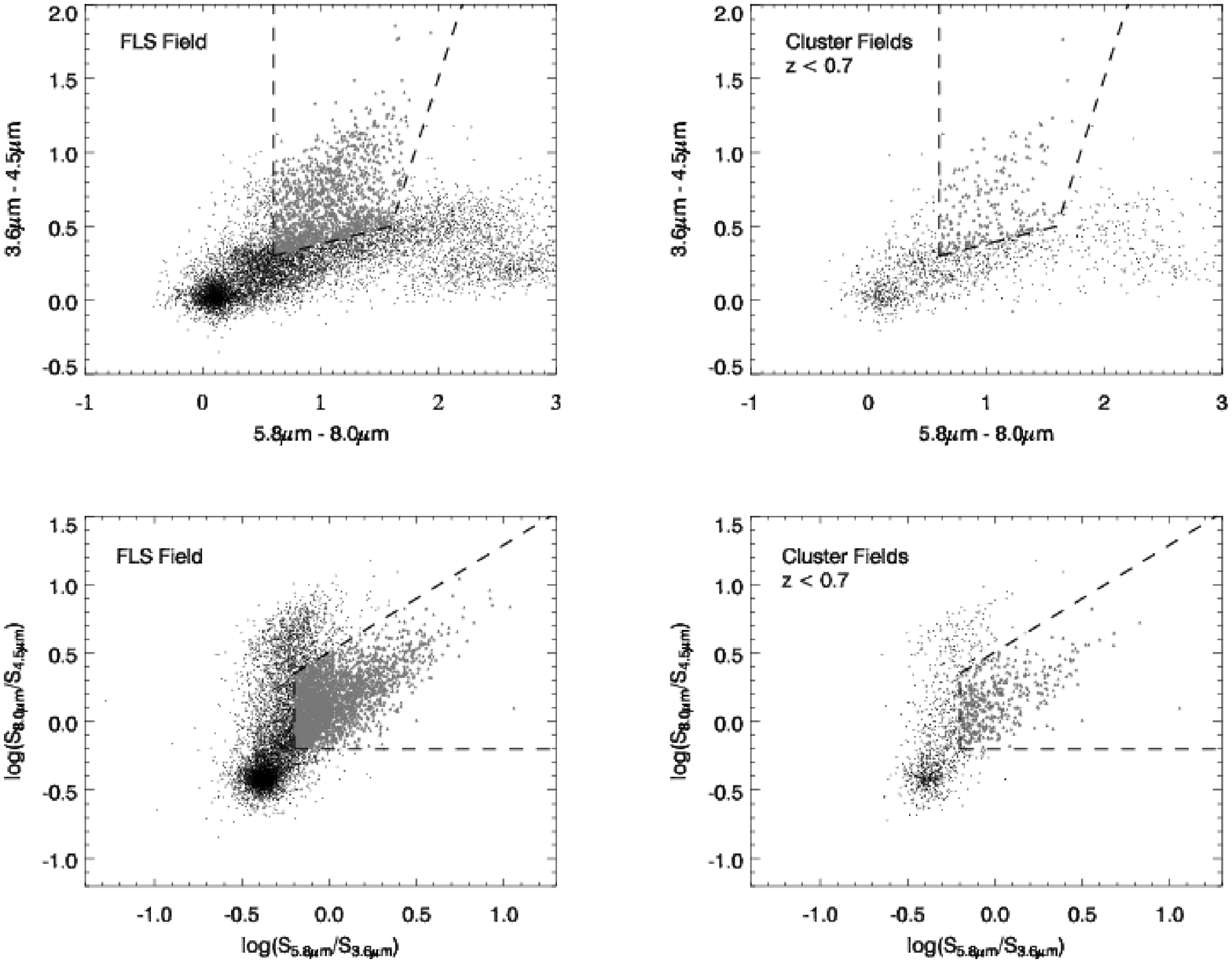}
\caption{\footnotesize Top Left Panel: Color-color plot of all galaxies in
  the FLS (small dots).  The
  dashed lines denote the region used to select AGN by Stern et
  al. (2005).  Bottom Left Panel: Same as top left, but for the
  Lacy et al. (2004) color space.   Right Panels: Color-color plots for galaxies at R $<$
  R$_{200}$ in the fields of clusters at $z <$ 0.7 (59 clusters).  The
  majority of these sources are foreground or background galaxies.  
  Background subtraction based on the surface density of sources in the left panels suggest that
  1$^{+1}_{-1}$\% of cluster
  galaxies detected at 3.6$\micron$ are  AGN and that $\sim$ 3$^{+3}_{-3}$\% of
  cluster galaxies detected at 5.8$\micron$ and 8.0$\micron$ are AGN.}
\end{figure*}
\subsubsection{Modeling the 5.8$\micron$ Luminosity Function}
\indent
The simplest fiducial model
that can be made for the MIR cluster galaxy population is to
assume that 
the bright end of the  LF is  dominated
by passive, dust-free, early type galaxies (i.e., the emission at
5.8$\micron$ and 8.0$\micron$ is completely stellar).  Although such a
model is unrealistic, it provides a baseline for predicting the
amount of emission in the MIR from stellar emission, and any excess
beyond this model is likely to be from dusty star formation in the cluster population.  Assuming
such a model,  the 
5.8$\micron$ LFs can be inferred from the 3.6$\micron$ LFs using
the 3.6$\micron$-5.8$\micron$ colors from the Bruzual \&
Charlot passive evolution model.  These predicted 5.8$\micron$ LFs are
overplotted on the LFs in Figure 15 as the solid red lines (Figure 15
also has additional models overplotted which are introduced in $\S$5.2.4).  
\newline\indent
Qualitatively, the 
3.6$\micron$ LFs and the passive evolution model
predict the 5.8$\micron$ LFs reasonably well at all redshifts.  This is perhaps not
surprising because due to k-corrections, 5.8$\micron$ is only
sensitive to emission from warm dust or PAHs
in star forming galaxies at $z < 0.3$ (see $\S$6.3).
For galaxies at higher redshift, 5.8$\micron$ probes
rest-frame wavelengths which, similar to the 3.6$\micron$ LFs, are dominated by stellar emission.
As a result, any dusty star forming cluster galaxies would only be visible
as a notable excess in the predicted 5.8$\micron$ LFs at $z <$ 0.3.
No such
excess is seen; however, the fraction of blue
star forming galaxies in clusters evolves rapidly (i.e., the
Butcher-Oemler Effect) and clusters at $z <$ 0.3 typically have low
blue fractions and relatively few
star forming galaxies (e.g., Ellingson et al. 2001; Balogh et al. 1999;
Margoniner et al. 2001).  This result confirms that
the fraction of star forming galaxies in clusters at $z <$ 0.3 low,
and, that furthermore there is no significant additional population of MIR
luminous dusty
star forming galaxies in clusters at these redshifts that are missing from optically-selected spectroscopic or photometric studies.
\subsubsection{Modeling the 8.0$\micron$ Luminosity Function}
\indent
Unlike the 5.8$\micron$ LFs, the cluster 8.0$\micron$ LFs are not
consistent with the passive evolution model predictions from the
3.6$\micron$ LFs illustrated by the solid red lines plotted
in Figure 16.  This model clearly underpredicts the number of galaxies in the 8.0$\micron$ LFs at all
redshifts.  
In order to construct a more useful model for the 8.0$\micron$ LF that  includes
the cluster star forming population, we use the
3.6$\micron$-8.0$\micron$ colors for different types of
star forming galaxies from J. Huang et al. (2008, in preparation).  These authors have
empirically extended the color/redshift models of Coleman et al. (1980) to
10$\micron$ using local galaxies with $ISO$ spectroscopy.  Some examples of
the colors from these models are presented in Wilson et al. (2007).
\newline\indent
Given the large number of permutations possible in the types of
star forming galaxies, we are interested in as simple a model as
possible  which will allow for a straightforward interpretation of the
data.  For this analysis we divide the cluster star forming population into
two populations: ``regular'' star forming cluster spirals, and dusty starburst galaxies.
Huang et al. (2008) have models for both Sbc and Scd galaxies; however,
the 3.6$\micron$-8.0$\micron$ colors of these models are
indistinguishable, and therefore we adopt their Sbc galaxy as the model for a
``regular'' star forming
cluster spiral.  Huang et al. also have colors for several ``canonical''
dusty starburst galaxies such as M82, Arp220,
and NGC 1068.  M82 is a moderate-strength dusty starburst, has no AGN component, and is classified as a
luminous infrared galaxy (LIRG).  By contrast, Arp220 and NGC 1068 are powerful
dusty starbursts with AGN components.  The IR luminosity of Arp220 is dominated by star formation from
a major merger, while the IR luminosity of NGC 1068 is dominated by a
powerful AGN (although both galaxies have AGN and starburst components).  Both are classified as ultra-luminous infrared
galaxies (ULIRGs).  Given that the majority of distant clusters studied thus far in the
MIR have shown a significant population of LIRGs but no population
of ULIRGs (e.g.,  Coia et al. 2005; Geach et
  al. 2006; Marcillac et al. 2007), 
we assume that any cluster dusty starbursts
will have colors similar to M82, rather than Arp220 or NGC 1068.  In
general, replacing M82 as the model for cluster dusty starbursts with
either Arp220 or NGC 1068 requires a smaller
fraction of dusty starbursts since they are more luminous.
\newline\indent
In order to ascertain the dominant mode of star formation present
in the cluster population  we can construct simple models for
the 8.0$\micron$ LFs from the 3.6$\micron$ LFs using various combinations of these
populations.  The purpose of the models is not to perfectly
reproduce the cluster 8.0$\micron$ LFs 
(this requires a much more detailed
knowledge of the populations in each cluster than can be obtained by
statistical background subtraction), but to demonstrate how the
8.0$\micron$ LFs should appear given different proportions of these
populations and thereby estimate the importance of each's contribution to
the 8.0$\micron$ LFs.  Hereafter we refer to the Sbc model as ``regular'',
the M82 model as ``dusty starburst'', and the Bruzual \&
Charlot passive evolution model as ``quiescent''.
\newline\indent
Beyond assuming that all cluster galaxies are quiescent, which clearly
underpredicts the 8.0$\micron$ LFs, the next most simple model
that can be made is to assume some fraction of the cluster galaxies
are ``regular'' star forming galaxies (hereafter we refer to this model as regular+quiescent).
In order to make such a model we require an approximation of the relative proportions of star forming and
quiescent galaxies in clusters as a function of redshift and
luminosity.  The best spectroscopically-classified data at these redshifts comes
from the MORPHS (Dressler et al. 1999; Poggianti et al. 1999) and
CNOC1 (Balogh et al. 1999; Ellingson et al. 2001)
projects.  Unfortunately, the number of cluster spectra per d$z$ is
relatively small in these samples and they cover only a modest
range in redshift (0.2 $< z <$ 0.5) and depth in terms of the cluster
M$^{*}$.  
\newline\indent
Although spectroscopic
classification would be the most reliable, the lack of data motivates the 
use of cluster blue fractions (f$_{b}$) as a function of redshift as a model for the
relative fractions of star forming/non-star forming galaxies.  Blue
fractions for reasonably large samples of clusters at different
redshifts have been calculated and it is fairly straightforward to
measure them as a function of magnitude within these clusters.  In
particular, using f$_{b}$ as an estimate of the star forming
fraction should predict the number of blue
star forming galaxies (i.e., those with colors similar to the Sbc model). If  a population of
red, dust-obscured starburst galaxies exists in clusters they should
be evident in the 8.0$\micron$ LFs as an excess of galaxies beyond
the regular+quiescent model.
\newline\indent
For f$_{b}$ as a function of redshift we use the data
of Ellingson et al. (2001) from the CNOC1 clusters which span the
redshift range $z$ = 0.2 to $z
=$ 0.4, and
for clusters at $z >$ 0.4 we use the data on RCS-1 clusters from Loh
et al. (2008).  Rough f$_{b}$ values for both these
samples were recomputed using only galaxies with M $<$ M$^{*}$ (D. Gilbank private communication), because this matches
the depth of the 5.8$\micron$ and 8.0$\micron$ LFs.  These f$_{b}$
values as a function of redshift are listed in Table 2.  
\newline\indent
The scatter in cluster f$_{b}$ values at a given redshift is large, and therefore different
studies find different mean values depending on sample.  The values we
have adopted are consistent with the majority of work in the field
(e.g., Butcher \& Oemler 1984; Smail et al. 1998;  Margoniner et
al. 2001; Andreon et al. 2004), although we have measured them using a brighter
luminosity cut.  Of course, the best way to infer the f$_{b}$ of the FLS
clusters would be to measure it from the clusters themselves;
however, we do not have the proper filter coverage at $z <$ 0.5 to
make this measurement properly nor a large enough sample
to make a measurement that would be statistically different from the
adopted values.  
\newline\indent
The cluster f$_{b}$ is also a function of limiting magnitude (e.g., Ellingson
et al. 2001), and
without incorporating some variation in f$_{b}$ as a function of
magnitude, all of the model LFs consistently overpredict the number of
bright galaxies in the 8.0$\micron$ LFs, and underpredict the number
of faint ones.  In order to estimate the variation of f$_{b}$ as a function of
magnitude we use the spectrally-typed LFs of Muzzin et
al. (2007a).  They measured the K-band
LF for cluster galaxies defined spectroscopically as either star forming or quiescent.
Comparing those LFs (their Figure 13) and assuming all star forming
galaxies are blue, and all quiescent galaxies are red, results in 
f$_{b}$ values of 0.19, 0.35, and 0.52 for galaxies with  M $<$ M$^{*}$, M$^{*}$ $<$ M $<$ M$^{*}$ + 1, and
M$^{*}$ +1 $<$ M $<$ M$^{*}$ + 2 respectively in clusters at $z
\sim$ 0.3.  Comparing these values shows that f$_{b}$ is 1.8 times larger at M$^{*}$ $<$ M
$<$ M$^{*}$ + 1 than at M $<$ M$^{*}$, and is 2.7  times larger at
M$^{*}$ +1 $<$ M $<$ M$^{*}$ + 2 than at M $<$ M$^{*}$.  We therefore adopt an f$_{b}$ that varies with
magnitude with the following conditions:  For galaxies with M $<$
M$^{*}$ in the 3.6$\micron$ LF we use the f$_{b}$ values from Table 2.  For
galaxies with M$^{*}$ $<$ M $<$ M$^{*}$ + 1, we assume that f$_{b}$ is
twice as large as the values in Table 2, and for galaxies with M$^{*}$
+ 1 $<$ M $<$ M$^{*}$ + 2 we assume that f$_{b}$ is
three times as large as the values in Table 2.  In cases where this causes
f$_{b}$ $>$ 1.0, it is set equal to 1.0.
\newline\indent
Combining the f$_{b}$ as a function of redshift and magnitude with the
3.6$\micron$ LFs assuming all ``blue'' galaxies have the color of
the Huang et al. Sbc galaxies and all ``red'' galaxies have the color
of the passive evolution model results in the models that are plotted as  
green dashed lines in Figures 15 and 16.
Comparing the data to these models shows that 
this simple model using only regular+quiescent galaxies predicts the cluster 8.0$\micron$ LFs fairly
well.  In particular, the z = 0.15, 0.25 and 0.33 LFs are well
described by this model.  For the higher redshift LFs this model is clearly
better than the purely quiescent model; however, it still does not
account for the entire 8.0$\micron$ population.  
\newline\indent
Most importantly, the regular+quiescent 
model shows that out to $z \sim$ 0.65, where 8.0$\micron$ still probes
rest-frame dust emission, there is no significant population of
bright (M $<$ M$^{*}$) galaxies in
clusters that cannot reasonably be accounted for by ``regular''
star forming cluster spirals.  This is significant because it suggests
that whatever processes responsible for transforming the
morphology and spectral-type of  bright cluster galaxies over the same redshift
range do not involve an ultra-luminous dusty starburst phase such as
those caused by major mergers of gas-rich galaxies (i.e., ``wet'' mergers).  We note that
there appears to be an overdensity of very bright galaxies
in the $z =$ 0.82 LF that cannot be accounted for by the
regular+quiescent model and
this suggests the possibility of an onset of luminous starbursts
(possibly from mergers) or AGN activity in bright galaxies at  higher redshift.    
\newline\indent
Although the regular+quiescent model predicts the bright end of the 8.0$\micron$
LFs  well at all redshifts, and the entire 8$\micron$ LF at lower
redshift, it fails to account for all of the LFs. In particular, this
model seems to underpredict the number of fainter galaxies in the
8.0$\micron$ LFs for clusters at $z >$ 0.4.  This suggests
a third component to the cluster 8.0$\micron$
population, possibly a red, dusty starburst population which is not
accounted for by the cluster f$_{b}$.  Such a population was suggested by
Wolf et al. (2005) who found that the SEDs of roughly 30\% of the red sequence galaxies in
the Abell 901/902 supercluster ($z$ = 0.17)
were better described by dusty templates rather than a dust-free, old
stellar population. In order to explore this possibility, we construct a new
model with the
same values of f$_{b}$ as a function of magnitude and redshift as for
the regular+quiescent model, but this time we assume that some of the
red quiescent galaxies are instead M82-like dusty starbursts.  M82
has optical-IR colors that are similar to quiescent galaxies (see
Huang et al. 2008 and
$\S$6.3) so it is reasonable to assume that any M82-like dusty starbursts would be part of
the population of red cluster galaxies rather than the blue cluster
galaxies.  
\newline\indent
If we assume that the dusty
starburst population is a constant fraction of the red cluster galaxies, this
would result in a varying ratio of dusty starburst to regular star forming
galaxies in clusters as a function of redshift.  In particular,
clusters at low redshift will have the highest fraction of dusty
starburst galaxies (because the f$_{b}$ is low and the red fraction is
high).  The LFs above have already suggested that there is no need for
a dusty starburst population at low redshift, so modeling the dusty
starbursts as a fixed fraction of the red galaxies seems
inappropriate.  Instead, a better way to model the population is to
assume that the cluster f$_{b}$ is a tracer of the total star
formation in the cluster and that ratio of dusty starburst to regular star forming
galaxies  is a constant.  Given this assumption we can predict the fraction of dusty
starbursts directly from the cluster f$_{b}$.  This fraction of dusty starbursts
is then removed from the fraction of red quiescent galaxies and a
model for the LFs can be made.  Hereafter we refer to this model as
starburst+regular+quiescent.
The fractions of the cluster galaxy populations in terms of f$_{b}$
are defined using the equations,
\begin{equation}
f_{dsb} = f_{b} \times f_{dsb/reg},
\end{equation}
\begin{equation}
f_{q} = 1 - f_{b} - f_{dsb},
\end{equation}
where f$_{dsb}$ is the fraction of dusty starburst galaxies, f$_{dsb/reg}$ is the assumed ratio of dusty starburst to regular
star forming galaxies, and f$_{q}$ is the fraction of quiescent
galaxies.  In cases where f$_{dsb}$ + f$_{b}$ $>$ 1 we set f$_{dsb}$ = 1
- f$_{b}$ and f$_{q}$ = 0.
\newline\indent
As of yet there are no good observational constraints
on the parameter f$_{dsb/reg}$. Therefore, as a first-order fiducial value we
assume that f$_{dsb/reg}$ = 0.5.  In general, we find that
allowing a range of values between 0.3 - 1.0 provides models that are
fairly similar.  More importantly, the differences in models that use  f$_{sb/reg}$ between
0.3 - 1.0 are much smaller than the difference between any of those models and
the regular+quiescent model. Therefore, the interpretation
of the data using these models
will not depend strongly on the assumed value of f$_{sb/reg}$.  
The starburst+regular+quiescent model with f$_{sb/reg}$ = 0.5 is
overplotted on Figures 15 and 16 as the dotted blue line.  
\newline\indent
This starburst+regular+quiescent model over-predicts the number of
bright galaxies in the $z <$ 0.4 8.0$\micron$ LFs
but it is better at describing the
LFs at $z >$ 0.4 than the regular+quiescent or purely quiescent
models.  This suggests that there is a population of dusty starbursts in
clusters at $z >$ 0.4 that does not exist at $z <$ 0.4, and that
these starbursts are consistent with being of an M82-type.  We discuss this in more
detail in $\S$6.1.
\section{Discussion}
\subsection{Evidence for a Change in Star Formation Properties of
  Cluster Galaxies?}
\indent
In order to better illustrate the differences in the model populations
described above, we subtract the quiescent model from the
5.8$\micron$ and 8.0$\micron$ LFs between 0.15 $< z <$ 0.65 and
plot the residuals in Figures 18 and 19.  The residuals
from the quiescent+regular model and starburst+regular+quiescent models
from $\S$5.2.4 are also plotted in Figures 18 and 19.  The solid vertical lines
in the plots represent the magnitude of M$^{*}$ inferred from the
3.6$\micron$ LF assuming the passive evolution model, and give some
indication of the depth of the LFs.
If we compare the data to the models and take the
results at face value, it suggests that the intensity of
star formation  in clusters is evolving with redshift and that it can be classified into
three types.  The
first type of star formation is ``weak'' and best describes the lowest
redshift clusters ($z <$ 0.15) which are consistent with the colors of an almost exclusively quiescent
population in
all IRAC bandpasses.  This result is consistent with numerous studies of nearby
clusters using spectroscopy which show few star forming galaxies
(e.g.,
Dressler et al. 1985, Popesso et al. 2007; Christlein \& Zabludoff 2005; Rines et al. 2005).  
\newline\indent
Between 0.2 $< z <$ 0.5 the 8.0$\micron$ LFs
are no longer well-described by the purely quiescent model and the
regular+quiescent model is the best model.  This shows that the
majority of star formation in clusters at this epoch is primarily
relegated to galaxies that have MIR colors similar to local late-type
star forming galaxies (i.e., the Sbc model).  This has direct implications for the SFRs of
these galaxies because Wu et al. (2005) showed that the
dust-obscured SFR of galaxies is proportional to their
8.0$\micron$ flux.  Although other authors have demonstrated that
there are caveats when using the
8.0$\micron$ flux to infer SFRs (i.e., the scatter can be as high as
a factor of 20-30, Dale et al. 2005), this still implies that the
average SFR or the average SFR per unit stellar mass (the average specific
star formation rate, SSFR) of star forming cluster galaxies at 0.2 $< z <$ 0.5 is similar to those in the
local universe (because they have 3.6$\micron$-8.0$\micron$ colors
similar to local Sbc galaxies).  This second mode of star formation in clusters is roughly what
would be considered ``regular'' star formation for galaxies in the
local universe.
\newline\indent
At $z >$ 0.5 the
starburst+regular+quiescent model becomes the best description of the LFs.
Again, assuming that 8.0$\micron$ flux is an indicator of SFR, the
M82 starburst model is approximately a factor 2.5 brighter at 8.0$\micron$
than the regular Sbc model for the same 3.6$\micron$ flux.
Given that our model suggests that regular star forming galaxies make
up $\sim$ 30-40\% of the cluster population at this redshift and M82 galaxies
make up $\sim$ 15-20\%, this implies that not only is the
  abundance of star forming galaxies in clusters higher at higher
redshift (i.e., the Butcher-Oemler Effect), but also the average SSFR of cluster galaxies
is approximately a factor of 1.5 higher at $z >$ 0.5 than it is at $z <$
0.5.  This increase in SSFR suggests a third mode of star formation in
cluster galaxies that could be 
considered a ``burst'' mode, at least relative to local star formation
rates.  Interestingly, this increase in the SSFR of cluster galaxies at
higher redshift is consistent with field studies of the
universal star formation density ($\rho_{*}$) which show  an
increase of roughly a factor of 2-5 between $z = 0.2$ and $z = 0.5$ (e.g.,
Lilly et al. 1996; Madau et al. 1996; Wilson et al. 2002;
Schiminovich et al. 2005; Le Floc'h et al. 2005). It suggests that the
increasing fraction of dusty starbursts in the cluster population could be interpreted as the
result of an increase in the universal SSFR of galaxies with redshift and the constant accretion
of these galaxies into clusters and is not necessarily because
starbursts are  triggered by the cluster environment.  Furthermore, these
galaxies might only be considered ``starbursts'' relative to the mean
SSFR locally, whereas at higher redshift their higher SSFR is simply
typical of galaxies at that redshift.  We compare the cluster
5.8$\micron$ and 8.0$\micron$ LFs to the field LFs in $\S$6.2 and
discuss this further in that section.
\newline\indent
It is interesting that the cluster star forming population transitions from
being best described  by
regular star forming galaxies to regular and dusty starburst galaxies
around a redshift of $z \sim $ 0.4.  This is notable because of the
discrepant abundances of k+a and a+k post-starburst galaxies found in clusters
by the MORPHS (Dressler et al. 1999) and CNOC1 (Balogh et al. 1999)
projects.  Dressler et al. (1999) found that approximately 18\% of cluster galaxy
spectra could be classified as k+a galaxies based on the equivalent
width of the H$\delta$ line, whereas Balogh et al. (1999) found that
only 2\% of the cluster population could be classified this way.
These results obviously lead to very different interpretations of the
role of starbursts in the evolution of cluster galaxies.  In
particular, Dressler et al. found that the number of k+a galaxies was an
order of magnitude higher in clusters than the coeval field, suggesting a
cluster-related process to the creation of these galaxies, while Balogh et
al. found roughly equal numbers, suggesting no environmental role.
\newline\indent
Although both Dressler et al. (2004) and Balogh et al. (1999) have pointed out that the different
methods of data analysis may be partly responsible for such discrepant results,
this study suggests that the slightly different redshift
range of
the MORPHS and CNOC1 sample may also play some role.
  Excluding the two
highest redshift clusters in the CNOC1 sample (MS 0451-03 and MS
0016+16, both at $z \sim$ 0.55) the mean redshift of the other 14/16 (88\%)
clusters in the sample is $z = 0.28$.  By contrast, the mean
redshift of the MORPHS sample is $z = 0.46$.  Our 8.0$\micron$
cluster LFs seem to indicate that $z \sim $ 0.4 represents a
transition redshift above which the dominant mode of star formation in clusters
is better described as starburst, as opposed to regular.  Given that
once star formation ceases, the
typical lifetime of the A star component of a starburst galaxy's
spectrum is $\sim$ 1.5 Gyr, and that the lookback time between $z =$
0.46 and $z =$ 0.28 is also 1.5 Gyr, it is possible that both dusty
starbursts, and k+a galaxies that are
in clusters at $z =$ 0.46 may have evolved to
quiescent ``k''-type galaxies by $z \sim$
0.28, provided that the dusty star formation is immediately truncated.  This would be consistent with the change in the
8.0$\micron$ LFs around this redshift and may explain why the
MORPHS and CNOC1 samples show different abundances of post-starburst 
galaxies.  Furthermore, 1.5 Gyr prior to $z =$ 0.46 is $z \sim$
0.65.  Our $z =$ 0.65 cluster LF has the largest abundance of dusty starburst
galaxies, and if a significant fraction of these had their star
formation truncated, these would be logical progenitors to the large
population of k+a galaxies seen at $z =$ 0.46 by Dressler et al. (1999).
\newline\indent
Our results, which show an increase in the strength of the dominant mode of star formation in
cluster galaxies (from weak to normal to starburst), as well as
an overall increase in the abundance of dusty star forming galaxies are 
also consistent with MIR observations of other
clusters at these redshift ranges.  In particular, Coia et al. (2005),
 Geach et al. (2006), Marcillac et
al. (2007), and Bai et al. (2007) have all shown that
clusters at higher redshifts have significantly more MIR sources than clusters at
lower-redshift, and that these sources are typically brighter than the
sources in lower-redshift clusters.  Taken at face value, our results and their
results show the equivalent of a Butcher-Oemler Effect in the MIR
where both the fraction, and SSFR of star forming galaxies is
increasing with increasing redshift.   Whether this increase is caused
by the increase in the universal SFR with redshift, and the constant infall of
such galaxies into the cluster environment, or by the triggering of
starbursts by the high-redshift cluster environment is still uncertain.
We investigate this point further in $\S$6.2 by comparing the cluster and
field IRAC LFs.
\begin{figure*}
\plotone{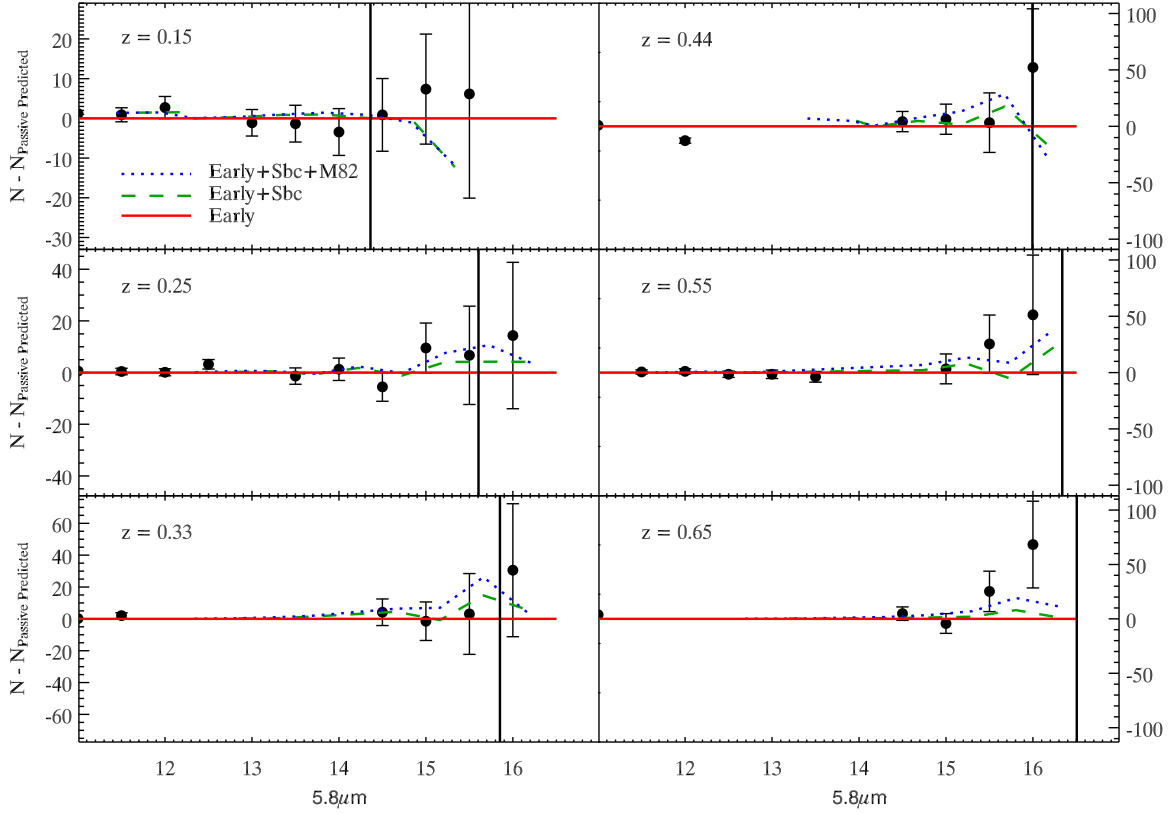}
\caption{\footnotesize Residuals of the cluster 5.8$\micron$
    LFs once the predictions from the 3.6$\micron$ LFs and the
    passive evolution model have been subtracted.  The solid red line
    shows the passive evolution model, the dashed green line shows the
    regular+quiescent model and the dotted blue line shows the
    starburst+regular+quiescent model.  The solid vertical line
    represents the location of M$^{*}$ from the 3.6$\micron$ LFs
    assuming the 3.6$\micron$ - 5.8$\micron$ color of the passive
    evolution model.}
\end{figure*}
\begin{figure*}
\plotone{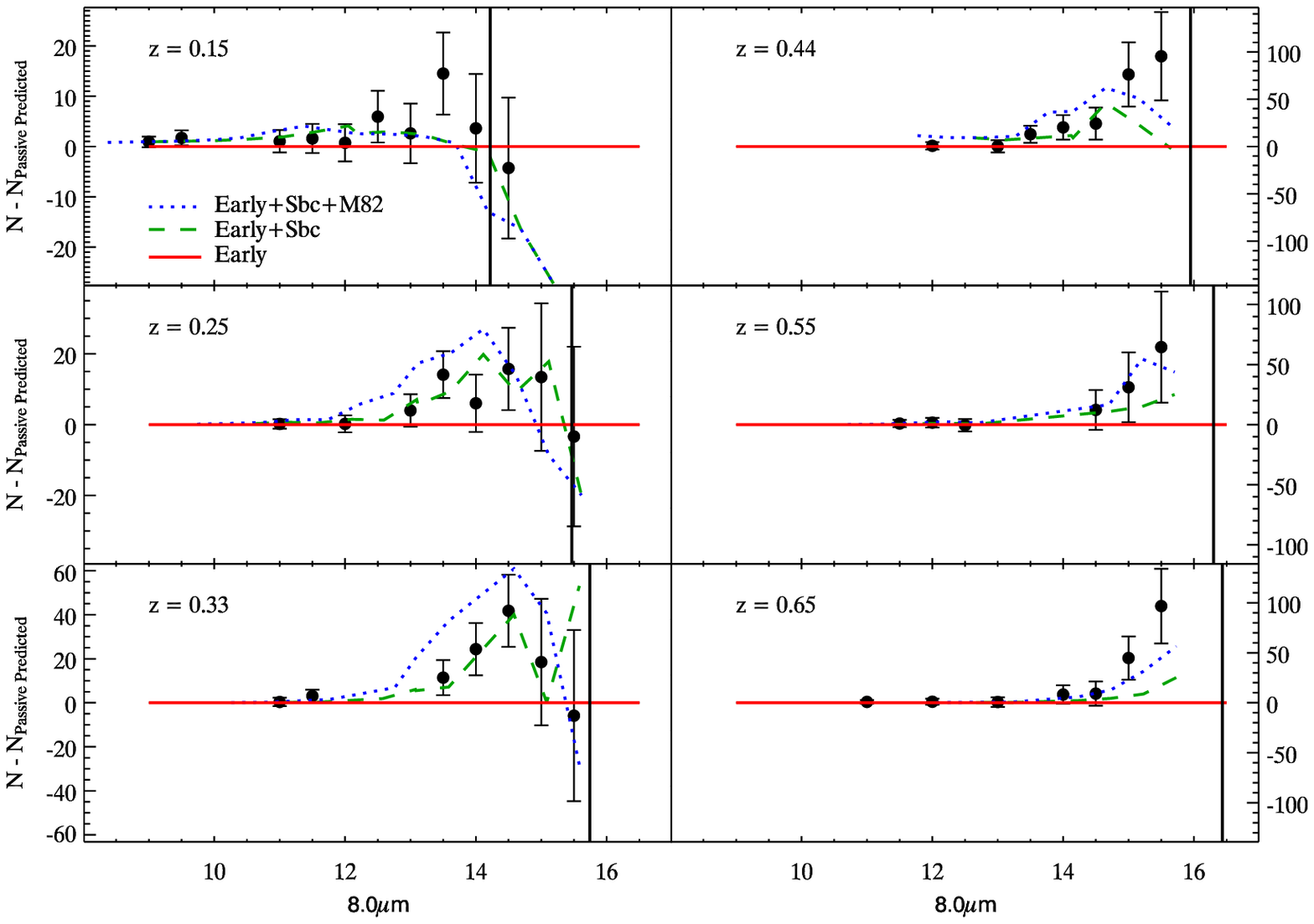}
\caption{\footnotesize Same as Figure 18 but for the 8.0$\micron$ LFs.}
\end{figure*}
\subsection{Is the Cluster Population Different From the Field
  Population?}
The most obvious way to understand if the cluster environment is
responsible for triggering starburst events is to
directly compare the field and cluster 5.8$\micron$ or 8.0$\micron$ LFs and look for an excess of
galaxies in
the cluster   LFs.  For 
this comparison we use the field LFs
measured by Babbedge et al. (2006, hereafter B06).  Their LFs are determined using photometric redshifts of
$\sim$ 100 000 galaxies from a 6.5 deg$^2$ patch of the
SWIRE survey.  The field LFs are measured in 5 redshift bins, and we
compare the cluster LFs to the three bins which overlap the redshift range of
the clusters (0.0 $<
z <$ 0.25, 0.25 $< z <$ 0.50, and 0.5 $< z <$ 1.0).  The corresponding
cluster LFs used for comparison are 
the $z = 0.15$, $z =$ 0.33, and $z = $ 0.65 LFs 
respectively.
\newline\indent
The B06 field LFs are determined using total luminosities, not
apparent magnitudes like
for the cluster LFs.  Converting the units of the cluster LFs to total
luminosities requires distance moduli and full k-corrections.  In
$\S$5.2.4 we showed that the cluster LFs can be well-described using 
three basic populations of galaxies: quiescent, regular star forming,
and dusty starburst.  We use the models of these three
spectral types for the k-corrections.  The
k-corrections for the quiescent galaxies are taken from the
single-burst model and the k-corrections for the regular and
dusty starburst galaxies are taken from the Huang et al. (2008) Sbc
and M82 models respectively.  Each LF is statistically k-corrected
using the relative proportions of the galaxies which best described the
LFs in $\S$5.2.4.  The apparent LF for each redshift is divided into the
three components by the fraction of galaxies of that type and are individually k-corrected and
shifted by the distance modulus. 
These LFs are then summed to provide the total cluster LF in terms of absolute
luminosities in units of $\nu$L$_{\nu}$/L$_{\odot}$.
\newline\indent
The cluster LFs are normalized  by the number of galaxies per
virial volume, whereas the field LFs are normalized by their actual number
density per Mpc$^{3}$.  The cluster normalization can be put in the
same units as the field LFs by dividing by the virial volume; however, this does not provide a fair comparison
because clusters have much higher volume densities of galaxies than the field. 
The most useful way to compare the cluster and field LFs
is on a per unit stellar mass basis.  We do not have 
stellar mass functions for either the field or cluster; however, we
can again assume
that the 3.6$\micron$ luminosity is roughly a proxy for stellar mass
and renormalize the LFs to a common normalization so that they
reproduce the same $\phi^{*}$ in the Schechter function fits.  The
renormalized 3.6$\micron$, 4.5$\micron$, 5.8$\micron$, and
8.0$\micron$ cluster LFs are plotted in Figures
20, 21, 22, and 23, respectively as the filled red circles.  The field
LFs are overplotted as blue squares.
\newline\indent
Figures 20 and 21 show that the overall shape of the cluster and field
3.6$\micron$ and 4.5$\micron$ LFs are similar at all
redshifts.  There is a slight, though not statistically
significant, excess in the number of
the brightest galaxies in the cluster LFs; however, these
are likely to be giant elliptical galaxies which
are common in clusters and typically do not follow the
distribution of the Schechter function.  Other than the giant ellipticals,
the shape of the 3.6$\micron$ and 4.5$\micron$ cluster and field LFs are
similar which shows that the distribution of galaxies as a function of
stellar mass is nearly identical in these environments.  This result
is consistent with K-band studies which have shown only small
differences in M$^{*}$ ($<$ 0.2 mag) between these environments (e.g.,
Balogh et al. 2001; Lin et al. 2004; Rines et al. 2004; Muzzin et al. 2007a).
\newline\indent
Conversely, there are significant differences in the 5.8$\micron$ and
8.0$\micron$ LFs of the cluster and field.  Both the 5.8$\micron$ and
8.0$\micron$ LFs follow
a sequence where the cluster LF is more abundant in MIR galaxies at
$z =$ 0.65, particularly moderate-luminosity galaxies, and thereafter the
abundance of MIR galaxies in clusters declines relative to the
field with decreasing redshift.  At $z
=$ 0.33, the cluster is slightly deficient in both 5.8$\micron$ and
8.0$\micron$ galaxies relative the field, reduced by a
factor of $\sim$ 2 for galaxies with $\nu$L$_{\nu}$ = 5 x 10$^9$ -  5 x 10$^{10}$ L$_{\odot}$.  At $z =$ 0.15, the cluster LF is significantly depleted
compared to the field, reduced by a factor of $\sim$ 5 for galaxies with $\nu$L$_{\nu}$ = 5 x 10$^8$ -  5 x 10$^{10}$ L$_{\odot}$.  
This trend not only indicates that the environment of 
dusty star forming galaxies affects their evolution, but that the
enironmental effects seem to evolve with redshift.  At $z =$ 0.15 dusty star forming galaxies
are more frequently found in the lower density field environment,
whereas at $z =$ 0.65 they are found more frequently in the
higher density cluster environment.
\newline\indent
Our results are similar to those from recent studies by Elbaz et al. (2007) and Cooper et
al. (2008) that have shown that the mean star formation rate of field galaxies in
higher density environments increases faster than those in low density
environments with increasing
redshift.  This differential increase leads to a remarkable reversal in the slope of the $<$SFR$>$ of
galaxies as a function of density at $z \sim$ 1 as compared to $z
\sim$ 0.  Field galaxies in
high density environments at $z \sim$ 1 actually have higher $<$SFR$>$ than those
in low density environments.  Although those studies compare $<$SFR$>$
of galaxies at a range of densities within the field and do not use
clusters per se, our comparison between the 8$\micron$
LFs of the cluster and field environments seem to at least
qualitatively suggest a similar trend.
\newline\indent
It is not entirely obvious why
starbursts should prefer the cluster environment over the field
environment at high ($z >$
0.5) redshift and then reject it at lower redshift ($z <$ 0.5).  We suggest that
starbursts could preferentially be triggered during the
initial formation and collapse of the cluster, and be quenched
thereafter by the high-density environment.  If this interpretation is
correct, it is likely that the parameter most
responsible for the change in star formation properties 
relative to the field is the degree of virialization of the
clusters.  
\newline\indent
Clusters that are unrelaxed, or in the process of collapsing, have two properties
that would permit increased numbers of dusty starbursts.  Firstly,
before virialization, the cluster gas has not yet been shock-heated to its maximum temperature.
This hot intracluster gas has long been considered the primary cause for
the quenching of
star formation in cluster galaxies because it prevents the cooling of
gas in the outer halo
of a galaxy, thereby ``strangling'' star formation.
Depending on the density/temperature threshold required for quenching,
it is possible that starbursts that would normally be quenched in virialized
clusters at lower redshifts may survive longer in
unvirialized clusters at high redshift.  Secondly, the velocity
dispersions in unrelaxed systems are lower and therefore mergers and
harassments should be more common at higher redshift (e.g., Tran et
al. 2005b).  It is
plausible that this more dynamically ``active'' environment preferentially triggers
star formation.  The combination of more triggered dusty starbursts
through harassment and mergers as well as a weaker quenching process
may be the reason for more dusty starbursts in clusters relative to
the field at higher redshift.
Once a cluster becomes virialized the interactions between galaxies
should become less frequent and the quenching of star formation by the hot
cluster gas will be more efficient.  In such a scenario the relative abundances of
dusty starbursts in clusters should decrease relative to the field. 
\newline\indent
If our interpretation is correct we might expect different results
from the 8.0$\micron$ LFs of X-ray selected samples of clusters (i.e., those which require a
hot virialized cluster gas component) compared to red sequence
selected samples, which, assuming the early type population is formed
prior to cluster collapse, do not require that clusters are fully virialized.
\begin{figure}
\plotone{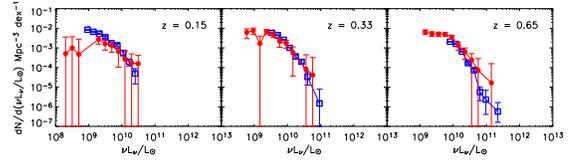}
\caption{\footnotesize Comparison between the cluster and field
  3.6$\micron$ LFs at different redshifts.  The field LFs are
  plotted as open blue squares and the cluster LFs are plotted as
  filled red circles.  The cluster LFs are renormalized so that the
  values of $\phi^{*}$ from the Schechter function fits ($\S$5.1)
  match the $\phi^{*}$ values from the Schechter function fits in B06. }
\end{figure}
\begin{figure}
\plotone{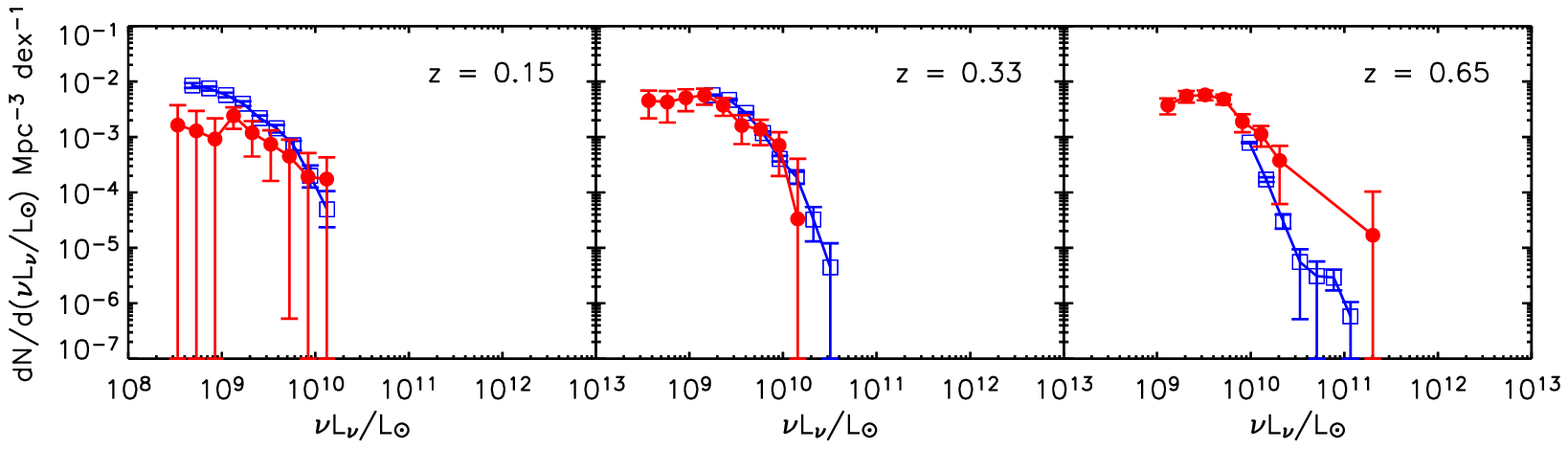}
\caption{\footnotesize Same as Figure 19 but for the 4.5$\micron$ LFs.}
\end{figure}
\begin{figure}
\plotone{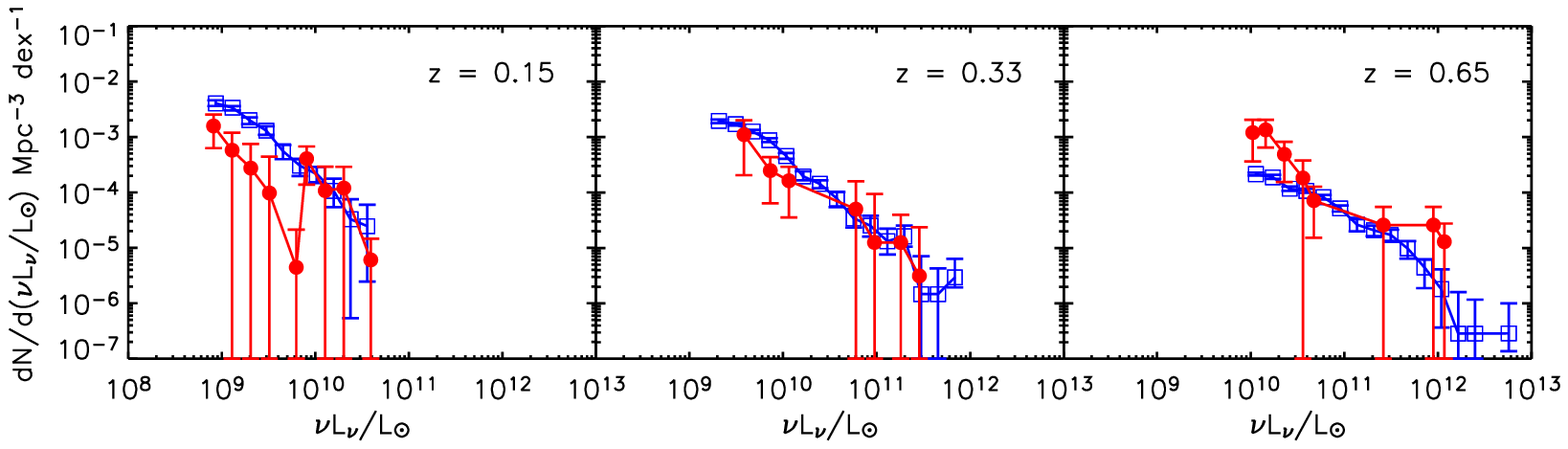}
\caption{\footnotesize Same as Figure 19 but for the 5.8$\micron$ LFs.}
\end{figure}
\begin{figure}
\plotone{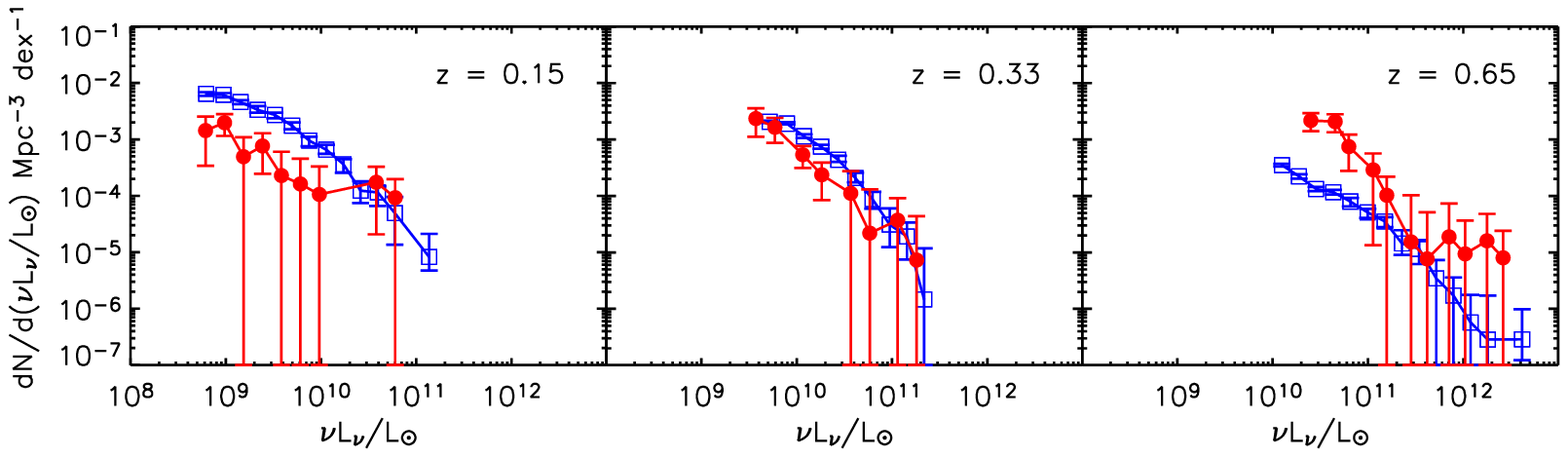}
\caption{\footnotesize Same as Figure 19 but for the 8.0$\micron$ LFs.
}
\end{figure}
\subsection{Are the Color Models Correct?}
\indent
The main conclusions from the cluster LFs presented in this paper
depend on interpreting color models that have been primarily calibrated or determined
using nearby galaxies.  If these models are not applicable at higher
redshift then this could cause incorrect conclusions to be drawn from
the LFs.   Using the spectroscopic redshifts
we can examine the colors of confirmed cluster galaxies as a function
of redshift to check if the models are reasonable.
\newline\indent
There are 55 spectroscopic redshifts available for
cluster galaxies (see $\S$2.4 \& $\S$2.5).
Using the spectra we can classify these galaxies into two basic types,
star forming and non-star forming.  For the Hectospec, SDSS, and WIYN
spectroscopy the best-fitting cross-correlation template is used for
the classification.  For the remaining galaxies the classification is made by eye-examining the spectra for any
evidence of the [OII], [OIII], or H$\alpha$ emission lines.  Galaxies
with any of these emission lines are classified as star forming, and
those without are classified as non-star forming.  Although this is a 
crude approach to classifying galaxies, we are only interested in a
rough classification and taking a more 
quantitative  approach, such as measuring EWs, is unnecessary.  Furthermore, in
all cases the cluster galaxies had spectra that were
typical of either normal star forming (several emission lines
including [OII] and H$\alpha$) or quiescent galaxies
(strong H and K lines and a 4000\AA$ $ break), and classification was
straightforward.  There were no hybrid
objects associated with clusters except two AGN from the Hectospec
data.  
\newline\indent
In Figure 24 we plot several of the colors of these galaxies as a function of
redshift.  Star forming galaxies are plotted as purple points and
non-star forming galaxies are plotted as red points.  The Bruzual \&
Charlot single-burst model is overplotted as the solid line, and the
Huang et al. (2008) Sbc and M82 models are overplotted as the dotted
and dash dotted lines, respectively.  In general, the non-star forming galaxies follow the single-burst model 
well at all redshifts.  There are a handful non-star forming galaxies which
appear to have some excess 8.0$\micron$ emission, and this may
be from either low-level  star formation or a low-luminosity
AGN.  
\newline\indent
There are fewer star forming than non-star forming galaxies in
the sample; however, their colors follow the Sbc and M82 models quite well.  At
8.0$\micron$, where the colors of the Sbc
and M82 models are most
different from the single-burst model, it is clear that galaxies with
emission lines have colors similar to those models, whereas those
without tend to follow the single-burst model.  Half of the star-forming
galaxies in Figure 24 (8/16) come from our spectroscopy of FLS
J172449+5921.3 (cluster \#10, $z =$ 0.252).  These galaxies were
selected for spectroscopy because they were detected at 24$\micron$.  Interestingly,
most of these galaxies (7/8) have a 3.6$\micron$ - 8.0$\micron$ color similar to
the Sbc model, yet they show a wide range in R - 3.6$\micron$
color.  A few have an R - 3.6$\micron$ color bluer than the
red sequence, typical of Sbc galaxies, whereas others have an R -
3.6$\micron$ color redder than the red sequence.  This illustrates
that there are both ``red'' and ``blue'' dusty star forming galaxies in
clusters, and that our approach of modeling the 5.8$\micron$ and
8.0$\micron$ LFs with populations of both is reasonable.
Furthermore, the fact that these are some of the brightest MIR
sources in the cluster field, and that most have colors similar to the
Sbc model, rather than the M82 model, is consistent with our conclusion 
that the 8.0$\micron$ LF at this redshift is best modeled using
the quiescent+regular model, with no need for a luminous dusty
starburst component. We defer a more detailed discussion of the
spectroscopy, including quantitative measurements of
star formation from line widths to a future paper
(Muzzin et al. 2008, in preparation).
\newline\indent
Overall, Figure 24 demonstrates that
the galaxy templates used to model the cluster LFs agree well with the colors of
spectroscopically confirmed cluster galaxies, and that they
are reasonable descriptions of star forming and non-star forming
galaxies between 0 $< z <$ 1.
\begin{figure*}
\plotone{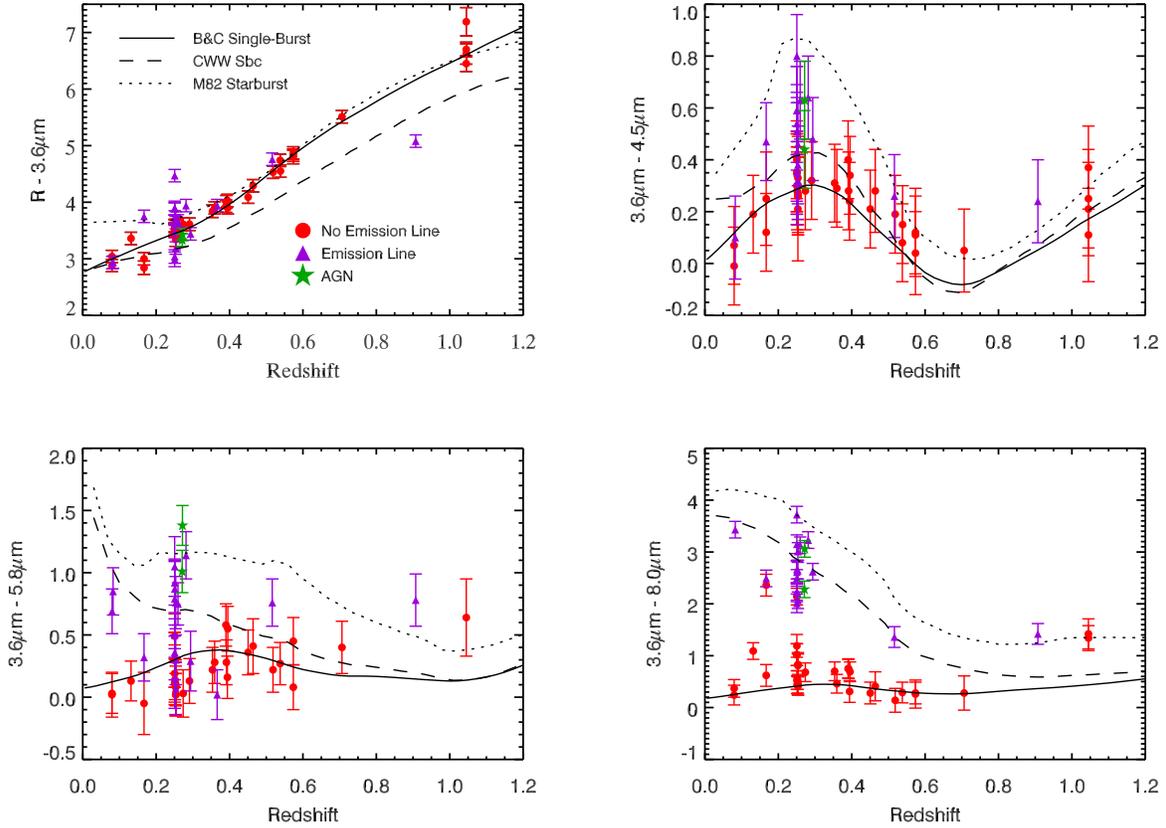}
\caption{\footnotesize Plot of optical - IRAC or IRAC - IRAC colors of
galaxies as a function of redshift.  The red, purple, and green points are spectroscopic
cluster members classified as non-star forming, star forming, and AGN
respectively.  The solid, dotted, and dashed lines are the model colors from the passive evolution
model, the Sbc model, and the M82 model respectively.}
\end{figure*}
\subsection{Systematic Uncertainties}
The data presented in this paper support a self-consistent
model of the evolution of stellar mass assembly and
dusty star formation in clusters; however, there are several details of this
analysis that have not been discussed and could potentially result in 
inappropriate conclusions being drawn from the data.  Although it is
difficult to quantify what effect, if any, these details will have on the
interpretation of the data, we believe it is important to at least
note these issues here.
\newline\indent
One worthwhile concern is the sample of clusters used in
the analysis.  Although this sample is much larger than the mere handful of clusters that have
been studied in the MIR thus far, it is still of modest size and
subject to cosmic variance.  In particular, given that the clusters
come from only 3.8 deg$^2$, it is
unclear whether the higher redshift clusters in the sample are
truly the progenitors of the lower redshift clusters.  
Unfortunately, a cosmologically significant sample of 
clusters covering of the order 100 degree$^2$ or more is likely needed
to avoid biases that might result from cosmic variance in the sample.
\newline\indent
Another potential problem is that  there are many
more low richness clusters in the sample than high richness clusters,
simply because of the nature of the cluster mass function.
Any effects that depend on cluster mass will clearly be missed by
combining these samples.  This could be important because processes
that could quench star formation (e.g., ram-pressure stripping, gas
strangulation) or incite starbursts (tidal effects, harassment) will
likely depend on cluster mass.  Using a much larger sample which can be
separated by both mass and redshift would be invaluable for studying
this issue further.
\newline\indent
Perhaps the most important concern is that there is a degeneracy between the intensity
of star formation in clusters and the fraction of star forming
galaxies.  We showed in $\S$6.3 that the color models used for the cluster
galaxies reproduce the colors of cluster galaxies with spectroscopic
redshifts very well; however, even though these colors are correct, the models of the
5.8$\micron$ and 8.0$\micron$ LFs still depend on the assumed
f$_{b}$ as a function of magnitude and redshift for the clusters.  If
the f$_{b}$ values are overestimated and need to be reduced, then a larger fraction of
dusty starburst galaxies than we have assumed will be required to
correctly model the cluster
5.8$\micron$ and 8.0$\micron$ LFs.  Likewise, if the f$_{b}$ is
underestimated, fewer dusty starbursts will be required.  The assumed f$_{b}$ are
consistent with most previous studies; however, optimally, if
more data were available the f$_{b}$ should be calculated from the
clusters themselves and this would avoid this degeneracy.
\newline\indent
Lastly, it is worth mentioning that much of the excess seen in
the 8.0$\micron$ LFs is near the limiting magnitude of the survey.
Problems with the background estimation could artificially inflate
these values.  It is unlikely that this is the case because if the excess of galaxies near the faint limit of
the survey were due to an undersubtraction of the background, it should
also be seen in the lower redshift LFs, which it is not.  Furthermore, 
undersubtraction of the background should be even more prevalent in the
lower redshift LFs because clusters have much larger angular sizes and therefore more
total area from which to undersubtract the background.  It is unlikely
that this is a problem; however, deeper data would be useful in
ensuring there are no errors due to completeness near the survey limit.
\section{Conclusions}
We have presented a catalogue of 99 candidate clusters and groups at 0.1 $< z_{phot} <$
1.3 discovered in
the $Spitzer$ First Look Survey using the cluster red sequence
technique.  Using spectroscopic redshifts from FLS followup campaigns
and our own spectroscopic followup of clusters we have shown that the
R - 3.6$\micron$ color of the
cluster red sequence is an accurate photometric redshift estimator at
the $\Delta$z = 0.04 level at $z <$ 1.0.  Furthermore, we demonstrated
that the properties of the FLS cluster
catalogue are similar to previous cluster surveys such as the RCS-1.  Using this cluster sample we studied the evolution of the cluster
3.6$\micron$, 4.5$\micron$, 5.8$\micron$, and 8.0$\micron$
LFs.  The main results from these LFs can be summarized as follows:
\begin{itemize}
\item In agreement with previous work, the evolution of the 3.6$\micron$
  and 4.5$\micron$ LFs between 0.1 $< z <$ 1.0 is
  consistent with a passively evolving population of galaxies formed
  in a single-burst at $z >$ 1.5.
  Given that the 3.6$\micron$ and 4.5$\micron$ bandpasses are
  reasonable proxies for stellar mass, this suggests that the majority
  of stellar mass in clusters is already assembled into massive galaxies by
  $z \sim$ 1.
\item The MIR color cuts used to select AGN by Lacy et al. (2004) and
  Stern et al. (2005) suggest that the fraction of cluster galaxies
  that host MIR-bright AGN at $z <$ 0.7 is low.  We estimate that the AGN fraction of
  cluster galaxies detected at 3.6$\micron$ is 1$^{+1}_{-1}$\%.  AGN
  are a larger, but still modest component of the 5.8$\micron$ and 8.0$\micron$ cluster
  population,  approximately
  3$^{+3}_{-3}$\% of these galaxies.
\item The cluster 5.8$\micron$ and 8.0$\micron$ LFs do not look
  similar to the 3.6$\micron$ and 4.5$\micron$ LFs, and this is
  due to the presence of  the cluster star forming galaxies. Star forming galaxies are much
  brighter in these bandpasses than early type galaxies and their
  varying fractions with redshift cause deviations from the shape of
  the 3.6$\micron$ and 4.5$\micron$ LFs.  The
  5.8$\micron$ and 8.0$\micron$ LFs are well-described using different
  fractions of three basic types of galaxies: quiescent, regular
  star forming, and dusty starburst by assuming that the fractions of
  the latter two are proportional to the cluster f$_{b}$.
\item The 8.0$\micron$ cluster LFs suggest that both the frequency and SSFR
  of star forming cluster galaxies is increasing with increasing redshift.  In
  particular it appears that when compared to star forming galaxies in
  the local universe, the
  intensity of star formation in clusters evolves from
  ``weak'' to ``regular'' to ``starburst'' with increasing
  redshift.  Qualitatively, this evolution mimics the evolution in the
  universal star formation density with redshift suggesting that
  this evolution is at least in part caused by the accretion of
  star forming galaxies into the cluster environment.
\item Comparing the 3.6$\micron$ and 4.5$\micron$ cluster and
  field LFs with similar normalization shows that the LFs in these
  environments are similar, with evidence for a small excess in the
  brightest galaxies in clusters, likely caused by the cluster giant ellipticals.
  In agreement with previous K-band studies this suggests that the distribution
  of galaxies as a function of stellar mass in both environments is roughly equivalent.
\item There is a significant differential evolution in the cluster and
  field 5.8$\micron$ and 8.0$\micron$ LFs with redshift.  At $z =$
  0.65 the cluster is more abundant in 8.0$\micron$ galaxies than
  the field; however, thereafter the
  relative number of 5.8$\micron$ and 8.0$\micron$ galaxies
  declines in clusters with decreasing redshift and by $z =$ 0.15 the
  cluster is underdense in these sources by roughly a factor of 5.
  This differential evolution could be explained if starbursts are preferentially triggered during
  the early formation stages of the cluster but then preferentially
  quenched thereafter by the high density environment.
\end{itemize}
\indent
A well-sampled spectroscopic study of several
  high-redshift clusters with MIR data would be extremely valuable
  for verifying our interpretation of the IRAC cluster LFs
  because it is always difficult to draw incontrovertible
  conclusions from LFs alone.  Still, the cluster LFs do show a strong increase in the
  number of 5.8$\micron$ and 8.0$\micron$ sources in clusters with
  increasing redshift which must almost certainly be attributed to
  increased amounts of dusty star formation in higher redshift clusters.
\newline\indent
One of the strengths of this analysis is that it is based on a
relatively large sample of galaxy clusters.  It has become
clear from the handful of clusters studied thus far by $ISO$ and $Spitzer$
that the MIR properties of cluster galaxies can be quite different from cluster to
cluster.  They may depend  on dynamical state, mass, f$_{b}$,
or other parameters (e.g., Coia et al. 2005; Geach et al. 2006).  The advantage of using many
clusters is that it provides a metric of how the ``average''
cluster is evolving as a function of redshift.  Detailed studies of individual
clusters with significant ancillary data will pave the way to a better
understanding of the physics behind the evolution of dusty star formation in
cluster galaxies; however, large statistical studies such as this one will indicate
whether the clusters studied in future work are representative of
the cluster population as a whole, or are potentially rare, biased
clusters with unusual properties caused by an ongoing merger or some other event.  
\newline\indent
It is worth noting that although the quality of the LFs
provided by the 99 clusters in the FLS is good, these LFs
would still
benefit from a larger statistical sample.  In particular, a larger
sample would allow for the separation of clusters by other
properties such as mass or morphology, and to understand if these
properties play a role in shaping the MIR cluster galaxy population.  We are currently
working on a survey to detect clusters in the much larger SWIRE survey: the
Spitzer Adaptation of the Red sequence Cluster Survey (SpARCS).  This
project has 13 times more area than the FLS and is a factor of 2 deeper in
integration time in the IRAC bands.  The analysis of that sample should provide a
significant improvement in the quality of the cluster LFs.
\acknowledgements
We thank the anonymous referee whose comments improved this manuscript significantly.
We would like to thank David Gilbank, Thomas Babbedge, Roberto De Propris, and
Stefano Andreon for graciously making their data available to us.
We thank David Gilbank for useful conversations which
helped improve the clarity of this analysis.  We also thank Dario
Fadda for recomputing the FLS R-band photometry using different apertures.
A.M. acknowledges support from the $Spitzer$ Visiting Graduate Student
Program during which much of this work was completed.  A.M. also
acknowledges support from the National Sciences and Engineering
Research Council (NSERC) in the form of PGS-A and PGSD2
fellowships. 
The work of H.K.C.Y. is supported by grants from the Canada Research
Chair Program, NSERC, and the University of Toronto.    This work is based in part on observations made with the Spitzer Space
Telescope, which is operated by the Jet Propulsion Laboratory, California
Institute of Technology under a contract with NASA.

\LongTables
\begin{deluxetable}{lcccrrrrcrc}
\tabletypesize{\footnotesize}
\scriptsize
\tablecolumns{11}
\tablecaption{FLS Cluster Catalogue}
\tablewidth{7.4in}
\tablehead{\colhead{\#} & \colhead{Name} & \colhead{ $z_{phot}$ } &  \colhead{ $z_{spec}$ } &
\colhead{R.A.} & \colhead{Decl.} & \colhead{B$_{gc,R}$} &
\colhead{$\epsilon$B$_{gc,R}$} & \colhead{M$_{200}$} & \colhead{ R$_{200}$} & \colhead{Centroid}
\\ 
\colhead{} & \colhead{} & \colhead{} & \colhead{} & \colhead{J2000} &
\colhead{J2000} & \colhead{Mpc$^{1.8}$} & \colhead{Mpc$^{1.8}$} &
\colhead{M$_{\odot}$ $\times$ 10$^{14}$} &   \colhead{Mpc} &  
  \colhead{}  \\
\colhead{(1)}& \colhead{(2)}& \colhead{(3)}& \colhead{(4)}&
\colhead{(5)}& \colhead{(6)} &
\colhead{(7)}& \colhead{(8)}& \colhead{(9)}& \colhead{(10)}& \colhead{(11)}
}
\startdata
0 & FLS J172321+5835.0 & 0.09 &  0.079(4)  & 17:23:21.5 & 58:35:03.5 & 237 & 133 & 
0.51 & 0.68 & BCG \\
1 & FLS J171059+5934.2 & 0.13 &  0.126(10)  & 17:10:59.8 & 59:34:16.4 & 521 & 196 & 
1.82 & 1.04 & BCG \\
2 & FLS J171639+5915.2 & 0.16 &  0.129(7)  & 17:16:39.3 & 59:15:13.5 & 326 & 155 & 
0.85 & 0.81 & BCG \\
3 & FLS J172319+6019.5 & 0.18 &  0.131(1)  & 17:23:19.7 & 60:19:33.7 & 358 & 162 & 
0.99 & 0.85 & BCG \\
4 & FLS J171233+5956.4 & 0.22 &  ---  & 17:12:33.0 & 59:56:28.2 & 534 & 199 & 
1.89 & 1.06 & RS-Flux \\
5 & FLS J172207+5943.8 & 0.24 &  0.271(2)  & 17:22:07.9 & 59:43:52.1 & 251 & 132 & 
0.55 & 0.71 & RS-Flux \\
6 & FLS J172618+5934.5 & 0.27 &  ---  & 17:26:18.8 & 59:34:32.3 & 386 & 168 & 
1.12 & 0.89 & BCG \\
7 & FLS J171618+5907.8 & 0.27 &  0.251(1)  & 17:16:18.5 & 59:07:53.0 & 251 & 132 & 
0.56 & 0.71 & RS-Flux \\
8 & FLS J171505+5859.6 & 0.29 &  0.252(9)  & 17:15:05.2 & 58:59:41.4 & 310 & 149 & 
0.78 & 0.79 & BCG \\
9 & FLS J171152+6007.7 & 0.29 &  0.293(1)  & 17:11:52.8 & 60:07:43.7 & 381 & 166 & 
1.09 & 0.88 & RS-Flux \\
10 & FLS J172449+5921.3 & 0.29 &  0.253(9)  & 17:24:49.0 & 59:21:22.9 & 861 & 252
 & 4.11 & 1.36 & BCG \\
11 & FLS J172454+5930.5 & 0.29 &  0.273(1)  & 17:24:54.4 & 59:30:32.8 & 447 & 181
 & 1.42 & 0.96 & BCG \\
12 & FLS J171431+5957.8 & 0.29 &  ---  & 17:14:31.1 & 59:57:52.2 & 378 & 165
 & 1.08 & 0.88 & RS-Flux \\
13 & FLS J171455+5836.5 & 0.30 &  0.291(1)  & 17:14:55.0 & 58:36:34.7 & 791 & 242
 & 3.58 & 1.30 & BCG \\
14 & FLS J172505+5932.3 & 0.34 &  ---  & 17:25:05.8 & 59:32:22.9 & 516 & 195
 & 1.79 & 1.04 & BCG \\
15 & FLS J172008+5949.9 & 0.36 &  0.359(2)  & 17:20:08.7 & 59:49:54.1 & 308 & 148
 & 0.77 & 0.79 & BCG \\
16 & FLS J171241+5855.9 & 0.38 &  0.390(2)  & 17:12:41.6 & 58:55:58.7 & 797 & 243
 & 3.63 & 1.31 & RS-Flux \\
17 & FLS J171537+5849.4 & 0.38 &  0.353(1)  & 17:15:37.0 & 58:49:24.4 & 590 & 209
 & 2.23 & 1.11 & RS-Flux \\
18 & FLS J172541+5929.9 & 0.38 &  0.366(1)  & 17:25:41.7 & 59:29:59.4 & 521 & 196
 & 1.82 & 1.04 & BCG \\
19 & FLS J171720+5920.0 & 0.39 &  0.395(1)  & 17:17:20.3 & 59:20:05.9 & 316 & 150
 & 0.81 & 0.80 & BCG \\
20 & FLS J171204+5855.6 & 0.41 &  ---  & 17:12:04.7 & 58:55:36.1 & 248 & 131
 & 0.54 & 0.70 & BCG \\
21 & FLS J172013+5925.4 & 0.41 &  ---  & 17:20:13.1 & 59:25:29.6 & 456 & 183
 & 1.47 & 0.97 & BCG \\
22 & FLS J171432+5915.9 & 0.41 &  0.394(1)  & 17:14:32.6 & 59:15:54.7 & 525 & 197
 & 1.84 & 1.05 & BCG \\
23 & FLS J171437+6002.8 & 0.42 &  ---  & 17:14:37.8 & 60:02:53.5 & 319 & 151
 & 0.82 & 0.80 & BCG \\
24 & FLS J172028+5922.6 & 0.42 &  0.281(1)  & 17:20:28.9 & 59:22:38.8 & 457 & 183
 & 1.47 & 0.97 & BCG \\
25 & FLS J172546+6011.5 & 0.43 &  0.450(1)  & 17:25:46.3 & 60:11:30.2 & 872 & 253
 & 4.20 & 1.37 & RS-Flux \\
26 & FLS J172026+5916.0 & 0.43 &  0.462(2)  & 17:20:26.9 & 59:16:05.0 & 804 & 243
 & 3.68 & 1.31 & RS-Flux \\
27 & FLS J171103+5839.9 & 0.43 &  ---  & 17:11:03.4 & 58:39:56.3 & 528 & 197
 & 1.86 & 1.05 & RS-Flux \\
28 & FLS J172418+5954.6 & 0.44 &  ---  & 17:24:18.5 & 59:54:37.4 & 391 & 169
 & 1.14 & 0.89 & BCG \\
29 & FLS J172158+6014.3 & 0.45 &  ---  & 17:21:58.3 & 60:14:20.2 & 323 & 152
 & 0.84 & 0.81 & BCG \\
30 & FLS J171153+5905.4 & 0.45 &  ---  & 17:11:53.6 & 59:05:28.2 & 530 & 198
 & 1.87 & 1.05 & RS-Flux \\
31 & FLS J171447+6018.9 & 0.48 &  0.464(1)  & 17:14:47.5 & 60:18:54.7 & 255 & 134
 & 0.57 & 0.71 & RS-Flux \\
32 & FLS J172540+5909.5 & 0.48 &  ---  & 17:25:40.5 & 59:09:34.5 & 600 & 211
 & 2.29 & 1.12 & RS-Flux \\
33 & FLS J172109+5939.2 & 0.49 &  ---  & 17:21:09.1 & 59:39:15.5 & 878 & 254
 & 4.24 & 1.38 & BCG \\
34 & FLS J172513+5923.6 & 0.49 &  0.518(1)  & 17:25:13.1 & 59:23:36.6 & 807 & 244
 & 3.70 & 1.32 & RS-Flux \\
35 & FLS J172142+5921.8 & 0.52 &  0.538(1)  & 17:21:42.9 & 59:21:49.1 & 597 & 210
 & 2.27 & 1.12 & BCG \\
36 & FLS J172342+5941.0 & 0.52 &  ---  & 17:23:42.2 & 59:41:03.5 & 320 & 152
 & 0.82 & 0.80 & BCG \\
37 & FLS J171622+5915.5 & 0.53 &  ---  & 17:16:22.7 & 59:15:30.7 & 250 & 132
 & 0.55 & 0.70 & BCG \\
38 & FLS J172122+5922.7 & 0.53 &  0.538(2)  & 17:21:22.0 & 59:22:46.3 & 1287 & 306
 & 7.89 & 1.69 & BCG \\
39 & FLS J172339+5937.2 & 0.53 &  ---  & 17:23:39.5 & 59:37:12.5 & 318 & 151
 & 0.82 & 0.80 & BCG \\
40 & FLS J171459+6016.7 & 0.55 &  ---  & 17:14:59.9 & 60:16:44.5 & 730 & 232
 & 3.15 & 1.25 & BCG \\
41 & FLS J171300+5919.4 & 0.55 &  ---  & 17:13:00.2 & 59:19:28.0 & 591 & 209
 & 2.23 & 1.11 & RS-Flux \\
42 & FLS J172228+6013.4 & 0.55 &  ---  & 17:22:28.8 & 60:13:24.2 & 453 & 182
 & 1.45 & 0.97 & RS-Flux \\
43 & FLS J171405+5900.6 & 0.55 &  0.516(1)  & 17:14:05.0 & 59:00:41.7 & 454 & 183
 & 1.46 & 0.97 & BCG \\
44 & FLS J171648+5838.6 & 0.56 &  0.573(2)  & 17:16:48.2 & 58:38:37.7 & 2040 & 383
 & 16.6 & 2.15 & BCG \\
45 & FLS J172037+5853.4 & 0.58 &  ---  & 17:20:37.2 & 58:53:26.2 & 518 & 195
 & 1.80 & 1.04 & RS-Flux \\
46 & FLS J171227+5904.8 & 0.61 &  ---  & 17:12:27.6 & 59:04:53.8 & 650 & 219
 & 2.61 & 1.17 & RS-Flux \\
47 & FLS J171452+5917.2 & 0.61 &  ---  & 17:14:52.8 & 59:17:12.9 & 719 & 230
 & 3.07 & 1.24 & BCG \\
48 & FLS J171104+5858.5 & 0.61 &  ---  & 17:11:04.6 & 58:58:32.7 & 926 & 261
 & 4.63 & 1.42 & BCG \\
49 & FLS J171634+6009.2 & 0.62 &  ---  & 17:16:34.4 & 60:09:15.2 & 621 & 214
 & 2.42 & 1.14 & BCG \\
50 & FLS J171420+6005.5 & 0.63 &  ---  & 17:14:20.1 & 60:05:35.3 & 1131 & 288
 & 6.40 & 1.57 & BCG \\
51 & FLS J171654+6004.8 & 0.63 &  ---  & 17:16:54.0 & 60:04:48.0 & 510 & 194
 & 1.76 & 1.03 & RS-Flux \\
52 & FLS J171628+5836.6 & 0.66 &  ---  & 17:16:28.5 & 58:36:40.9 & 229 & 124
 & 0.48 & 0.67 & BCG \\
53 & FLS J171523+5858.7 & 0.68 &  ---  & 17:15:23.9 & 58:58:47.2 & 916 & 260
 & 4.55 & 1.41 & BCG \\
54 & FLS J171633+5920.9 & 0.68 &  ---  & 17:16:33.9 & 59:20:54.6 & 292 & 143
 & 0.71 & 0.77 & BCG \\
55 & FLS J172601+5945.7 & 0.69 &  ---  & 17:26:01.1 & 59:45:47.0 & 637 & 217
 & 2.52 & 1.16 & BCG \\
56 & FLS J172013+5845.4 & 0.69 &  ---  & 17:20:13.0 & 58:45:26.9 & 1051 & 278
 & 5.68 & 1.51 & BCG \\
57 & FLS J171836+6006.7 & 0.70 &  ---  & 17:18:36.7 & 60:06:43.3 & 430 & 177
 & 1.33 & 0.94 & BCG \\
58 & FLS J171903+5851.8 & 0.70 &  ---  & 17:19:03.7 & 58:51:51.1 & 430 & 177
 & 1.33 & 0.94 & RS-Flux \\
59 & FLS J172246+5843.7 & 0.71 &  ---  & 17:22:46.3 & 58:43:43.3 & 429 & 176
 & 1.33 & 0.94 & BCG \\
60 & FLS J171703+5857.9 & 0.72 &  ---  & 17:17:03.5 & 58:57:57.8 & 912 & 259
 & 4.51 & 1.40 & BCG \\
61 & FLS J172431+5928.3 & 0.72 &  ---  & 17:24:31.8 & 59:28:23.4 & 220 & 120
 & 0.44 & 0.66 & BCG \\
62 & FLS J171203+6006.6 & 0.73 &  ---  & 17:12:03.9 & 60:06:38.8 & 289 & 142
 & 0.70 & 0.76 & RS-Flux \\
63 & FLS J171430+5901.7 & 0.73 &  ---  & 17:14:30.1 & 59:01:47.1 & 978 & 269
 & 5.06 & 1.46 & RS-Flux \\
64 & FLS J171834+5844.6 & 0.73 &  ---  & 17:18:34.1 & 58:44:39.4 & 359 & 160
 & 0.99 & 0.85 & BCG \\
65 & FLS J172009+6008.0 & 0.73 &  0.706(1)  & 17:20:09.7 & 60:08:02.6 & 426 & 176
 & 1.31 & 0.94 & BCG \\
66 & FLS J172319+5922.2 & 0.73 &  ---  & 17:23:19.5 & 59:22:15.9 & 356 & 159
 & 0.98 & 0.85 & BCG \\
67 & FLS J172525+5924.7 & 0.74 &  ---  & 17:25:25.8 & 59:24:46.4 & 633 & 216
 & 2.49 & 1.16 & RS-Flux \\
68 & FLS J171508+5845.4 & 0.75 &  ---  & 17:15:08.8 & 58:45:27.2 & 1116 & 287
 & 6.27 & 1.56 & BCG \\
69 & FLS J172148+6016.1 & 0.77 &  0.907(1)  & 17:21:48.5 & 60:16:07.7 & 774 & 239
 & 3.46 & 1.29 & BCG \\
70 & FLS J171454+5958.3 & 0.77 &  ---  & 17:14:54.6 & 59:58:18.4 & 360 & 160
 & 1.00 & 0.86 & BCG \\
71 & FLS J171511+6028.0 & 0.77 &  ---  & 17:15:11.3 & 60:28:01.4 & 704 & 228
 & 2.97 & 1.22 & RS-Flux \\
72 & FLS J172012+5958.3 & 0.78 &  ---  & 17:20:12.7 & 59:58:19.9 & 705 & 228
 & 2.97 & 1.22 & RS-Flux \\
73 & FLS J172209+5935.2 & 0.78 &  ---  & 17:22:09.4 & 59:35:16.6 & 360 & 160
 & 1.00 & 0.86 & RS-Flux \\
74 & FLS J172035+5928.6 & 0.78 &  ---  & 17:20:35.5 & 59:28:40.4 & 428 & 176
 & 1.32 & 0.94 & BCG \\
75 & FLS J171411+6027.7 & 0.78 &  ---  & 17:14:11.7 & 60:27:44.3 & 705 & 228
 & 2.97 & 1.22 & BCG \\
76 & FLS J171545+5853.8 & 0.78 &  ---  & 17:15:45.9 & 58:53:48.6 & 291 & 142
 & 0.71 & 0.76 & BCG \\
77 & FLS J171556+5859.9 & 0.79 &  ---  & 17:15:56.1 & 58:59:54.3 & 636 & 217
 & 2.52 & 1.16 & BCG \\
78 & FLS J171932+5929.3 & 0.79 &  ---  & 17:19:32.0 & 59:29:18.5 & 499 & 191
 & 1.70 & 1.02 & RS-Flux \\
79 & FLS J172019+5926.6 & 0.79 &  ---  & 17:20:19.8 & 59:26:41.4 & 291 & 142
 & 0.71 & 0.76 & RS-Flux \\
80 & FLS J171828+5836.2 & 0.79 &  ---  & 17:18:28.7 & 58:36:13.8 & 498 & 191
 & 1.69 & 1.02 & RS-Flux \\
81 & FLS J172304+5832.3 & 0.81 &  ---  & 17:23:04.5 & 58:32:18.6 & 363 & 161
 & 1.01 & 0.86 & BCG \\
82 & FLS J171657+6004.8 & 0.82 &  ---  & 17:16:57.9 & 60:04:49.3 & 711 & 229
 & 3.02 & 1.23 & BCG \\
83 & FLS J171945+5909.1 & 0.84 &  ---  & 17:19:45.4 & 59:09:09.1 & 507 & 193
 & 1.74 & 1.03 & RS-Flux \\
84 & FLS J171155+6013.1 & 0.90 &  ---  & 17:11:55.1 & 60:13:08.5 & 522 & 196
 & 1.82 & 1.04 & RS-Flux \\
85 & FLS J171808+5915.8 & 0.91 &  ---  & 17:18:08.7 & 59:15:50.7 & 387 & 168
 & 1.12 & 0.89 & RS-Flux \\
86 & FLS J171223+6015.1 & 0.95 &  ---  & 17:12:23.7 & 60:15:09.4 & 605 & 211
 & 2.32 & 1.13 & RS-Flux \\
87 & FLS J171051+5930.8 & 1.02 &  ---  & 17:10:51.8 & 59:30:50.5 & 760 & 237
 & 3.36 & 1.27 & BCG \\
88 & FLS J172147+6011.5 & 1.02 &  ---  & 17:21:47.3 & 60:11:35.7 & 277 & 141
 & 0.65 & 0.74 & BCG \\
89 & FLS J171852+6009.9 & 1.02 &  ---  & 17:18:52.7 & 60:09:56.9 & 485 & 189
 & 1.62 & 1.00 & RS-Flux \\
90 & FLS J171221+6010.6 & 1.03 &  ---  & 17:12:21.2 & 60:10:41.0 & 349 & 160
 & 0.95 & 0.84 & BCG \\
91 & FLS J171431+5946.9 & 1.06 &  ---  & 17:14:31.9 & 59:46:59.5 & 425 & 177
 & 1.31 & 0.94 & RS-Flux \\
92 & FLS J171117+5902.8 & 1.06 &  ---  & 17:11:17.5 & 59:02:48.6 & 287 & 144
 & 0.69 & 0.76 & BCG \\
93 & FLS J172126+5856.6 & 1.11 &  1.045(4)  & 17:21:26.4 & 58:56:41.7 & 646 & 218
 & 2.58 & 1.17 & BCG \\
94 & FLS J171227+6015.2 & 1.14 &  ---  & 17:12:27.0 & 60:15:16.7 & 448 & 182
 & 1.43 & 0.96 & BCG \\
95 & FLS J172045+5834.8 & 1.17 &  ---  & 17:20:45.3 & 58:34:50.9 & 386 & 169
 & 1.12 & 0.89 & RS-Flux \\
96 & FLS J172113+5901.0 & 1.24 &  ---  & 17:21:13.8 & 59:01:05.7 & 338 & 158
 & 0.90 & 0.83 & RS-Flux \\
97 & FLS J171223+6006.9 & 1.27 &  ---  & 17:12:23.6 & 60:06:56.4 & 208 & 124
 & 0.41 & 0.64 & RS-Flux \\
98 & FLS J171942+5938.3 & 1.38 &  ---  & 17:19:42.8 & 59:38:23.2 & 374 & 165
 & 1.06 & 0.87 & RS-Flux \\
\enddata
\tablecomments{(3) Photometric redshift estimated from red sequence
  color, (4) Mean spectroscopic redshift of galaxies with red sequence
  weights $>$ 0.2, the number of spectroscopic redshifts is included in
  brackets, (7) Cluster richness parameterized by B$_{gc,R}$, (8)
  Error in B$_{gc,R}$, (9) M$_{200}$ estimated from B$_{gc,R}$ using
  eq. 1.  (10) R$_{200}$ estimated from B$_{gc,R}$ using eq. 2., (11)
  Best centroid of the cluster.
}
\end{deluxetable}

\begin{deluxetable}{cc}
\tabletypesize{\footnotesize}
\scriptsize
\tablecolumns{2}
\tablecaption{Assumed Blue Fractions}
\tablewidth{1.3in}
\tablehead{\colhead{$z$} & \colhead{ F$_{b}$ (M $<$ M$^{*}$)} 
\\ 
\colhead{(1)}& \colhead{(2)}
}
\startdata
0.15 & 0.05\\
0.25 & 0.15\\
0.33 & 0.20\\
0.44 & 0.25\\
0.55 & 0.30\\
0.65 & 0.40\\
0.76 & 0.50\\
0.82 & 0.50\\
1.01 & 0.60\\
1.21 & 0.60
\enddata
\end{deluxetable}

\end{document}